\documentstyle[rmp,aps,floats,amssymb,graphicx]{revtex}

\makeatletter
\def\p@subsection{}
\def\p@subsubsection{}
\makeatother

\newcommand{\Hethree}{$^3$He}
\newcommand{\s}{\mbox{\bf s}}
\def\v{\mbox{\bf v}}
\def\r{\mbox{\bf r}}

\begin{document}


\title{VORTEX FORMATION AND DYNAMICS IN SUPERFLUID $^3$He\\
AND ANALOGIES IN QUANTUM FIELD THEORY}




\author{V.B.~Eltsov}
\address{Low Temperature Laboratory, Helsinki
  University of Technology, Box 2200, FIN-02015 HUT, Finland\\
and P.L.~Kapitza Institute for Physical Problems, 119334 Moscow,
  Russia}
\author{M.~Krusius}
\address{Low Temperature Laboratory, Helsinki
  University of Technology, Box 2200, FIN-02015 HUT, Finland}
\author{G.E.~Volovik}
\address{Low Temperature Laboratory, Helsinki
  University of Technology, Box 2200, FIN-02015 HUT, Finland\\
and L.D.~Landau Institute for Theoretical Physics, 119334 Moscow,
  Russia
}
\date{\today}

\maketitle

\begin{abstract}

  The formation and dynamics of topological defects of different structure
  has been of central interest in the study of the $^3$He superfluids.
  Compared to superfluid $^4$He-II, the variability of the important
  parameters with temperature and pressure is wider and many features are more ideal. This
  has made experimentally new approaches possible. An example is the formation of quantized
  vortices and other topological defects in a rapid time-dependent phase
  transition from the normal state to $^3$He-B. This vortex formation process is known as the
  Kibble-Zurek mechanism and is one the central topics of this review. It demonstrates the use
  $^3$He-B as a model system for quantum fields, with detailed control from the laboratory.

\vspace{5mm}


\end{abstract}

\newpage
\twocolumn

\tableofcontents
\newpage

\section{Superfluid \Hethree\ and quantum field theory}

In recent times condensed matter and elementary particle physics have been
experiencing remarkable convergence in their developments, as many-body
aspects have become increasingly more important in particle physics: The
Early Universe is the ultimate application field for the theories on
interacting particle systems which can be worked out for different energy
regimes, or epochs of the expansion after the Big Bang.  Compact
astro-physical objects provide other ``laboratories'' with a narrower range
of conditions in which to test theories.  Actual collider experiments are
set up to study interacting particle systems, such as the quark-gluon
plasma in heavy ion collisions or the pion condensate.

Collective phenomena in interacting many-body systems is what condensed matter
physics is about, but elementary particle systems are adding to the picture extreme
quantum behavior plus relativistic motion. So far systems, where all of these
features would be of importance, are not available for direct observation. During
the nineties it turned out that valuable analogues can nevertheless be
constructed from comparisons with non-relativistic many-body quantum systems of
condensed matter, such as superconductors and helium superfluids. More recently optically
cooled alkali atom clouds have been added to this list. Here we limit our discussion
to the fermionic $^3$He superfluids.

The liquid $^3$He phases provide attractive advantages as model systems for
the study of various general concepts in quantum field theory: The liquid
is composed of neutral particles in an inherently isotropic environment ---
no complications arise from electrical charges or a symmetry constrained by
a crystalline lattice. This allows one to concentrate on the consequences
from a most complex symmetry breaking, which gives rise to a
multi-dimensional order parameter space, but is well described by a
detailed microscopic theory. Experimentally superfluid $^3$He is
devoid of extrinsic imperfections. In fact, with respect to impurities and dirt it is  one of the purest of all condensed matter systems, excepting optically cooled atom clouds. The superfluid coherence length is large such that even surface
roughness can be reduced sufficiently to transform the container walls to
almost ideally behaving boundaries. There are topologically stable defects
of different dimensionality and type, which often can be detected with NMR
methods with single-defect sensitivity. The study of the different
intrinsic mechanisms, by which these defects are formed, provide one of the
important parallels to other systems. Phase transitions of both first and
second order exist, which can be utilized in the
investigation of defect formation.  Another example is the use of the
zero-temperature $^3$He systems to model the complicated physical vacuum of
quantum field theory --- the modern ether: The bosonic and fermionic
excitations in $^3$He (in the dilute limit) are in many respects similar to
the excitations of the physical vacuum --- the elementary particles. This
approach has been quite successful in constructing a physical picture of
the interactions of elementary particles with strings and domain walls, as
described in a recent monograph by one of the present authors
(Volovik 2003).

This review describes some examples from superfluid $^3$He research where
strong connections exist to quantum field theory. In fact, some of these
studies were only taken up because they can be used as laboratory models,
to answer the question whether a proposed principle is physically correct
or not. Our discussion has been split in two parts: The first part Chap.~\ref{ExpChap} deals
with a detailed description of defect formation and evolution in a rapid
non-equilibrium phase transition of second order, caused by the localized heating from an
absorption event in ionizing radiation. This phenomenon originally became interesting because
it can be thought to model phase transitions in the early Universe. Defect
formation in these transitions has been suggested as the origin for the
inhomogeneous mass distribution in the present Universe. The second part Chap.~\ref{SecOtherAnalogs}  discusses concepts from vortex dynamics which have been used in Chap.~\ref{ExpChap} to analyze vortex formation in rapid phase transitions. This discussion proceeds in more general terms and continues to highlight connections to quantum field theory.

\section{Defect formation in quench-cooled superfluid transition} \label{ExpChap}

\subsection{Non-equilibrium phase transitions} \label{Intro}

A rapid phase transition is generally associated with a large
degree of inhomogeneity.  After all, this is the process by which
materials like steel or amorphous solids are prepared. This
disorder we attribute to heterogeneous extrinsic influence which
is usually present in any system which we study: impurities, grain
boundaries, and other defects depending on the particular system.
To avoid disorder and domain formation, we generally examine phase
transitions in the adiabatic limit, as close to equilibrium as
possible.

But suppose we would have an ideally homogeneous infinite system with no boundaries. It is
rapidly cooling from a symmetric high temperature state to a low temperature phase
of lower symmetry, which we call the broken-symmetry phase, for example by uniform
expansion. What would happen in such a transition? Are defects also formed in this
case? Such a measurement of a homogeneous transition as a function of the transition
speed, is difficult. As we shall see, ultimately it also raises the question whether
the source of the precipitated inhomogeneity has been reliably identified.

\subsection{Cosmic large-scale structure} \label{L-S_Structure}

According to the standard theory of cosmology the Universe started off in the ``Big
Bang'' in a homogeneous state. It then rapidly cooled through a sequence of phase
transitions, in which the four fundamental forces of nature separated out. Today the
Universe exists in a state with inhomogeneous large-scale structure. The clumped
distribution of visible mass has become most evident from galaxy surveys --- maps
which show that galaxies form clusters and these in turn super clusters, such as the
``Great Wall'', which are the largest structures discovered to date (Geller and Huchra 1989).  The
clustering takes the form of long chains or filaments, which are separated by large
voids, regions empty of visible mass. Recent extended galaxy surveys indicate that
the length scale of large-scale structure is of order 100 Mpc\footnote{1 Mpc =
$10^6$ parsec = 3.262$\cdot 10^6$ light years.} (Peacock et~al.\ 2001).

Another image of large-scale structure has been preserved in the cosmic microwave
background radiation, from the time when the Universe had cooled to a few eV, when
nuclei and electrons combined to form atoms and the Universe became transparent to
photons. Since then the background radiation has cooled in the expanding Universe.
It now matches to within 3 parts in $10^5$ the spectrum of a black body at 2.728 K,
as measured eg. with the spectrometers on board of the satellite Cosmic Background
Explorer (COBE) in 1990. Later balloon-borne measurements examined the spatial
distribution of the residual anisotropy with high angular resolution and amplitude
sensitivity. The recent satellite Wilkinson Microwave Anisotropy Probe (WMAP) has
extended this work, mapping all-sky surveys with an angular resolution better than
$1^\circ$ on different frequency bands, to separate out the interfering signal from
our galaxy (Bennett et~al.\ 2003; Spergel et~al.\ 2003). Combined with other information, this anisotropy
of the cosmic microwave background radiation, 30\,$\mu$K in amplitude, which
represents the density fluctuations in the structure of the Universe when it was
only $300\,000$ years old, is expected to explain the mechanisms which originally
seeded and led to the formation of the large-scale structure, as we observe it
today.

\subsection{Kibble-Zurek mechanism} \label{KW_mechanism}

One of the early explanations for the origin of large-scale structure was
offered by Tom Kibble (1976). He suggested that the inhomogeneity
was created in rapid phase transitions of the early expanding Universe.
Even in a perfectly homogeneous transition of second order defects can be
expected to form, if the transition proceeds faster than the order
parameter of the broken-symmetry phase is able to relax. In such a
non-equilibrium transition the new low-temperature phase starts to form,
due to fluctuations of the order parameter, simultaneously and
independently in many parts of the system. Subsequently during further
cooling, these regions grow together to form the new broken-symmetry phase.
At the boundaries, where different causally disconnected regions meet, the
order parameter does not necessarily match and a domain structure is
formed.

If the broken symmetry is the U(1) gauge symmetry, then domains with different value
of the order parameter phase are formed. Such a random domain structure reduces to a
network of linear defects, which are vortex lines in superfluids and superconductors
or perhaps cosmic strings in the Early Universe. If the symmetry break is more
complicated, as is the case in $^3$He superfluids, then defects of different
dimensionality and structure may be formed.

Subsequent numerical simulations of rapidly cooled second order phase transitions
confirmed that defects were indeed forming (Vachaspati and Vilenkin 1984). In 1985 Wojciech
Zurek proposed a conceptually powerful phenomenological approach how to understand a
phase transition far out of equilibrium, when the outcome from the transition
becomes time dependent (Zurek 1985, 1996). He characterizes the transition speed
with a quench time
\begin{equation}
        \tau_{\rm Q} = \left( \frac{1}{T_{\rm c}}
          \left|{\frac{dT}{dt}}\right|_{T=T_{\rm c}} \right)^{-1}~~,
\label{QuenchTime}
\end{equation}
which allows him to approximate temperature and time with a linear
dependence during the thermal quench.

The quench time $\tau_{\rm Q}$ is compared to the order parameter
relaxation time $\tau$, which in a Ginzburg-Landau system at a second
order phase transition is of the general form
\begin{equation}
     \tau (T) = \tau_0 (1 - T/T_{\rm c})^{-1}~~.
\label{RelaxTime}
\end{equation}
In superfluid $^3$He, $\tau_0$ is on the order of
$\tau_0 \sim \xi_0/v_{\rm F}$, where $\xi_0$ is the zero temperature
limiting value of the temperature $(T)$ and pressure $(P)$ dependent
superfluid coherence length $\xi(T,P)$. Close to $T_{\rm c}$, it is of
the form $\xi(T,P) = \xi_0(P) (1-T/T_{\rm c})^{-1/2}$. The second
quantity, $v_{\rm F}$, is the velocity of the thermal quasiparticle
excitations which are excited above the superfluid energy gap.


\begin{figure}[!!!tb]
  \centerline{\includegraphics[width=0.90\columnwidth]{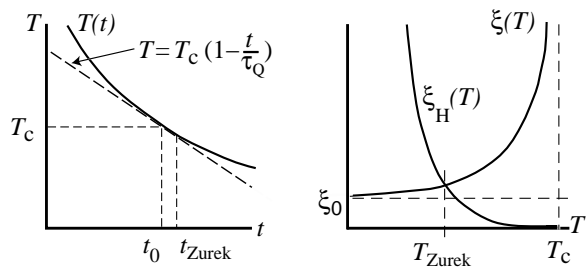}}
  \bigskip
\caption[KZ&Quench] {Principle of KZ mechanism. Rapid thermal
  quench through $T_{\rm c}$: {\it (Left)} Temperature $T(t)$ and its
  linear approximation $T =
  T_{\rm c} (1 - t/\tau_{\rm Q})$ during the quench, as a function of
  time $t$. {\it (Right)} Superfluid coherence length $\xi(T)$ and
  order parameter relaxation time $\tau(T)$ diverge at $T_{\rm c}$. At
  the freeze-out point $t_{\rm Zurek}$, when phase equilibrium is
  achieved, the edge of the correlated region, the causal horizon, has
  moved out to a distance $\xi_{\rm H}(t_{\rm Zurek})$, which has to
  equal the coherence length $\xi(t_{\rm Zurek})$. }
\label{KZ&Quench}
\end{figure}


As sketched in Fig.~\ref{KZ&Quench}, this means that below $T_{\rm
c}$ the order parameter coherence spreads out with the velocity
$c(T) \sim \xi/\tau = \xi_0 \, (1-T/T_{\rm c})^{1/2}/ \tau_0$. The
freeze-out of defects occurs at $t=t_{\rm Zurek}$, when the
causally disconnected regions have grown together and superfluid
coherence becomes established in the whole volume. At the
freeze-out temperature $T(t_{\rm Zurek}) < T_{\rm c}$, the causal
horizon has travelled the distance $\xi_{\rm H} (t_{\rm Zurek}) =
{\int}_0^{t_{\rm Zurek}} \, c(T) \, dt = \xi_0 \tau_{\rm Q}
(1-T_{\rm Zurek}/T_{\rm c})^{3/2}\, / \tau_0$ which has to be
equal to the coherence length $\xi(t_{\rm Zurek})$. This condition
establishes the freeze-out temperature $T_{\rm Zurek}/T_{\rm c} =
1 - \sqrt{\tau_0/\tau_{\rm Q}} \;$ at the freeze-out time $t_{\rm
Zurek} = \sqrt{\tau_0 \tau_{\rm Q}}\;$, when the domain size has
reached the value
\begin{equation}
  \xi_{\rm v} = \xi_{\rm H}(t_{\rm Zurek}) =
  \xi_0 \; (\tau_{\rm Q}/\tau_0)^{1/4}~.
\label{eq:xi-initial}
\end{equation}

In superfluid $^3$He we may have $\xi_0 \sim 20$ nm, $\tau_0 \sim 1$~ns, and in the
best cases a cool-down time of $\tau_{\rm Q} \sim 1$ $\mu$s can be reached. From
these values we expect the domain structure to display a characteristic length scale
of order $\xi_{\rm v} \sim 0.1$ $\mu$m. Assuming a U(1)-symmetry-breaking
transition, where vortex lines are formed at the domain boundaries, the average
inter-vortex distance and radius of curvature in the randomly organized vortex
network is on the order of the domain size $\xi_{\rm v}$. In general, we can expect
a rapid quench to lead to an initial defect density (defined as vortex length per
unit volume)
\begin{equation}
  l_{\rm v} = \frac{1}{a_l \xi_{\rm v}^2} ~,
\label{eq:n_v-initial}
\end{equation}
where the numerical factor $a_l \sim 1$\,--\,100 depends on the details of the model
system. Numerical simulation calculations (Ruutu et~al.\ 1998a) give $a_l\approx 2$ for
$^3$He-B. In this system with global symmetry and negligible thermal fluctuations
the Kibble-Zurek (KZ) model of defect formation can be expected to work well.
Moreover, Ginzburg-Landau theory provides a good description close to $T_{\rm c}$
and might lead via microscopic calculations to more rigorous predictions even on a
time-dependent phase transition.

\subsection{Experimental verification of KZ mechanism} \label{ExpVerific}

The predictions of the KZ model have been tested in numerous numerical experiments
under varying conditions (cf. Bettencourt et~al.\ 2000). Laboratory experiments are
fewer and their results are less conclusive.

The first experiments were performed on liquid crystals (Chuang et~al.\ 1991; Bowick et~al.\ 1994), where the
transition from the isotropic to the nematic phase was examined. This is a weakly
first order transition of rod-like molecules from a disordered to an ordered state,
where defects of different dimensionalities can be formed. The results were
generally found to be consistent with the KZ interpretation. However, a
liquid-crystal transition at room temperature is rather different from the ideal KZ
case.

The first experiments on coherent quantum systems were started with liquid $^4$He
(Hendry et~al.\ 1994), by releasing the pressure on a convoluted phosphor-bronze
bellows so that the sample expanded from the normal liquid through the $\lambda$
transition into the superfluid phase. For a mechanical pressure system the quench
time is of order $\tau_Q \sim 10$ ms. The depressurization is followed by a dead
time of similar length, which is required for damping down the vibrations after the
mechanical shock. The vortex density after the transition is determined from the
attenuation in second sound propagation through bulk liquid. Expansion of $^4$He
across the $T_{\lambda}(P)$ line (which has negative slope in the $(P,T)$ phase
diagram) requires careful elimination of all residual flow which might result from
the expansion itself, foremost the flow out of the filling capillary of the sample
cell, but also that around the convolutions of the bellows. Such flow will
inevitably exceed the critical velocities, which approach zero at $T_{\lambda}$, and
leads to vortex nucleation (Sec.~\ref{SpontVorFormation}). After the initial
attempts, where vortices were detected but a fill line was still present, later much
improved measurements failed to yield any evidence of vortex line production, in
apparent contradiction with the KZ prediction (Dodd et~al.\ 1998).

The early $^4$He expansion experiments were analyzed by Gill and Kibble (1996) who concluded
that the initially detected vortex densities were unreasonably large and must have
originated from extrinsic effects, most likely created by the flow out the fill
line. The absence of vorticity in the later results was shown not to be evidence
against the KZ mechanism (Karra and Rivers 1998; Rivers 2000): Although the pressure drop in
expansion cooling is large, 20 -- 30 bar, the change in relative temperature
$\epsilon = 1-T/T_{\lambda}$ is small, $|\Delta \epsilon| < 0.1$. The final
temperature $\epsilon_f < 0.06$ is below $T_\lambda$ but still above the Ginzburg
temperature $T_{\rm G}$, ie. in the regime where thermal fluctuations might be
sufficiently effective to wipe out domain formation. Here $T_{\rm G}$ is defined as
the temperature at which $k_{\rm B} T_{\rm G}$ equals the energy of a vortex loop of
the size $\xi_{\rm v}$, which is expected after the quench according to
Eq.~(\ref{eq:xi-initial}). Thus the current view about the $^4$He expansion
experiments holds that the vortices decay away before the observation switches on
(Rivers 2000; Hendry et~al.\ 2000).

Simultaneously with the $^4$He expansion experiments, measurements of quite
different nature were carried out in $^3$He-B (Ruutu et~al.\ 1996a; B{\"a}uerle et~al.\ 1996). Here a
detailed comparison with the KZ predictions was achieved. These efforts were
followed by measurements of thermal quenches through the superconducting transition
in different superconducting geometries.

The most conclusive results have been obtained with high-T$_{\rm c}$ YBCO films
(Maniv et~al.\ 2003).  The film is heated above its transition temperature of 90\,K
with a laser pulse. The hot film is then cooled from below back to the
ambient temperature of 77\,K. This happens laterally uniformly through the
contact with an optically transparent substrate of much larger thickness
and heat capacity. The maximum cooling rate was measured to be $10^8$\,K/s,
by monitoring the electrical resistance of the film. The total net flux,
which remains frozen in the pinning sites of the YBCO film after the
transition, was measured with a sensing coil below the substrate and
connected to a SQUID magnetometer. This flux was found to display very
weak dependence on the cooling rate, as predicted by the KZ scaling theory.

As discussed by Kibble and Rajantie (2003), the important information to obtain from
such measurements would be the spatial distribution of the flux quanta in the film.
Spatial correlations between the fluxoids would support the KZ mechanism, if the
vortices were found to display negative correlations, ie. neighboring vortices would
prefer to have opposite signs. A competing mechanism involves thermal fluctuations
of the magnetic field which generates vortices during the quench cooling through $T_{\rm c}$ where the critical field vanishes: $H_{\rm c2} \rightarrow 0$. This second
mechanism is predicted to display positive correlations, ie. vortices of same sign
would tend to cluster. An analogous situation arises in the $^3$He-B experiments which will be described below. Here at temperatures below 3\,mK thermal fluctuations are not important since any energy barriers will generally automatically be orders of magnitude higher. Nevertheless, even in these conditions a phenomenon can be identified which competes with the KZ mechanism (which is a volume effect). This involves the phase boundary between the normal liquid and $^3$He-B (a surface effect) if a phase boundary exists in the presence of flow. The reason is again that at $T_{\rm c}$ the critical velocity for vortex formation vanishes: $v_{\rm c} \rightarrow 0$.

In superconductors a measurement with spatial resolution and single-fluxoid
sensitivity has been recently accomplished (Kirtley et~al.\ 2003). Here a scanning SQUID
microprobe is used to map the flux quanta trapped in an amorphous Mo$_{3}$Si film
which had been lithographically patterned in ring-like elements. Unfortunately, in
this measurement the cooling rate was limited to $\sim 20\,$K/s at the $T_{\rm c} =
7.8\,$K of the film and the detected quantized flux was concluded to originate from
a thermally activated process.

These examples illustrate the difficulties in attempting to find experimental proof
for the KZ process. The KZ mechanism is intuitively easy to accept, but its
experimental verification is complicated and involves careful considerations. In a
phase transition defect formation is more usual than its absence. In most
experiments both extrinsic and intrinsic sources of vortex formation are present and compete in actual vortex generation. For
instance, even slow cooling of liquid $^4$He across the $\lambda$ transition is
known to produce primordial vortices which remain pinned as remanent vorticity on
the walls of the container (Awschalom and  Schwarz 1984). There are no claims in the
literature yet that an experiment would have produced a perfectly vortex-free bulk
sample of superfluid $^4$He.

A second point to note is that the KZ model describes a second order transition
where the energy barrier, which separates the symmetric high-temperature phase from
the  broken-symmetry states at low-temperatures, vanishes at $T_{\rm c}$. Thus the
transition becomes an instability. A first order transition is different: Here the
barrier remains finite and the low temperature phase has to be nucleated, usually by
overcoming the barrier via thermal activation. Nucleation may occur in different
parts of the system nearly simultaneously, depending on the properties of the
system, the fluctuations, the quench time $\tau_{\rm Q}$, and the nucleation
mechanism. Such a situation, which often is called bubble nucleation, also leads to
domain formation and to a final state which is qualitatively similar to that
expected after the second order transition and the KZ mechanism.

In all of the experimental examples listed above, other mechanisms are also possible
which might replace or operate in parallel to the KZ process. The differences are
subtle, as we shall see in Sec.~\ref{MechVorForm} for $^3$He-B or as has been
emphasized for superconductors by Rajantie (2001). In practice measurements
differ from the original KZ model, ie. from a homogeneous transition, which occurs
simultaneously in the whole system and where there are no gradients in density or
temperature. In practical measurements on bulk material a rapid quench through the transition is often forced by strong gradients. On some level gradients will always appear in any laboratory
system and their influence has to be investigated.

Superfluid $^3$He measurements have so far established the most quantitative
comparison to the KZ scaling theory (Ruutu et~al.\ 1996a; B{\"a}uerle et~al.\ 1996). A happy coincidence of
many valuable features has made this possible: A reduced likelihood of defect
formation from extrinsic sources, the virtual absence of remanent vorticity, the
possibility to perform thermal quenches in microseconds, the presence of a pure
second order phase transition, and the availability of measuring methods to detect
different types of defects, often with a resolution of one single defect. A critical discussion of these questions is presented below.

The search for firm proof of the KZ mechanism has grown to an important topic in condensed matter physics. The reason is that it provides (at least superficially) easy interpretation of the fastest non-equilibrium phase transitions. Amusingly this question does not enjoy similar interest any more in cosmology where the problem was born. Measurements of the anisotropy in the cosmic microwave background radiation with the WMAP satellite (with an angular resolution $\sim 0.25^{\circ}$ and an amplitude sensitivity $\sim 1\, \mu$K) have concluded that the angular distribution of the anisotropy expanded in its harmonic components gives a multipole expansion with a sharp dominating peak at $\ell \sim 200$, which corresponds to an angular scale $\sim 1^{\circ}$ (Bennett et~al.\ 2003; Spergel et~al.\ 2003). This distribution can be fit with models which are based on inflationary expansion of the Early Universe, {\it i.e.} with models with an early period of accelerated expansion. In contrast, topological defects, such as cosmic strings, are expected to generate a clearly different broader peak. These two different scenarios have been the ruling contenders as the explanation for the origin of large-scale structure. Thus at present time the KZ mechanism appears to survive only in condensed matter physics.

\begin{figure}[!!!tb]
  \centerline{\includegraphics[width=0.90\columnwidth]{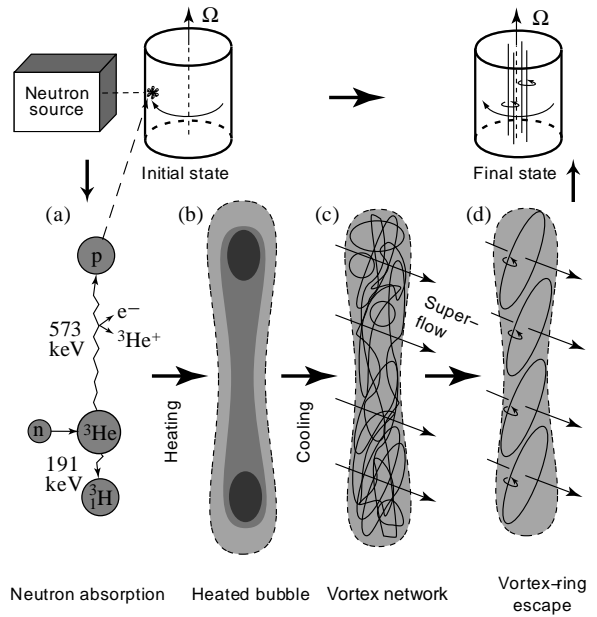}}
  \bigskip
\caption[ExpPrinciple]
  {Principle of the quench-cooled rotating experiment in superfluid
  $^3$He-B: {\it (Top)} A cylindrical sample container with superfluid
  $^3$He-B is rotated at constant angular
  velocity $\Omega$ and temperature $T$, while the NMR
  absorption is monitored continuously. When the sample is irradiated
  with neutrons, vortex lines are observed to form. {\it (Bottom)}
  Interpretation of the processes following a neutron absorption event
  in bulk superflow; see the text for details. (From
  Ruutu et~al.\ 1996a). }
\label{ExpPrinciple}
\end{figure}

\subsection{Principle of superfluid $^3$He experiments} \label{ExpPrincipleSec}

To study reliably defect formation in a quench-cooled transition,
two basic requirements are the following: First, the transition
can be repeated reproducibly, and second, a measurement is
required to detect the defects after their formation, either
before they annihilate or by stabilizing their presence with a
bias field. Both of these requirements can be satisfied in $^3$He
experiments.

\subsubsection{Outline of experimental method} \label{ExpOutline}

A schematic illustration of one set of experiments (Ruutu et~al.\ 1996a) is depicted in
Fig.~\ref{ExpPrinciple}. The top part displays the events which are controlled by
the observer in the laboratory. A cylindrical container filled with superfluid
$^3$He-B is rotated at constant velocity $\Omega$. The velocity of rotation is
maintained below the critical value at which quantized vortex lines are
spontaneously formed. In other words, the initial state is one of metastable
vortex-free counterflow of the superfluid and normal components in constant
conditions.

\begin{figure}[!!!tb]
  \centerline{\includegraphics[width=0.90\columnwidth]{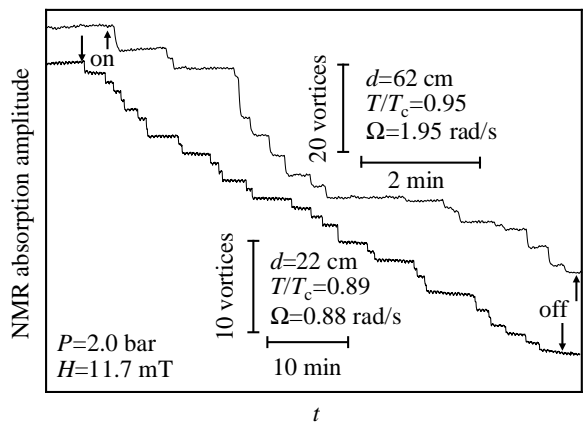}}
  \bigskip
  \caption[ExpSignature]
  {Peak height of the NMR signal which monitors the velocity of the bias
  flow from rotation (the so-called counterflow peak in the NMR absorption
  spectrum of $^3$He-B), shown as a function of time $t$ during neutron
  irradiation. The {\it lower trace} is for a low and the {\it upper
  trace} for a high value of rotation velocity $\Omega$. The initial
  state in both cases is metastable vortex-free counterflow.
  The vertical arrows indicate when the neutron source was turned on/off.
  Each step corresponds to one neutron absorption event and its height,
  when compared to the adjacent vertical calibration bar, gives the
  number of vortex lines formed per event. The distance of the neutron
  source from the sample is denoted with $d$. (From
    Ruutu et~al.\ 1996b). }
   \label{ExpSignature}
\end{figure}

Next a weak source of thermal neutrons is placed in the vicinity of the $^3$He
sample. If the rotation velocity is sufficiently high, then vortices start to appear
at a rate which is proportional to the neutron flux.  The neutron source is
positioned at a convenient distance $d$ from the cryostat so that vortex lines are
observed to form in well-resolved individual events. The experimental signal for the
appearance of a new vortex line is an abrupt jump in NMR absorption. An example,
measured at low rotation, is shown by the lower recorder trace in
Fig.~\ref{ExpSignature}.

The lower part of Fig.~\ref{ExpPrinciple} shows in more detail
what is thought to happen within the superfluid.  Liquid $^3$He
can be locally heated with the absorption reaction of a thermal
neutron: n + $^3_2$He $\rightarrow$ p + $^3_1$H + $E_0$, where
$E_0 = 764$ keV. The reaction products, a proton and a triton, are
stopped by the liquid and produce two collinear ionization tracks
(Meyer and Sloan 1997).  The ionized particles, electrons and $^3$He ions,
diffuse in the liquid to recombine such that 80 \% or more of
$E_0$ is spent to heat a small volume with a radius $R_{\rm b}
\sim 50$ $\mu$m from the superfluid into the normal phase.  The
rest of the reaction energy escapes in the form of ultraviolet
emission~(Stockton et~al.\ 1971; Adams et~al.\ 1995) and, possibly, in the form
of long-living molecular $^3$He$^*_2$ excitations
(Keto et~al.\ 1974; Kafanov et~al.\ 2000), which partially relax at the
walls of the container.

For measurements in rotating $^3$He, neutron absorption is an ideal heating
process. For slow neutrons from a room temperature source with a Maxwellian distribution (at $20.4\, ^\circ$C or 0.0253\,eV) the velocity at the peak of the distribution is 2200\,m/s. The capture cross section of the $^3$He nucleus for these neutrons is huge:  $5327 \pm 10\,$barn =
5.3$\cdot 10^{-21}$ cm$^2$ (Beckurts and Wirtz 1964). This means a short mean free path of about 100
$\mu$m. Thus most neutrons are absorbed within a close distance from the
wall, although still within the bulk liquid, but  such that the velocity of the applied counterflow is accurately specified.  In other parts of the refrigerator neutron
absorption is negligible and thus the temperature of the $^3$He sample can
be maintained stable during the irradiation. Finally, the cooling is
fast, as the small volume heated by the neutron absorption
event is embedded in the cold bulk system.

Subsequently, the normal liquid in the heated neutron bubble cools back
through $T_{\rm c}$ in microseconds. The measurements in
Fig.~\ref{ExpSignature} demonstrate that vortex lines are indeed created in
this process.  Vortex loops, which form within a cooling neutron bubble in
the bulk superfluid, would normally contract and disappear in the absence
of rotation. This decay is brought about by the inter-vortex interactions
between the vortex filaments in the presence of dissipation from mutual
friction between the normal and superfluid components. In rotation the
externally applied counterflow provides a bias which causes sufficiently
large loops to expand to rectilinear vortex lines and thus maintains them
for later detection. In the rotating experiments in Helsinki these
rectilinear vortex lines are counted with NMR methods
(Ruutu et~al.\ 1996a, 1998a; Finne et~al.\ 2004a).

In a second series of $^3$He experiments, performed in Grenoble
(B{\"a}uerle et~al.\ 1996, 1998a) and Lancaster (Bradley et~al.\ 1998), the vortices formed
in a neutron absorption event are detected calorimetricly with
very-low-temperature techniques. In the zero temperature limit mutual
friction becomes exponentially small and the life time of the vorticity
might be very long, even if there is no flow. In this situation the
existence of the vorticity can be resolved as a deficit in the energy
balance of the neutron absorption reaction.

\subsubsection{Interpretation of $^3$He experiments} \label{Interpretation}

Several alternative suggestions can be offered on how to explain these observations.
The KZ model is one of them. Whatever the mechanism, the superfluid transition in a
localized bubble represents a new process for creating vortex rings. This phenomenon
is associated with one of the fastest 2nd order phase transition, which has been
probed. It takes place in the bulk superfluid and not in immediate contact with a
solid wall, where most other vortex formation processes occur in the presence of
applied counterflow.

The KZ interpretation of the later stages of the experiment is contained in
the two illustrations marked as (c) and (d) in Fig.~\ref{ExpPrinciple}.
First a random vortex network (c) is created while cooling through $T_{\rm
  c}$. Next the network starts to evolve under the influence of
inter-vortex interactions and the applied flow. At high temperatures the
process is highly dissipative, owing to the large mutual friction force
which connects the superfluid and normal fractions and acts on a moving
vortex. The average inter-vortex distance increases, small loops are
smoothed out, and the network becomes rarefied or ``coarse grained''. Also
reconnections, which take place when two lines cross, contribute to the
rarefaction. The applied flow favors the growth of loops with the right
winding direction and orientation. It causes sufficiently large loops to
expand, while others contract or reorient themselves with respect to the
flow. The final outcome is that correctly oriented large loops (d), which
exceed a critical threshold size, start expanding spontaneously as vortex
rings, until they meet the chamber walls. There the superfluous sections of
the ring will annihilate and only a rectilinear vortex line will finally be
left over. It is pulled to the center of the container, where it remains in
stationary state parallel to the rotation axis, stretched between the top
and bottom surfaces.

This picture applies at high temperatures in $^3$He-B where Kelvin-wave excitations
on vortices are exponentially damped and spontaneous loop formation on existing
vortices does therefore not occur. In this situation the number of vortex rings,
which are extracted from the neutron bubble (Fig.~\ref{ExpPrinciple}d), remains
conserved during the later expansion. Thus the number of rectilinear vortex lines,
the end result from the neutron absorption event, characterizes the extraction
process as a function of the applied bias velocity. This situation is very different
from that in $^4$He-II where mutual friction damping never reaches this high values
(except extremely close to $T_{\lambda}$).

Very different time scales are at work in this process of vortex formation. Except
for the retarded molecular excitations, the heating from the initial ionization and
subsequent recombination is limited by the diffusion of charges in the liquid and
takes place much faster than the thermal recovery. The initial vortex network forms
during cooling through $T_{\rm c}$, for which the relevant time scale is
microseconds. The later evolution of the network and the loop escape happen again on
a much slower time scale, namely from milliseconds up to seconds, since here the
vortex motion is governed by the mutual-friction-dependent superfluid hydrodynamics.

\subsection{Measurement of vortex lines in $^3$He-B} \label{VorMeasurement}

There are two major phases of superfluid $^3$He, the A and B
phases (Fig.~\ref{PresDepend}). The neutron measurements have
been performed in the quasi-isotropic $^3$He-B. In the present
context we may think of vortices in $^3$He-B as being similar to
those in superfluid $^4$He-II, where only the U(1) symmetry is
broken and the order parameter is of the general form $\Psi =
|\Psi(T)| \; e^{i \Phi({\bf r})}$: The vortices are topologically
stable, have a singular core, inside of which $|\Psi|$ deviates
from its bulk value, while outside the phase $\Phi$ changes by
$2\pi\nu$ on a closed path which encircles the core.  A persistent
superfluid current is trapped as a single-quantum circulation
($\nu=1$) around the core, with the circulation quantum $\kappa =
h/(2m_3) \simeq 0.0661$ mm$^2$/s and a superflow velocity ${\bf
v}_{\rm s,vortex} = \kappa /(2\pi) \; {\bf \nabla} \Phi =
\kappa/(2\pi r) \, {\hat {\mbox{\boldmath$\phi$}}}$.

\subsubsection{Critical velocity of vortex formation}
\label{SpontVorFormation}

The energy of a single rectilinear vortex line, aligned along the symmetry axis of a
rotating cylindrical container of radius $R$ and height $Z$, consists primarily of
the hydrodynamic kinetic energy stored in the superfluid circulation trapped around
the core,
\begin{equation}
  E_{\rm v} = {1 \over 2} \int{ \, \rho_{\rm s} v_{\rm s,vortex}^2 \; dV} =
  {{\rho_{\rm s} \kappa^2 Z} \over {4 \pi}} \; \ln{R \over \xi}~~.
\label{eq:E_vortex}
\end{equation}
Here the logarithmic ultraviolet divergence has been cut off with
the core radius, which has been approximated with the coherence
length $\xi$. If the container rotates at an angular velocity
$\Omega$, the state with the first vortex line becomes
energetically preferred when the free energy $E_{\rm v} - \Omega
L_{\rm z}$ becomes negative. The hydrodynamic angular
momentum from the superflow circulating around the vortex core is
given by
\begin{equation}
  L_{\rm z} = \int{ \, \rho_{\rm s} r v_{\rm s,vortex} \; dV} =
  \rho_{\rm s} \kappa R^2 Z~~.
\label{eq:L_vortex}
\end{equation}
One finds that $E_{\rm v} - \Omega L_{\rm z} < 0$, when the
velocity $\Omega R$ of the cylinder wall exceeds the Feynman
critical velocity, $v_{\rm c1} = \kappa /(2\pi R) \,
\ln{(R/\xi)}$. With a container radius $R$ of a few mm, the
Feynman velocity is only $\sim \! 10^{-2}$ mm/s. However, if we
can exclude remanent vorticity and other extrinsic mechanisms of
vortex formation (which is generally the case in $^3$He-B at temperatures  $T>0.6\,T_{\rm c}$, but not in $^4$He-II), then vortex-free superflow will persist as a
meta-stable state to much higher velocities because of the
existence of a finite nucleation energy barrier.

The height of the nucleation barrier decreases with increasing velocity. In
$^4$He-II intrinsic vortex formation is observed in superflow through
submicron-size orifices. Here the barrier is ultimately at sufficiently
high velocities overcome by thermal activation or at low temperatures by
quantum tunnelling (Packard 1998; Varoquaux et~al.\ 1998). The other possibility is
that the thermodynamic stability limit of the superfluid is reached
(Andreev and Melnikovsky 2003). In $^3$He-B the barrier is
unpenetrably high in practically any situation and the flow velocity has to
be increased to the point where the barrier approaches zero. Vortex
formation then occurs in the form of an instability which takes
place locally at some sharp asperity on the cylindrical wall of the rotating sample
where the stability limit of superflow is first reached (Parts et~al.\ 1995; Ruutu et~al.\ 1997a).  The average
velocity at the cylindrical container wall we call the spontaneous critical velocity $v_{\rm c}(T,P)$ (Fig.~\ref{CritVel&NeutrThreshold}), which now depends on surface roughness and is therefore container specific. When the instability occurs and a vortex is formed, the actual local velocity at the critical dominating asperity equals the intrinsic
bulk instability velocity $v_{\rm cb}(T,P)$.

When the flow reaches the instability value $v_{\rm cb}$ at the dominating
sharpest protrusion, a segment of a vortex ring is formed there. The energy of the
smallest possible vortex ring, which has a radius comparable to
the coherence length, is of order $E_{\rm v} \sim \rho_{\rm s}
\kappa^2 \xi$. This energy constitutes the nucleation barrier. On
dimensional grounds we may estimate that $E_{\rm v}/(k_{\rm B}T) \sim (\xi/a)
\, (T_{\rm F}/T)$, where $T_{\rm F} = \hbar^2/(2 m_3 a^2) \sim 1$K
is the degeneracy temperature of the quantum fluid and $a$ the
interatomic spacing. In $^4$He-II the coherence length is of order
$\xi \sim a$ and the temperature $T \sim T_{\rm F}$, which means
that $E_{\rm v}/(k_{\rm B}T) \sim 1$ and thermally activated nucleation
becomes possible.

In contrast in $^3$He-B the coherence length is more than 10 nm and we find
$E_{\rm v}/(k_{\rm B}T) \sim 10^5$ at temperatures of a few mK or less. The
consequence from such an enormous barrier is that both thermal activation
and quantum tunnelling are out of question as nucleation mechanisms. The
only remaining intrinsic mechanism is a hydrodynamic instability, which
develops at rather high velocity. This fact has made possible the
measurement of neutron-induced vortex formation as a function of the
applied flow velocity $v < v_{\rm c}$.

\subsubsection{Rotating states of the superfluid} \label{RotStates}

\begin{figure}[!!!tb]
  \centerline{\includegraphics[width=0.55\columnwidth]{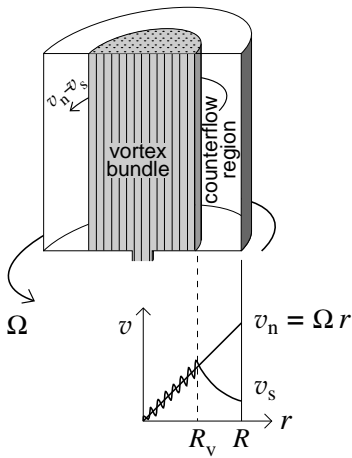}}
  \bigskip
\caption[VorCluster] {{\it (Top)} Metastable state with a central
   vortex cluster in the rotating cylinder. This axially symmetric
   arrangement consists of rectilinear vortex lines surrounded by vortex-free
   counterflow. {\it (Bottom)} Radial distribution of the velocities of the
   superfluid component, $v_{\rm s}$, and the normal component,
   $v_{\rm n}$. Because of its large viscosity, the normal fraction
   corotates with the container. (From Krusius et~al.\ 1994). }
\label{VorCluster}
\end{figure}

When a container with superfluid is set into rotation and the formation of
vortex lines is inhibited by a high energy barrier, then the superfluid
component remains at rest in the laboratory frame. Its velocity is zero in
the whole container: $v_{\rm s}= 0$. The normal component, in contrast,
corotate with the container and $v_{\rm n}= \Omega r$. This state of
vortex-free counterflow is called the Landau state. It corresponds to the
Meissner state in superconductors, with complete flux expulsion.

In $^4$He-II vortex-free flow with sufficiently high velocity is out of
reach in most situations, because even at very low velocities remanent
vorticity leads to efficient vortex formation. In $^3$He-B the superfluid
coherence length is several orders of magnitude larger and remanent
vorticity can be avoided if the container walls are sufficiently smooth.
Coupled with a high nucleation barrier, metastable rotating states then
become possible. These can include an arbitrary number $N$ of rectilinear
vortex lines, where $N \leq N_{\rm max}$. In most experimental situations
the maximum possible vortex number, $N_{\rm max}$, equals that in the
equilibrium vortex state, $N_{\rm eq}\approx 2\pi \Omega R^2/\kappa$. The
equilibrium vortex state is obtained by cooling the container slowly at
constant $\Omega$ through $T_{\rm c}$.

If the number of lines is smaller than that in the equilibrium state, then
the existing vortex lines are confined within a central vortex cluster, as
shown in Fig.~\ref{VorCluster}. The confinement comes about through the
Magnus force from the normal-superfluid counterflow, $v = v_{\rm n} -
v_{\rm s}$, which circulates around the central cluster in the vortex-free
region. Within the cluster mutual repulsion keeps the lines apart such that
they form a triangular array at the equilibrium density, $n_{\rm v} =
2\Omega/ \kappa$, which is constant over the transverse cross section of
the cluster. At this line density the cluster corotates with the container
at constant $\Omega$, like a solid body. The total number of lines is given
by $N = \pi n_{\rm v} R_{\rm v}^2$, where $R_{\rm v}$ is the radius of the
cluster.

With $N$ vortex lines in the cluster, the superflow velocity
outside has the value $v_{\rm s} = \kappa N/(2\pi r)$, which is
equivalent to that around a giant vortex with $N$ circulation
quanta $\kappa$ (Fig.~\ref{VorCluster}). The counterflow velocity
$v$ vanishes inside the cluster (measured on length scales which
exceed the inter-vortex distance), while outside it increases from
zero at $r=R_{\rm v}$ to its maximum value at the cylinder wall,
$r=R$:
\begin{equation}
v(R) = \Omega R - {{\kappa N} \over {2 \pi R}}\;. \label{CFvel}
\end{equation}
In the absence of the neutron source, new vortex lines are only formed if $v(R)$ is
increased to its container-dependent spontaneous critical value $v_{\rm c}(T,P)$.


\begin{figure}[tb]
\centerline{\includegraphics[width=0.7\columnwidth]{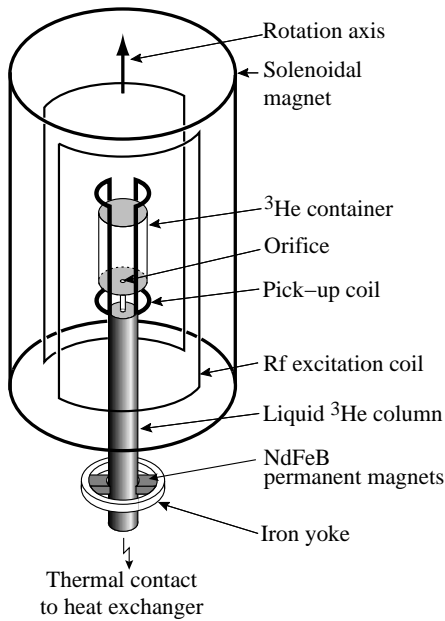}}
\bigskip
\caption[NMR-Cell] {Liquid $^3$He sample container with coils for NMR measurement.
  The cylindrical sample on the top is connected via a narrow aperture with
  a long liquid $^3$He column which serves as a thermal path to the
  sintered heat exchanger on the nuclear refrigeration stage. The nuclear
  stage is located below this structure while the mixing chamber of the
  pre-cooling dilution refrigerator is above. The excitation coil and the
  outer magnet have not been drawn to scale. (From Ruutu et~al.\ 1997a).}
\label{NMR-Cell}
\end{figure}


\subsubsection{Experimental setup} \label{SetUp}

Fig.~\ref{NMR-Cell} shows a measuring setup for rotating NMR measurements
in $^3$He-B. The cylindrical sample is connected via an orifice to a long
liquid $^3$He column which is needed to fill the sample volume, to
provide the thermal contact with the nuclear refrigeration stage, and to isolate the sample in a metal-free environment for NMR measurement. The major resistances in the thermal path between the sample and the nuclear cooling stage are the orifice, the liquid column, and the Kapitza surface resistance of the porous sintered heat exchanger in a large liquid volume below the column. The most important characteristic of the sample container itself is its spontaneous critical velocity $v_{\rm c}(T,P)$, which may depend in complicated ways on
temperature and pressure, since different sources contribute to vortex formation. These arise
from (i) the surface roughness of the cylindrical walls in the volume above the orifice, (ii) sites in which remanent vortices can be trapped, and (iii) the
leakage of vortices through the orifice (since the volume in contact with the rough heat exchanger surfaces will be filled with the equilibrium number of vortices already at low $\Omega$). In practice for a clean quartz cylinder the most important source is some isolated surface protrusion, a localized defect or a piece of dirt, which increases the flow velocity at a sharp
asperity by up to an order of magnitude (Parts et~al.\ 1995; Ruutu et~al.\ 1997a). To achieve a high critical velocity, the geometry and surface quality of the sample container are of utmost importance.

The sample container structure in Fig.~\ref{NMR-Cell} is prepared from fused quartz.
The sample cylinder itself has a radius $R = 2.5\,$mm, length $Z=7\,$mm, and wall
thickness 0.5\,mm. An orifice of 0.5\,mm diameter links it with the 60\,mm long
connecting column. The long length is needed to place the sample in the middle of
the NMR polarization magnet. The container is fused together from tubular and flat
plate parts in an oxygen-acetylene flame. After assembly the structure is annealed
in an oven, etched, and cleaned with solvents. All indications point to an average
surface roughness well below $1\,\mu$m, such that it is dirt particles and perhaps
isolated localized defects on the glass surface which ultimately control the
critical velocity. For the quartz glass cylinder in Fig.~\ref{NMR-Cell} the
spontaneous critical velocity $v_{\rm c}$ was found to be a factor of three or more
larger than the lowest velocity $v_{\rm cn} (T,P)$ at which vortices start to appear
in neutron irradiation (cf. Fig.~\ref{CritVel&NeutrThreshold}).


\begin{figure*}
  \centerline{\includegraphics[width=0.8\textwidth]{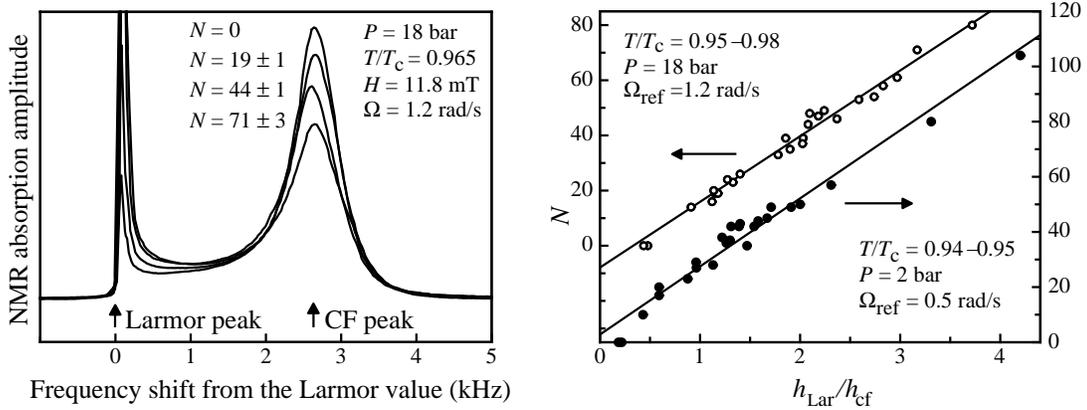}} \bigskip
\caption[VorNMR] {
  NMR measurement of vortex lines in superfluid $^3$He-B, when rotated
  in a long cylinder with the magnetic field oriented axially: {\it
    (Left)} NMR spectra recorded at a reference velocity $\Omega_{\rm
    ref} = 1.2$ rad/s with different number of vortex lines $N$. The
  Larmor frequency is at the left and the counterflow peak at the
  right vertical arrow. {\it (Right)} Two calibration measurements of
  the number of vortex lines $N$ as a function of the ratio of the
  Larmor and counterflow peak heights, $h_{\rm Lar}/h_{\rm cf}$,
  measured at two different rotation velocities $\Omega_{\rm ref}$. (From
  Xu et~al.\ 1996).}
\label{VorNMR}
\end{figure*}


The non-invasive continuous-wave NMR measurement is conducted from the outside with
a system of three orthogonal superconducting coils. The innermost is a saddle-shaped
detector coil, wound from $25\,\mu$m solid Nb wire on a thin epoxy shell and mounted
directly on the sample cylinder. It is thermally anchored with a Cu strip to the
nuclear cooling stage. The outermost coil is a magnet which produces the axially
oriented homogeneous polarization field. This end-compensated solenoid on a bronze
body is fixed mechanically and thermally to the mixing chamber. The coil in the
middle is used for rf excitation. It consists of two turns in each half and is fixed
inside the polarization magnet. All parts shown here are located inside a
superconducting Nb shield, to avoid interference from the demagnetization field
which is required for cooling and temperature stabilization. The Nb jacket is part
of the heat shield which is fixed to the mixing chamber.

The neutron source is a weak $^{241}$Am/Be specimen which is sandwiched between two paraffin moderator tiles so that a cubic box results with sides of 25\,cm length. For irradiation the box is carried to the cryostat and placed at room temperature at a desired distance from the $^3$He sample. By varying this distance the neutron flux can be adjusted to the required low value, as determined from the vortex formation rate $\dot N$ in Fig.~\ref{SourceDistance}. Between measurements the source is moved a fair distance away from the cryostat ($\sim 25\,$m).

\subsubsection{NMR measurement} \label{NMR_measurement}

The unusual NMR properties of the $^3$He superfluids are a direct consequence from
the Cooper pairing in states with spin $S=1$ and orbital momentum $L=1$. In $^3$He-B
the NMR absorption spectrum, recorded at low rf excitation with traditional
transverse cw methods, is related to the spatial distribution of the order parameter
orientation in the rotating cylinder. Spin-orbit coupling, although weak, is
responsible for the large frequency shifts which give rise to absorption peaks
shifted from the Larmor frequency (Vollhardt and W{\"o}lfle 1990). The height of the so-called {\it
counterflow} (CF) peak, which depends on the CF velocity around the vortex cluster,
can be used to determine the number of rectilinear vortex lines in the cluster.

In the left panel of Fig.~\ref{VorNMR} NMR absorption spectra are shown,
which have been recorded at the same $\Omega$, but with a different number
$N$ of vortex lines in the vortex cluster. In the vortex-free state the
large CF peak on the right increases rapidly with $\Omega$ when the
orienting effect from the CF grows. The absorption intensity for this
growth in peak height is shifted from the asymmetric peak on the left close
to the {\it Larmor} frequency. This second peak reflects the order
parameter orientation in the center of the container in the region of the
vortex cluster. At fixed
$\Omega$, if more vortex lines are added, the central vortex cluster
expands, the CF velocity outside the cluster is reduced, and the absorption
intensity from the CF peak is shifted back to the Larmor peak
(Kopu et~al.\ 2000).

The frequency shift of the CF peak increases monotonically with decreasing
temperature and is used to measure the temperature. The total integrated
absorption in the NMR spectrum is proportional to the B phase
susceptibility $\chi_{\rm B}(T,P)$ and decreases rapidly with temperature.
However, by maintaining the CF peak at fixed frequency shift, the
temperature is kept stable, and the relative amount of absorption in the
Larmor and CF peaks can then be used to determine the number of vortex
lines. By monitoring the peak height of
either of the two absorption maxima, one can detect a change in $N$, as was
illustrated in Fig.~\ref{ExpSignature}.

In the right panel of Fig.~\ref{VorNMR} the ratio of the two peak heights
has been calibrated to give the vortex number at two different angular
velocities $\Omega_{\rm ref}$. In both cases only a relatively small number
of vortex lines $N \ll N_{\rm eq}$ is present and the CF peak is large
(Xu et~al.\ 1996).  These calibration plots were measured by starting with
an initially vortex-free sample rotating at $\Omega_{\rm ref}$, into which
a given number of vortex lines $N$ was introduced with neutron absorption
reactions, as in Fig.~\ref{ExpSignature}. After that the irradiation was
stopped and the entire NMR spectrum was recorded (by sweeping the magnetic
polarization field $H$). From the spectrum the ratio of the peak heights of
the two absorption maxima is worked out and plotted versus $N$. This gives
the linear relationships in Fig.~\ref{VorNMR}.

\begin{figure}[tb]
  \centerline{\includegraphics[width=0.9\columnwidth]{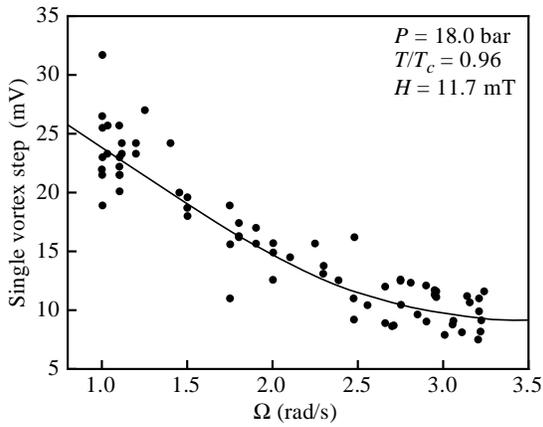}}
  \bigskip
\caption[VorSignal] {
  Absolute magnitude of the single-vortex signal (as in
  Fig.~\protect\ref{ExpSignature}), measured in the vortex-free
  counterflow state at different $\Omega$. The reduction in the height
  of the counterflow peak is given here in millivolts at the output of
  the cooled preamplifier which has a gain of $\approx 10$ and is located
  inside the vacuum jacket of the cryostat. The scatter in the
  data arises from residual differences in the measuring temperature, perhaps
  from small differences in the order-parameter texture from one cool down to the next, and from an electronic
  interference signal which is generated by the rotation of the cryostat. More details
  on the NMR spectrometer are given by
  Parts et~al.\ (1995) and Ruutu et~al.\ (1997a). }
\label{VorSignal}
\end{figure}

The reduction in the CF peak height from the addition of one single vortex line can
be discerned with good resolution in favorable conditions (Fig.~\ref{ExpSignature}).
The optimization of this measurement has been analyzed by Kopu et~al.\ (2000). With a
small-size superconducting magnet and its modest field homogeneity $(\Delta H/H \sim
10^{-4})$, the best conditions are usually achieved at low magnetic field $(H \sim
10$ -- 20 mT) close to $T_{\rm c} $ $(T \geq 0.8\, T_{\rm c})$ and at relatively
high CF velocity ($\Omega > 0.6$ rad/s), where the CF peak is well developed
(Korhonen et~al.\ 1990). The single-vortex signal, i.e. the change in peak height per vortex
line, decreases with increasing $\Omega$ and CF peak height, as shown in
Fig.~\ref{VorSignal}. In neutron irradiation measurements, the rate of vortex line
creation $\dot N$ is determined directly from records like the two traces in
Fig.~\ref{ExpSignature}.  The plot in Fig.~\ref{VorSignal} provides a yard stick for
the single-vortex signal, to estimate the number of new lines if multiple lines are
created in one neutron absorption event.

\subsection{Vortex formation in neutron irradiation} \label{VorForm}

\begin{figure}[tb]
  \centerline{\includegraphics[width=1.0\columnwidth]{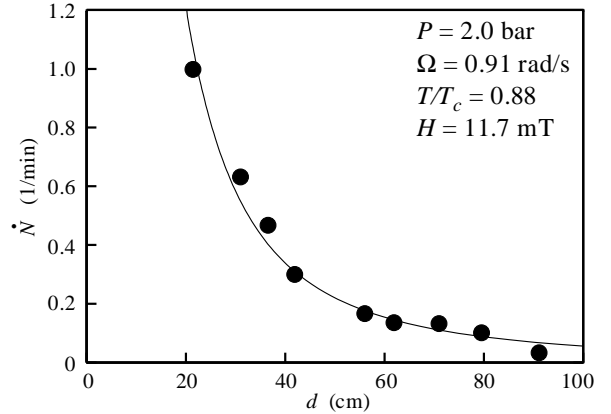}} \bigskip
\caption[SourceDistance]{
  Rate of vortex-line formation in neutron irradiated counterflow, $\dot
  N$, as a function of the distance $d$ between neutron source and $^3$He-B
  sample. The initial state for all data points is vortex-free counterflow
  at a rotation velocity of 0.91 rad/s. The neutron irradiation time is 30
  min.  The fitted curve is of the form $\ln{[1+(R_{\rm s}/d)^2]}$, where
  $R_{\rm s}$ is the radius of the front surface of the paraffin
  moderator box (perpendicular to ${\bf d}$), in which the Am-Be source is
  embedded. }
\label{SourceDistance}
\end{figure}

A measurement of vortex-line formation in neutron irradiated normal-superfluid
counterflow is performed at constant ambient conditions. The externally controlled
variables include the rotation velocity $\Omega$, temperature $T$, pressure $P$,
magnetic field $H$, and neutron flux $\phi_{\rm n}$. The initial state is one of
vortex-free counterflow ($N=0$). When stable conditions have been reached, a weak
neutron source is placed at a distance $d$ from the $^3$He-B sample and the output
from the NMR spectrometer is monitored, as shown in Fig.~\ref{ExpSignature}. From
this NMR absorption record as a function of time, the vortex lines can be counted
which are formed during a given irradiation period.

The process evidently exhibits stochastic variation. To measure the vortex-formation
rate $dN/dt = {\dot N}$, records with a sufficiently large number of detected
neutron absorption events are analyzed to obtain a representative result. After that
one of the experimental parameters is changed and a new run is performed. In this
way the dependence of the vortex formation process on the external variables can be
studied. It was found that ${\dot N}$ varies as a function of all the external
parameters, i.e. neutron flux, rotation velocity, temperature, pressure, and
magnetic field. In the following we shall first describe these empirically
established dependences.

The rate of vortex-line formation is proportional to the neutron
flux. The latter is varied by changing the distance of the source
from the cryostat. In this way individual absorption events can be
studied, which are well separated in time. A calibration plot of
the measured rate ${\dot N}$, as a function of the distance
between sample and source, is shown in Fig.~\ref{SourceDistance}.
By means of this plot the results can be scaled to correspond to
the same incident neutron flux, for example with the source at the
minimum distance $d=22$ cm, which is given by the outer radius of
the liquid He dewar.

A most informative feature is the dependence of ${\dot N}$ on rotation velocity
(Ruutu et~al.\ 1996a,b, 1998a). Rotation produces the applied bias velocity at
the neutron absorption site. Since the absorption happens close to the wall
of the container, the velocity of this bias flow is approximately equal to
$v(R)$ (Eq.~(\ref{CFvel})), the counterflow velocity at the cylinder wall.
We shall denote this velocity simply with $v$. It depends both on the
angular velocity $\Omega$ and the number of vortex lines $N(t)$ which are
already present in a central cluster.  The bias flow provides the force
which allows vortex rings to escape from the heated neutron bubble and to
expand to rectilinear vortex lines, which are then preserved in the central
cluster.

\begin{figure}[!!!tb]
  \centerline{\includegraphics[width=0.90\columnwidth]{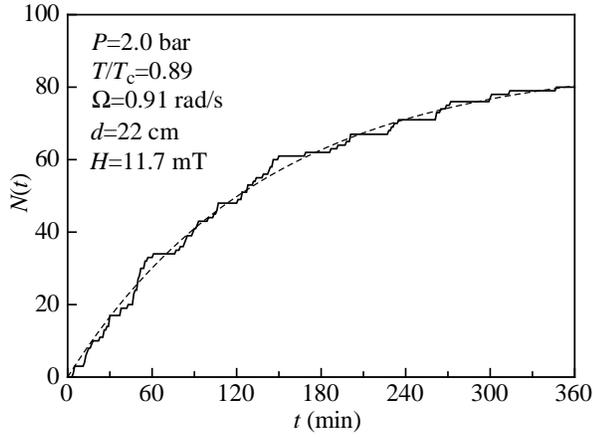}}
  \bigskip
\caption[Saturation] {
  Cumulative number of vortex lines $N(t)$ as a function of time $t$, after
  turning on the neutron irradiation at constant flux on an initially
  vortex-free $^3$He-B sample rotating at constant $\Omega$.  The dashed
  curve represents a fit to Eq.~(\protect\ref{N(t)curve}) with $\gamma =
  1.1$ min$^{-1}$ and $v_{\rm cn} = 1.9$ mm/s. The result demonstrates
  that the rate of vortex-line formation is controlled by the counterflow
  velocity $v(N)$ and not by that of the normal component
  $v_{\rm n}(R) = \Omega R$, which is constant during this entire measurement.
  (From Ruutu et~al.\ 1996b).}
   \label{Saturation}
\end{figure}

These features can be seen in Fig.~\ref{Saturation}. It demonstrates that it is the
counterflow velocity $v(N)$, which is the important variable governing the rate
${\dot N}$ and not, for instance, that of the normal component, $v_{\rm n}(R) =
\Omega R$. Here the total vortex number $N(t)$ is recorded during neutron
irradiation at constant external conditions over a time span of six hours. The
irradiation is started from the vortex-free state: $N(t=0)=0$. The rate ${\dot
N}(t)$, at which vortex lines accumulate in the center of the container, is not
constant: Initially both the counterflow velocity and the rate of vortex formation
are the highest. When more vortex lines collect in the central cluster, both the
velocity $v$ and the rate $\dot N$ fall off. Finally the vortex number $N$
approaches a saturation value, beyond which no more lines are formed. This means
that there exists a lower limit, a threshold value for the counterflow velocity $v$,
below which the neutron absorption events produce no vortex lines at all.

\begin{figure}[!!!tb]
  \centerline{\includegraphics[width=0.95\columnwidth]{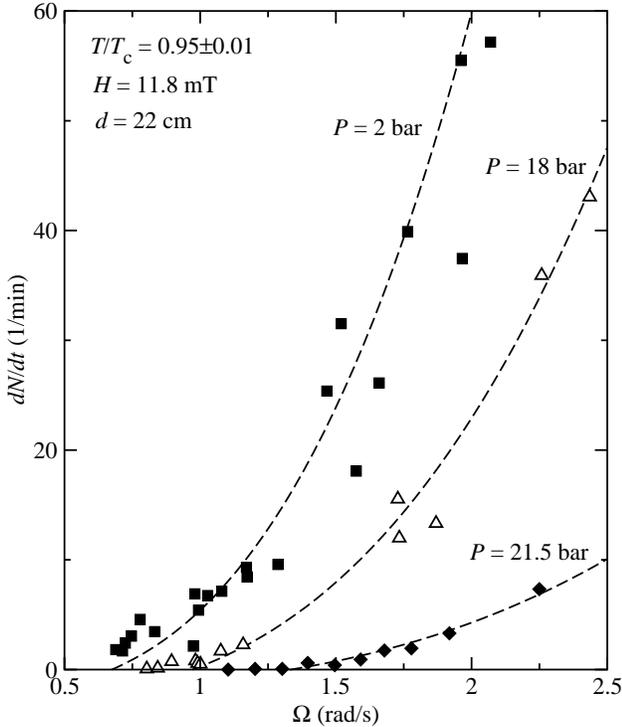}}
  \bigskip
\caption[Ndotraw]{
  Initial rate $\dot N$ of vortex line formation at different rotation
  velocities $\Omega$. $\dot N$ has been measured as the average of
  all vortex lines formed during an accumulation period of 15 -- 30
  min.  The fitted curves are of the form
  $\gamma \, [(\Omega/\Omega_{\rm cn})^3 - 1]$ (Eq.~(\protect\ref{Ndot})). Their low
  velocity end points determine the threshold velocity $v_{\rm cn} =
  \Omega_{\rm cn} R$.  }
   \label{Ndotraw}
\end{figure}

The dependence of $\dot N$ on the counterflow velocity $v$ is obtained from
experiments like that in Fig.~\ref{Saturation}. However, more efficient is
a measurement of only the initial slope of the $N(t)$ record. Typically
15\,--\,30 min accumulation periods are then enough. Only a small number of
vortices form during such a run compared to the number of vortices in the
equilibrium state. Thus the total decrease in the counterflow velocity is
small and can be accounted for by assigning the average of the velocities
before and after irradiation as the appropriate value of the applied bias
velocity. An example of the measured rates is shown in Fig.~\ref{Ndotraw}.

These measurements reveal a vortex-formation rate ${\dot N}$ as a function of $v$
which has an onset at a threshold velocity $v_{\rm cn}$, followed by a rapid
non-linear increase. As shown in Fig.~\ref{ExpSignature}, close above the threshold
a successful neutron absorption event produces one vortex line, but at high flow
rates many lines may result. Therefore the nonlinear dependence ${\dot N}(\Omega)$
arises in the following manner: Initially the fraction of successful absorption
events increases, ie. of those which produce one new vortex line. Eventually this
effect is limited by the neutron flux, when essentially all absorbed neutrons
produce at least one new vortex. The second part of the increase is brought about by
the fact that more and more lines are produced, on an average, in each absorption
event.

\begin{figure}[!!!tb]
  \centerline{\includegraphics[width=0.90\columnwidth]{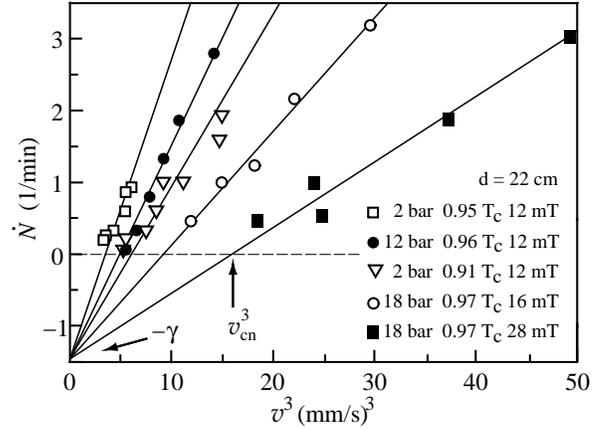}}
  \bigskip
\caption[CritVel]{
  Vortex-formation rate $\dot N$ plotted versus the cube of the
  bias velocity $v$. This plot is used to determine the
  threshold velocity $v_{\rm cn}$. A line has been fit to each set of
  data, measured under different external conditions, to identify the
  horizontal intercept $v^3_{\rm cn}$ and the common vertical intercept
  $-\gamma = -1.4$ min$^{-1}$. (From Ruutu et~al.\ 1996a).}
   \label{CritVel}
\end{figure}

As demonstrated in Fig.~\ref{CritVel} the dependence of the rate $\dot N$ on
the counterflow velocity $v$ can be approximated by the
empirical expression
\begin{equation}
   {\dot N} (v) = \gamma \left[ \left( {v \over
      v_{\rm cn}} \right)^3 - 1 \right]  \label{Ndot}
\end{equation}
with $\gamma$ and $v_{\rm cn}$ as parameters.
This equation also describes the results in Fig.~\ref{Saturation}: If initially $v$
in the vortex-free state is only slightly larger than the critical velocity $v_{\rm
cn}$, then Eq.~(\ref{Ndot}) can be linearized to $\dot N \simeq 3 \gamma (v/v_{\rm
cn} - 1)$. Thus $N(t)$ is obtained by integration from
\begin{equation}
  \dot{N}(t) = 3 \gamma \left(
    \frac{v(0) - \kappa N(t)/(2\pi R)}{v_{\rm cn}} -1
    \right)
\end{equation}
with the solution
\begin{equation}
  N(t) = \frac {2\pi R} {\kappa} \; \left[ v(0) -
  v_{\rm cn} \right] \; \left[ 1 - \exp{ \left( - \frac {3\gamma
        \kappa t } {2 \pi R v_{\rm cn}} \right) } \right] \;.
\label{N(t)curve}
\end{equation}
This equation has been fit to the measurements in Fig.~\ref{Saturation}, to give the
two parameters $\gamma$ and $v_{\rm cn}$. The resulting values agree with those of
the horizontal and vertical zero intercepts in Fig.~\ref{CritVel}.

\begin{figure}[!!!tb]
  \centerline{\includegraphics[width=1.0\columnwidth]{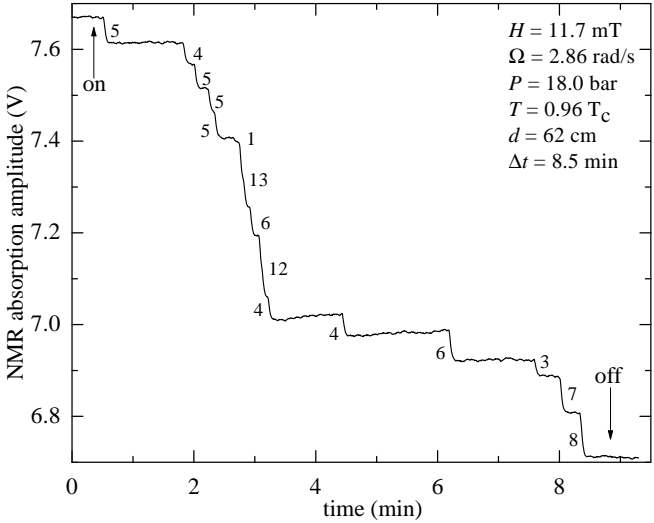}}
  \bigskip
  \caption[VortexLineYield] {Neutron irradiation and
    vortex-line yield at high rotation velocity. The total number of lines,
    which are created in one irradiation session, can be determined in
    several ways: 1) In this example the direct count of the steps in
    signal amplitude yields 88 lines (cf.  Fig.~\ref{ExpSignature}). The
    number next to each step denotes its equivalent in rectilinear
    vortex lines. 2) From Fig.~\ref{VorSignal} one finds that the drop in
    signal amplitude per vortex line is 11 mV in the present measuring
    conditions. The total amplitude drop of 960 mV corresponds thus to 87
    vortex lines. 3) A measurement of the annihilation threshold, by
    successive deceleration to lower and lower $\Omega$ (Ruutu et~al.\ 1997a),
    gives the total number of lines in the vortex cluster. Here this
    measurement yields $\Omega_{\rm v} = 0.245$ rad/s. The annihilation
    threshold, which at $\Omega_{\rm v} \gtrsim 0.2$ rad/s coincides with
    the equilibrium vortex state, corresponds to $N = \pi R^2 2 \Omega_{\rm
      v} \kappa^{-1} (1-0.18\sqrt{1\,\mbox{rad\,s}^{-1}/\Omega_{\rm v}})$,
    as measured by Ruutu et~al.\ (1998b). With $R= 2.45$ mm, this gives 88
    vortex lines.  4) The rate equation (\ref{Ndot}), with the measured
    values of $\gamma$ and $v_{\rm cn}$ from Fig.~\ref{CritVel}, gives 88 lines for an
    irradiation time of 8.5 min (with
    $\Omega_{\rm cn} = 0.75$ rad/s). Thus all of the reduction in the height
    of the counterflow peak (cf. Fig.~\ref{VorNMR}) can be
    reliably accounted for in many different ways. }
  \label{VortexLineYield}
\end{figure}
At high rotation velocities a large number of vortex lines is
produced in each neutron absorption event. The accumulation record
for one neutron irradiation session (similar to the upper trace in
Fig.~\ref{ExpSignature}) then consists of many large steps where,
in the worst case, two absorption events might even be
overlapping. Nevertheless, the final total number of lines, which
have been collected into the central cluster, can be determined in
a few different ways and these give consistent answers. These
tests have been described in Fig.~\ref{VortexLineYield}. One of
them includes the use of the rate equation (\ref{Ndot}) where the
multiplier $\gamma$ has the value from Fig.~\ref{CritVel}.

In fact, it can be guessed from Fig.~\ref{CritVel} that Eq.~(\ref{Ndot})
has wider applicability than is apparent from the present examples: The
same equation holds universally under different externally applied
conditions with all dependence on external variables contained in $v_{\rm
  cn}(T,P,H)$, while $\gamma$ is proportional to the neutron flux but does
not depend on temperature, pressure, or magnetic field. Hence ${\dot
  N}(v/v_{\rm cn})$ appears to be a universal function for all measurements
which have been performed in the temperature regime $T > 0.8 \, T_{\rm c}$.

To summarize, the experimental result displays two distinguishing features: 1) the
cubic dependence on the bias $v$ and 2) the universality that all dependence of the
vortex formation properties on the experimental variables $T$, $P$, and $H$ is
contained in the threshold velocity $v_{\rm cn}$.

\subsection{Volume or surface mechanism?} \label{MechVorForm}

In the following sections we examine the various experimental
properties of the neutron-induced vortex formation. The analysis
demonstrates that most features are consistent with the KZ mechanism -- none have so far been found contradictory. But does this fact constitute solid proof of the mechanism?

Neutron-induced vortex formation in a rotating superfluid does not exactly fit the ideal KZ model --- a rapid second order transition in an infinite homogeneous system. In the heated neutron bubble there is a strong thermal gradient and a strict boundary condition
applies at its exterior, imposed by the bulk superfluid state outside. The
comparison of experiment and model is further complicated by the fact that
the cool down occurs so fast that an extrapolation from the equilibrium
state theories becomes uncertain, whether it concerns the hydrodynamics or
even the superfluid state itself. (The applicability of the KZ mechanism to
inhomogeneous transitions is considered in more detail in
Sec.~\ref{TransMoveFront}.) Perhaps, other processes of hydrodynamic origin
can be suggested which account for the experimental observations?

The most viable alternative arises from the boundary condition in the presence of the applied bias flow: It is the instability of superflow along the normal-superfluid interface in the outer peripheral regions of the neutron bubble.  In the Ginzburg-Landau temperature regime the container-dependent spontaneous critical velocity $v_{\rm c}(T,P)$ decreases with increasing temperature and vanishes at $T_{\rm c}$ (dashed curves in
Fig.~\ref{CritVel&NeutrThreshold}). It is related (via surface roughness
enhancement) to the intrinsic instability velocity $v_{\rm cb}(T,P)$ of the
bulk superfluid (Parts et~al.\ 1995; Ruutu et~al.\ 1997a), which has qualitatively a similar
temperature and pressure dependence, but is larger in magnitude.  In the outermost region of the neutron bubble the fluid remains in the B phase, but is heated above the surrounding bulk
temperature $T_0$. Consequently, if no other process intervenes, the superflow instability has
to occur within a peripheral shell surrounding the hot neutron bubble where
$v_{\rm cb}(T,P)$ drops below the applied bias $v$. This process was discussed by Ruutu et~al.\ (1998a) and compared to rotating measurements, but was found incompatible with the measured results. As explanation it was suggested that the KZ mechanism is inherently the fastest process for creating vortices.

The interplay between the interior volume of the neutron bubble and its boundary was examined by Aranson et~al.\ (1999) in a rapidly cooling model system with a scalar order parameter, using
the thermal diffusion equation to account for cooling and the
time-dependent Ginzburg-Landau equation for order-parameter relaxation. Rigorous analytic calculations showed that the normal-superfluid interface becomes unstable in the
presence of superflow along the interface. Numerical calculations
(Aranson et~al.\ 1999, 2001) further confirmed that the instability develops such that vortex rings are formed. The rings encircle the bubble perpendicular to the applied flow and screen the superflow, so that the superfluid velocity (in the rotating frame) is zero inside the bubble. The number of rings depends on the shape and size of the bubble and on the bias velocity. In this process it does not matter what the state of the liquid is inside the bubble:
The rings are formed within an external cooler layer, which remains throughout the
process in the B phase. After their formation, the rings start to expand and might
eventually be pulled away by the Magnus force. Thus these rings could be the source for the
vortex lines which are observed in the rotating measurements.

However, the calculations also confirmed that in addition to the superflow
instability at the bubble boundary, the KZ mechanism in the bubble interior is also
present -- thus in this calculation both processes produce vortices. In the simulation the loops, which eventually manage to escape into the bulk bias flow, originate from the boundary while the random vortex network in the interior collapses and decays away. Thus Aranson et al. confirmed the presence of the KZ mechanism, but concluded that the vortices observed in the rotating measurements are  produced by the boundary instability. The time-dependent Ginzburg-Landau treatment with one common relaxation time for both the phase and the magnitude of the order parameter is rigorously obeyed only within a narrow temperature interval from $T_{\rm c}$ down to $1-T/T_{\rm c} \sim 10^{-3}$. The bulk liquid temperature in the rotating measurements is at least an order of magnitude below this range (see Fig.~\ref{CritVel&NeutrThreshold}).  Another difference between experiment and calculation lies in the size of the neutron bubble: The numerical calculation is carried out on a grid with unit length smaller than  the coherence length $\xi(T_0)$. To make this tractable, the reaction energy $E_0$ is reduced by one or two orders of magnitude. This scales down the neutron bubble radius $R_{\rm b}$ via the thermal diffusion equation, which is used to model the neutron bubble (see Sec.~\ref{ThreshVelocity}).  A smaller bubble (see Eq.~(\ref{e.2})) means a faster velocity of the phase front $( \propto 1/R_{\rm b})$. At present time it is not known how reliably the calculation with its differences extrapolates into the range of the measurements.

How do we, in fact, distinguish from measurements whether it is the KZ mechanism or the boundary instability which is responsible for the observed vortex lines? Clearly the critical velocity $v_{\rm cn}$ does not discriminate between them. In both cases $v_{\rm cn}$ is the velocity of the
bias flow at which the largest vortex ring which fits in the bubble is
stable, as will be shown in Sec.~\ref{LoopEscape}.  However, the different nature
of the two processes has important consequences for the vortex formation
rate ${\dot N}(v)$. The row of vortex rings, which is produced by the boundary instability to screen the interior of the neutron bubble from the bias flow, resembles a vortex sheet. The sheet is formed in the heated peripheral shell of the neutron bubble, where
the outside superfluid moves with velocity $v$ along the bubble interface and the
superfluid inside is stationary. The density of vorticity in the sheet is $v/\kappa$
and the number of loops produced by one neutron absorption event is $N \propto v
R_{\rm b}/\kappa$, where $R_{\rm b}$ is the size of the bubble along the flow
direction. Thus $N(v)$ grows linearly with the applied bias flow velocity, $N(v) \propto v$, which agrees with the result of Aranson et~al.\ (1999). In contrast, the KZ mechanism is a volume effect, which results in the cubic dependence expressed in Eq.~(\ref{Ndot}) (see also Sec.~\ref{AnalVorEsc}). The measured result in the figures of the previous section is not linear, but is consistent with the cubic dependence.

Are there other arguments emerging from the measurements which would support the KZ origin of the observed vortices? Several more can be listed in addition to the rate dependence ${\dot N}(v)$, which favor a volume effect: (i) The boundary instability is expected to be a deterministic process which in every neutron event produces the same surface density of
vortex rings.  Variations in the number of rings arise only owing to
variations in the shape of the neutron bubble and its orientation with
respect to the bias flow. (Unfortunately, the distribution of the number of
rings was not studied in detail in the simulations by Aranson et~al.\ 2001.) In
contrast, the KZ mechanism produces a random vortex network from which the
number of loops extracted by the bias flow varies and follows a stochastic
distribution (Fig.~\ref{histo}). (ii) The
boundary instability should not be particularly sensitive to the processes
inside the heated neutron bubble and only regular mass-flow
vortices are expected to form in such an instability. The KZ mechanism,
on the other hand, can be expected to produce all possible defects. Their
presence might be either directly observed in the final state or via their
influence on the evolution of the vortex network inside the neutron bubble. Such observations are discussed in Sec.~\ref{OtherDefects}.

\subsection{Threshold velocity for vortex loop escape} \label{LoopEscape}

\begin{figure}[!!!tb]

\centerline{\includegraphics[width=0.9\columnwidth]{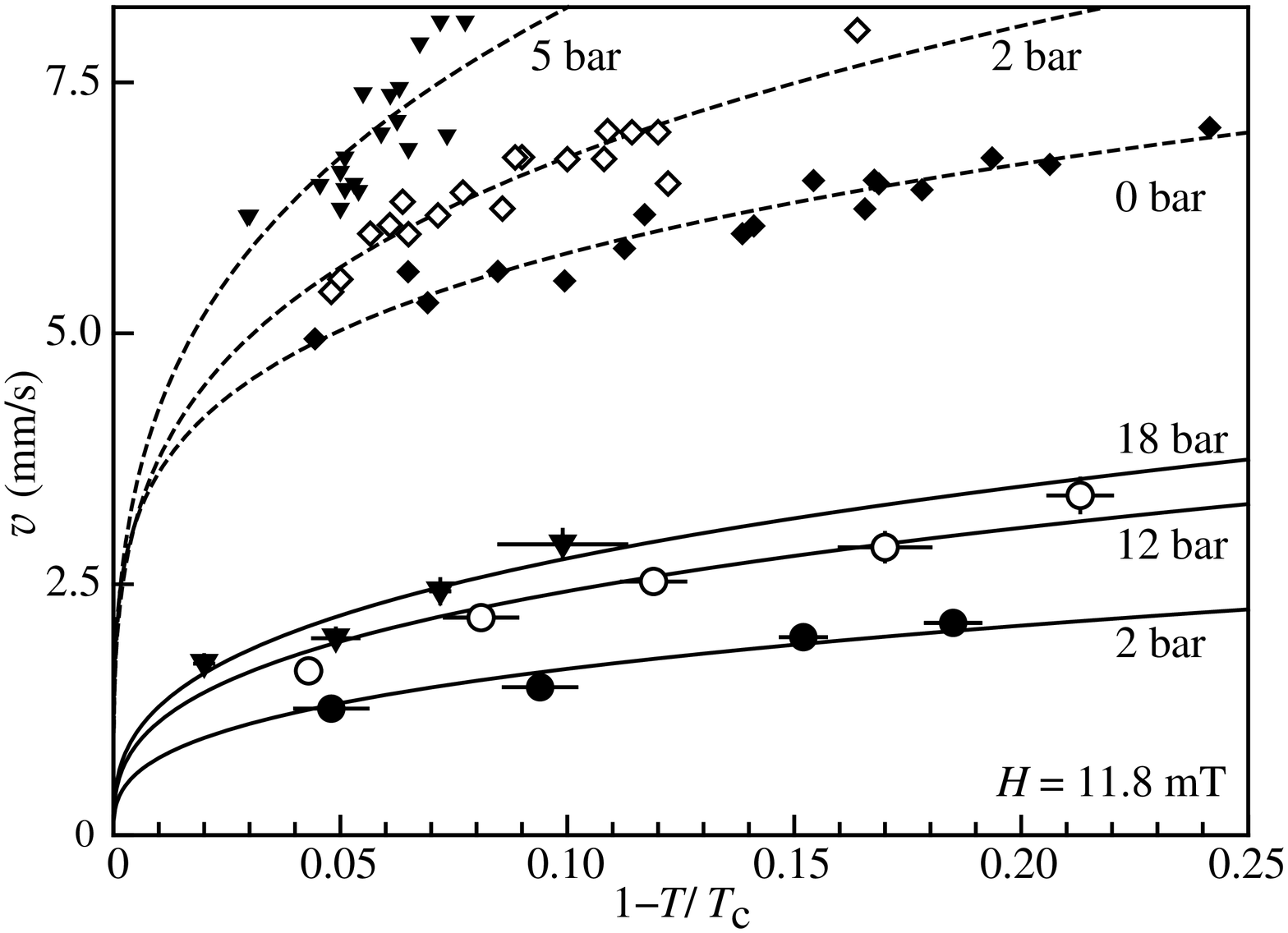}}
  \bigskip
\caption[CritVel&NeutrThreshold]{
  Critical values of the applied bias velocity $v$ vs.
  temperature at different pressures:
  ({\it Solid lines})
  Threshold velocity $v_{\rm cn} \propto$ $(1-T/T_{\rm c})^{1/3}$ for
  vortex formation in neutron irradiation
  (Ruutu et~al.\ 1996a). ({\it Dashed lines}) Critical velocity
  $v_{\rm c} \propto (1-T/T_{\rm c})^{1/4}$ of spontaneous vortex
  formation in the same quartz glass container in the
  absence of the neutron source. (From Parts et~al.\ 1995; Ruutu et~al.\ 1997a). }
   \label{CritVel&NeutrThreshold}
\end{figure}

The dependence on the counterflow bias $v$ can be studied from the threshold $v_{\rm
cn}$ up to the critical limit $v_{\rm c}$ at which a vortex is spontaneously
nucleated at the cylindrical wall in the absence of the neutron flux
(Parts et~al.\ 1995; Ruutu et~al.\ 1997a). The threshold velocity $v_{\rm cn}(T,P,H)$ is one of the
features which can be examined to learn more about neutron-induced vortex formation.
Measurements of both critical velocities, $v_{\rm cn}$ and $v_{\rm c}$, are shown in
Fig.~\ref{CritVel&NeutrThreshold} as a function of temperature.

\subsubsection{Properties of threshold velocity} \label{ThreshVelocity}

By definition the threshold $v_{\rm cn}$ represents the smallest bias velocity at
which a vortex ring can escape from the neutron bubble after the absorption event.
It can be connected with the bubble size in the following manner. A vortex ring of
radius $r_\circ$ is in equilibrium in the applied bias flow at $v$ if it satisfies
the equation
\begin{equation}
  r_\circ(v) = {\kappa \over {4\pi v}} \; \ln{\left(
    {r_\circ \over \xi(T,P)} \right)} \, .
\label{VorRing}
\end{equation}
As explained in more detail in Sec.~\ref{AnalVorEsc}, a ring with a
radius larger than $r_\circ$ will expand in the flow while a smaller one will
contract. Thus the threshold or minimum velocity at which a vortex ring can start to
expand towards a rectilinear vortex line corresponds to the maximum possible
vortex-ring size. This must be comparable to the diameter of the heated bubble. For
a simple estimate we set the vortex ring radius equal to that of a spherical neutron
bubble: $r_\circ (v_{\rm cn}) \sim R_{\rm b}$. (In fact, the numerical simulations
to be described in Sec.~\ref{EvolSimSec} suggest that $r_\circ (v_{\rm cn}) \approx
2R_{\rm b}$ because of the complex convoluted shape of the largest rings in the
random vortex network.)

A simple thermal diffusion model can be used to yield an order of magnitude estimate
for the radius $R_{\rm b}$ of the bubble which originally was heated above $T_{\rm
c}$. In the temperature range close to $T_{\rm c}$ the cooling occurs via diffusion
of quasiparticle excitations out into the surrounding superfluid with a diffusion
constant $D \approx v_{\rm F} l$, where $v_{\rm F}$ is their Fermi velocity and $l$
their mean free path.  The difference from the surrounding bulk temperature $T_0$ as
a function of the radial distance $r$ from the centre of the bubble can be
calculated from the diffusion equation
\begin{equation}
{{\partial T(r,t)} \over
  {\partial t}} = D \left({{\partial^2 T} \over {\partial r^2}} + {2 \over
  r} {{\partial T} \over {\partial r}}\right) \; .
\label{TherDiff}
\end{equation}
With the assumption that at $t=0$ the reaction energy $E_0$ is deposited at
$r=0$, the solution is given by
\begin{equation}
  T(r,t) - T_0 \approx {E_0\over C_{\rm v}} \;
  {1\over (4 \pi D t )^{3/2}} \; \exp \Biggl ({-r^2\over 4Dt} \Biggr ),
\label{e.1}
\end{equation}
where $C_{\rm v}$ is the specific heat. For now, we may assume that all
the energy of the neutron absorption reaction is deposited as
heat. The bubble of normal fluid, $T(r) > T_{\rm c}$, first
expands and reaches a maximum radius
\begin{equation}
R_{\rm b} = \sqrt {3 \over {2 \pi e}} \;
\left( {E_0 \over {C_{\rm v} T_{\rm c}}}
\right)^{1/3} \;  (1-T_0/T_{\rm c})^{-1/3} \; .
\label{e.2}
\end{equation}
It then starts cooling and rapidly shrinks with the characteristic
time $\tau_{\rm Q} \sim R_{\rm b}^2/D \sim 1 \mu$s. Since $v_{\rm cn}$
is inversely proportional to $r_\circ \sim R_{\rm b}$, it has the
temperature dependence $v_{\rm cn} \propto (1-T_{0}/T_{\rm c})^{1/3}$. This
is in agreement with the solid curves in Fig.~\ref{CritVel&NeutrThreshold}
which
have been fitted to measurements on $v_{\rm cn}$. The prefactor of these curves
is in agreement with that from Eqs.~(\ref{e.2}) and (\ref{VorRing}) within a
factor of $\sim 2$, and its increase with increasing pressure is well
described by the decrease in bubble size according to Eq.~(\ref{e.2}),
where $C_{\rm v}$ and $T_{\rm c}$ increase with pressure (Greywall 1986).

\begin{figure}[!!!tb]
  \centerline{\includegraphics[width=1.0\columnwidth]{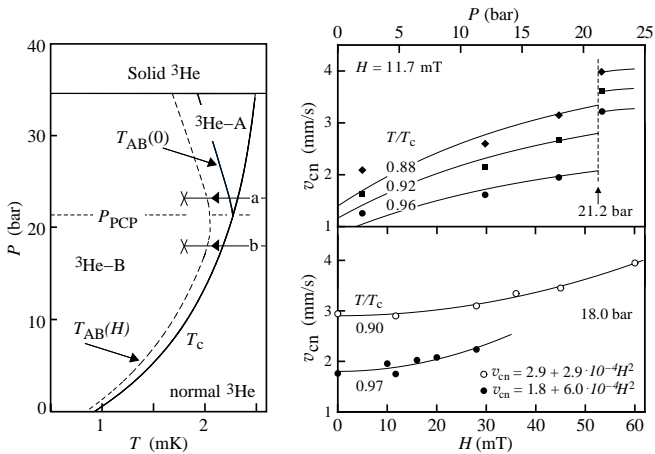}}
  \bigskip
\caption[PresDepend]
{Threshold velocity $v_{\rm cn}$ for the onset of vortex formation
  during neutron irradiation: {\it (Left)} Phase diagram of $^3$He
  superfluids in the pressure vs.  temperature plane, with the
  A$\rightarrow$B transition at $T_{\rm AB}(0)$ in zero field (solid line)
  and at $T_{\rm AB}(H)$ in nonzero field (dashed line). Two quench
  trajectories, distinguishing different types of measurements on
  the right, are marked with (a) and (b). {\it (Right top)} The pressure
  dependence of $v_{\rm cn}$ displays a steep change at the pressure
  $P_{\rm PCP}$ of the polycritical point.  {\it (Right bottom)} The
  magnetic field dependence of $v_{\rm cn}$ is parabolic, similar to
  that of the equilibrium state A$\rightarrow$B transition
  $T_{\rm AB}(H)$. (From Ruutu et~al.\ 1998a).  }
   \label{PresDepend}
\end{figure}

\subsubsection{Influence of $^3$He-A on the threshold velocity $v_{\rm cn}$} \label{A-phase}

Measurements on the threshold velocity $v_{\rm cn}$ offer interesting possibilities
to study whether the interior of the heated bubble participates in vortex formation
or not. In the left panel of Fig.~\ref{PresDepend} the phase diagram of $^3$He is
shown at temperatures below 2.5 mK. In the measurements which we discussed so far,
the liquid pressure has been below 21.2 bar, so that the heated bubble cools from
the normal phase directly into the B phase (along trajectory b). On the top right in
Fig.~\ref{PresDepend} the measured $v_{\rm cn}$ is plotted as a function of pressure
$P$ at constant reduced temperature $T/T_{\rm c}$. Here the pressure dependence
displays an abrupt increase at about 21.2 bar, the pressure $P_{\rm PCP}$ of the
polycritical point: It thus makes an unexpectedly large difference whether the
quench trajectory follows a path denoted with a or with b! The two cooling
trajectories differ in that above $P_{\rm PCP}$ a new phase, $^3$He-A, is a stable
intermediate phase between the normal and B phases. Consequently, although the bulk
liquid is well in the B phase in all of the measurements of Fig.~\ref{PresDepend},
vortex formation is less than expected on the basis of extrapolations from lower
pressures, when the quench trajectory crosses the stable A-phase regime.

The fitted curves in Fig.~\ref{PresDepend} represent $v_{\rm cn}(P)$ at pressures
below $P_{\rm PCP}$
\begin{equation}
v_{\rm cn}= {\cal A} \kappa/(4\pi R_{\rm b}) \; \ln{(R_{\rm b}/\xi)} \; ,
\label{e.vcn}
\end{equation}
where $R_{\rm b}$ is obtained from Eq.~(\ref{e.2}). If we assume that
all of the reaction energy is transformed to heat ($E_0=$ 764 keV), then the common
scaling factor ${\cal A}$ of the three curves has the value ${\cal A} = 2.1$.
This is the only fitting parameter in Figs.~\ref{CritVel&NeutrThreshold} and \ref{PresDepend}
where the same fit is compared to measurements as a function of both temperature and pressure. The agreement is reasonable and suggests that the spherical thermal diffusion model is
not too far off. However, as seen in Fig.~\ref{PresDepend}, $v_{\rm cn}$ is clearly
above the fit at $P > P_{\rm PCP}$. Also this offset is largest at the highest
temperature, which in Fig.~\ref{PresDepend} is $0.96\; T_{\rm c}$. Both features
suggest that vortex formation is reduced when the relative range of A-phase
stability increases over the quench trajectory below $T_{\rm c}$.

The lower right panel of Fig.~\ref{PresDepend} shows the dependence of $v_{\rm cn}$
on the applied magnetic field $H$. Below $P_{\rm PCP}$ the magnetic field acts to
stabilize $^3$He-A in a narrow interval from $T_{\rm c}$ down to the first order
A$\rightarrow$B transition at $T_{\rm AB}(P,H)$. This result confirms our previous
conclusion: $v_{\rm cn}$ again increases when the range of A-phase stability
increases. The parabolic magnetic field dependence of $v_{\rm cn}(H)$ in the lower
right panel is reminiscent of that of the equilibrium $T_{\rm AB} (H)$ transition
temperature.\footnote{According to Tang et~al.\ (1991), below $P_{\rm PCP}$ the
equilibrium A$\leftrightarrow$B transition temperature is in first order of the form
$T_{\rm AB} (P,H) = T_{\rm c}(P) \, (1- b H^2)$, where $b (P) \sim (0.5$
-- $10) \cdot 10^{-6}$ (mT)$^{-2}$.} It should be noted that the measurements in the
upper right panel of Fig.~\ref{PresDepend} have not been carried out in zero
magnetic field but in 11.7\,mT. However, as shown in the lower panel, the field
dependence in the range 0\,--\,12\,mT is not visible within the experimental
precision.

Thus by increasing the pressure above $P_{\rm PCP}$ or by increasing the magnetic
field we reduce the A-phase energy minimum relative to that of the B phase and both
operations act to increase $v_{\rm cn}$. If we go back to Figs.~\ref{Ndotraw} and
\ref{CritVel}, we note that both operations seem to leave the rate equation
(\ref{Ndot}) and its rate constant $\gamma$ unchanged. To be more careful, we conclude that in first order the changes from increased A-phase stability appear to affect only $v_{\rm
cn}$. This conclusion is compatible with the KZ model, as will be discussed in the
next sections. What about the competing model, the superflow instability at
the boundary of the heated neutron bubble, can it also account for the observations in Fig.~\ref{PresDepend}? At first glance this does not appear to be the case.

The superflow instability occurs in B phase in the outer periphery of
the heated bubble in a shell where the temperature is close to $T_{\rm c}$.
In principle this phenomenon should not be influenced by the appearance of
A phase in the interior of the bubble during its cool down to the bath
temperature $T_0$. However, it can be affected by the presence of an A
phase shell around the hot bubble, bounded by an inner surface $T = T_{\rm c}$ and an outer $T = T_{\rm AB}$, {\it i.e.} a region where the temperature
exceeds the AB transition temperature $T_{\rm AB}(P,H)$. The B-phase bulk critical velocity
is approximately equal to the pair-breaking velocity
and in the Ginzburg-Landau temperature regime it is of the form
(Vollhardt et~al.\ 1980; Kleinert 1980)
$$v_{\rm cb}(T,P) \approx v_{\rm c0}(P) \; \sqrt{1-T/T_{\rm
c}}\, ,$$ where $ v_{\rm c0}(P) = 1.61 \; (1 + F_1^s /3) \; k_{\rm B} T_{\rm
c}/p_{\rm F}$. This velocity is a smooth function of pressure and cannot account for
the steep rise of $v_{\rm cn}$ at $P_{\rm PCP}$ in Fig.~\ref{PresDepend} (upper right panel). The
B-phase boundary instability occurs within an outer shell where the inner surface is defined by $T = T_{\rm c}$ and the outer by the condition $v_{\rm cb}(T,P) = v$. For the
data in Fig.~\ref{PresDepend} the latter condition puts the outer surface of this shell at $1- T/T_{\rm c} \sim 10^{-3}$ -- $10^{-5}$, which falls inside the shell within which A-phase is stable. Thus here it is the A-phase shell within which the boundary instability has to occur. Such a situation is complicated (see Blaauwgeers et~al.\ 2002; Parts et~al.\ 1993) and possibly one where the escape of vorticity from the A-phase shell into B-phase is reduced. In first order these considerations suggest that $v_{\rm cn}$ increases with increasing pressure and magnetic field, when the range of the stable A-phase regime increases over the length of the quench trajectory as a function of temperature. However, it is unclear whether in a 3-phase problem --- where an outer region of  warm superfluid surrounds a hot interior with normal liquid, while both are embedded in a cold bath of B-phase ---  coherence is established such that we can apply shell structure considerations. Therefore, in the absence of more quantitative estimates, we must conclude that the measurements in Fig.~\ref{PresDepend} do not yet allow a preference between the KZ model or the superflow instability at the neutron bubble boundary. Clearly more extensive measurements are needed to study rapid thermal quenches in the 3-phase situation, as will be emphasized in Sec.~\ref{MeasureRate}.

\subsection{Other defect structures formed in neutron irradiation}
\label{OtherDefects}

\subsubsection{Radiation-induced supercooled A $\rightarrow$ B transition}
\label{ABtransition}

The first order transition of supercooled $^3$He-A to $^3$He-B is another
experimentally confirmed phenomenon which is catalyzed by the interior of
the neutron bubble. Usual arguments about homogeneous first-order phase
transitions show that the A$\rightarrow$B transition is forbidden
(Leggett 1984). The normal phase symmetry $SO(3)\times SO(3) \times U(1)$
of liquid $^3$He can be broken to the $U(1) \times U(1)$ symmetry of the A
phase or to the $ SO(3)$ symmetry of the B-phase, with only a small energy
difference between these two states, but separated by an extremely high
energy barrier ($\sim 10^6 \, k_{\rm B}T_{\rm c}$). Nevertheless, so far it has not
been reported that the transition would not have taken place ultimately, if
only one cools to sufficiently low temperature.

The A$\rightarrow$B transition has not been found to be affected by rotation (Hakonen et~al.\ 1985) and in the absence of ionizing radiation it is generally believed to be caused by
extrinsic sources, perhaps at solid surfaces. A dramatic increase in the transition
probability has been shown to take place in the bulk liquid in the presence of
$\gamma$ radiation or thermal neutrons (Schiffer et~al.\ 1992; Schiffer and Osheroff 1995). A generally accepted understanding of this feature is still missing, because two quite different models, the ``baked Alaska'' (Leggett 1992; Schiffer et~al.\ 1995; Leggett 2002) and the
KZ mechanism (Volovik 1996; Bunkov and Timofeevskaya 1998a,b), are discussed in this
context. Since only the latter is appropriate in radiation-induced vortex formation,
we briefly describe it in the context of the supercooled A$\rightarrow$B transition.

When $^3$He-A is supercooled to low temperatures, below the thermodynamic
A$\leftrightarrow$B transition, the sample stays in the A phase for long times,
perhaps even indefinitely, depending on temperature and other largely unknown
requirements. The deeper the supercooling, the faster the
transition into the B phase follows. With increasing magnetic field the transition
probability decreases, since a magnetic field acts to stabilize the A phase. However, when neutron irradiation is turned on, the transition probability increases sharply. In such measurements the initial state is supercooled $^3$He-A at varying temperatures and magnetic fields. Consequently here the A-phase state is enforced by the boundary condition around the hot neutron bubble: The boundary condition is not the source for the A$\rightarrow$B transition! Rather the new
B-phase state has to emerge from the interior of the bubble.


\begin{figure}[!!!tb]
  \centerline{\includegraphics[width=1.0\columnwidth]{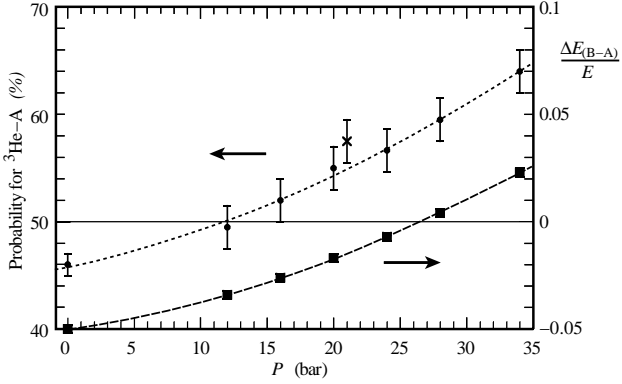}}
  \bigskip
\caption[AB_EnergyDifference] {Normalized energy difference, $\Delta E_{\rm (B-A)} /E(P)$,
  between A and B phases {\it (right vertical axis)} and probability of A-phase formation in a
  rapid cool-down {\it (left vertical axis)}, plotted as a function of pressure at
  temperatures close to $T_{\rm c}$ in the Ginzburg-Landau regime. The
  results have been calculated with the weak coupling $\beta$
  parameters $(\beta_i = -1,2,2,2,-2)$, except for the data point
  marked with a cross $(\times)$ on the upper curve, for which
  strong-coupling parameter values have been used according to the
  spin fluctuation model. (From Bunkov and Timofeevskaya 1998a,b).  }
\label{AB_EnergyDifference}
\end{figure}


According to the KZ model, in different parts of the rapidly cooling neutron bubble
the order parameter may fall initially either in the A or B phase energy minima.
Thus a domain structure of size $\xi_{\rm v}$ (Eq.~\ref{eq:xi-initial}) is laid
down, where A and B-phase blobs form a patch work. When these blobs later grow
together, then AB interfaces are formed. The lower the temperature, the lower is the
B-phase energy minimum relative to that of the A phase, the larger is the proportion
of B-phase blobs in the patch work, and the more likely it is that many of them
manage to merge together to form one large bubble, where the AB interface exceeds
the critical diameter of about 1 $\mu$m, needed for spontaneous A-phase expansion to
start. This one seed is then sufficient to initiate the A$\rightarrow$B transition
in the whole sample.

The time dependent Ginzburg-Landau calculations of
Bunkov and Timofeevskaya (1998a,b) are consistent with this model. In the
calculations there is no spatial dependence, an initial fluctuation is
imposed at $T_{\rm c}$ into a random direction of the phase space, and the
evolution towards the final state is followed. As seen in in
Fig.~\ref{AB_EnergyDifference}, the energy difference between the A and B
phases is small, but pressure dependent. During rapid cool down through
$T_{\rm c}$ the order parameter may settle into either of these two energy
minima with a pressure-dependent probability, which is also shown in
Fig.~\ref{AB_EnergyDifference}.

Thus the KZ mechanism provides a simple explanation of the A$\rightarrow$B
transition probability as a function temperature and magnetic field in
supercooled $^3$He-A. This model is independent of the boundary condition,
which dictates $^3$He-A. Furthermore, in ambient conditions, where only B
phase is stable, A-phase blobs ultimately shrink away and only B phase
remains. Before that, however, a network of AB interfaces is created which
interact with the formation of other topological defects. This feature can
be used to explain the sharp increase in $v_{\rm cn}$ at $P_{\rm PCP}$ in
Fig.~\ref{PresDepend} (top right panel).

\subsubsection{Vortex formation, AB interfaces, and KZ mechanism}
\label{ThresholdVelocity+AB_Interfaces}

The KZ model predicts that in different parts of the cooling neutron bubble the
order parameter may fall initially either in A or B phase energy minima, and a patch
work of AB interfaces is laid down. The relative number of A and B-phase blobs
of size $\xi_{\rm v}$ is not only determined by the difference in the A and
B-phase energies: It also depends on the trajectory from the normal phase
to the new energy minimum in the phase space spanned by the order parameter
components. The latter aspect gives more preference to the higher energy A
state: Although the A-phase is energetically not
favorable in a wide pressure range, the probability of its formation has a
value close to 50 \% even at lower pressures, as shown in Fig.~\ref{AB_EnergyDifference}. The AB interfaces interfere with the
simultaneous formation of a random vortex network, which is laid down separately in
both the A and B phases. It is known from experiments on both a moving AB interface
(Krusius et~al.\ 1994) and a stationary AB interface (Blaauwgeers et~al.\ 2002) that the penetration of
vortex lines through the AB phase boundary is suppressed. We thus expect that the
presence of A-phase blobs reduces the combined volume which remains available for B
phase blobs and their vortex network inside a neutron bubble. Consequently, when
B-phase blobs merge to larger units, the overall volume, into which the B-phase
vortex network is confined, is smaller, and B-phase vortex formation becomes
impeded. This process increases the value of $v_{\rm cn}$.

To conclude, the KZ mechanism provides a unified view on the problems of
pressure and magnetic field dependences of $v_{\rm cn}$
(Fig.~\ref{PresDepend}) and also the
A$\rightarrow$B transition in supercooled $^3$He-A.  In both cases AB
interfaces appear as a new type of defect in the rapidly cooling neutron
bubble.  Normally, to create a sizeable bubble of B phase within bulk A
liquid requires that an AB interface of large size is formed.  This is a
slow and energy consuming process. In quench cooling through $T_{\rm c}$
the freeze-out of disconnected blobs with A and B-phase like fluctuations
happens first and these are initially separated by supercooled normal
liquid. The AB interfaces appear later as metastable defects when the
blobs grow together. Such an interpretation suggests that the KZ mechanism
is the fastest process by which defects are created, before other effects
manage to switch on.

\subsubsection{Spin-mass vortex}
\label{SpinMass}

Unlike vortex lines, which are produced in neutron irradiation and are then partly
stabilized by the applied flow, no bias is provided for maintaining AB interfaces in
Fig.~\ref{PresDepend}. Thus in these rotating experiments, the evidence for AB
interfaces remains indirect. However, there exists another topologically stable
defect, which is formed in a neutron absorption event and is preserved by the bias
flow for later examination. This is a combined object called the spin-mass vortex
(Kondo et~al.\ 1992; Korhonen et~al.\ 1993).

\begin{figure}[!!!tb]
  \centerline{\includegraphics[width=1.0\columnwidth]{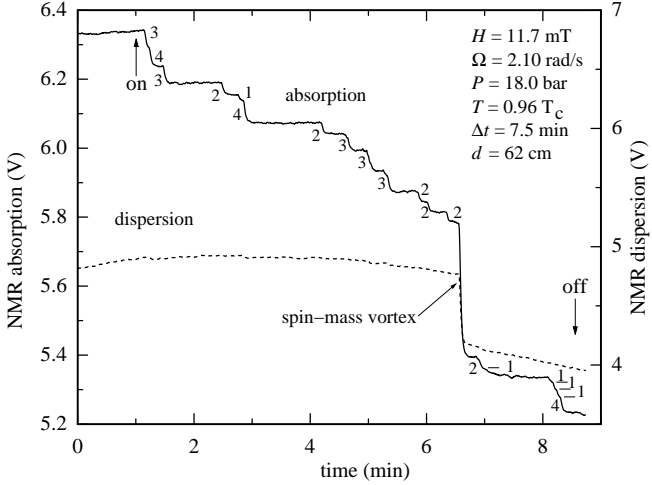}}
  \bigskip
\caption[SpinMassStepSize] {Signal from spin-mass vortex in
  neutron irradiation. This irradiation session includes one large jump in
  signal amplitude. Like in Fig.~\protect\ref{VortexLineYield}, the
  various methods for extracting the total number of lines give: 1) A
  direct count of the steps in signal amplitude yields 45 lines plus the
  large jump.  The combined amplitude drop for these 45 lines amounts to
  716 mV which translates to 16 mV/vortex, in agreement with
  Fig.~\protect\ref{VorSignal}. With this step size the large jump of 390
  mV would correspond to 25 additional lines. 2) A measurement of the
  annihilation threshold gives $\Omega_{\rm v} = 0.155$ rad/s, which
  corresponds to 48 lines. 3) The rate Eq.~(\ref{Ndot}) gives ${\dot
    N} = 6.2$ lines/min or a total of 48 lines for an irradiation of 7.5
  min duration with the measured threshold value of $\Omega_{\rm cn} =
  0.75$ rad/s. For these three estimates to be consistent, the large jump
  cannot be 25 lines.  Instead it is interpreted to represent one spin-mass
  vortex line.  In this case the size of the jump is not related to the
  number of lines, but to the counterflow velocity $v(R)$
  (Eq.~(\protect\ref{CFvel})). It controls the length $v(R)/2\Omega$ of the
  soliton sheet (Fig.~\protect\ref{SpinMassStructure}). The dispersion
  signal also displays a simultaneous large jump, while it behaves smoothly
  when usual mass-current vortex lines are formed. Outside the
  discontinuity the dispersion signal is slowly drifting, because it is
  very sensitive to a residual
  temperature creep. (From Eltsov et~al.\ 2000).} \label{SpinMassStepSize}
\end{figure}

The signature from the spin-mass vortex is explained in
Fig.~\ref{SpinMassStepSize}, where the counterflow peak height is
plotted as a function of time during neutron irradiation (as in
Figs.~\ref{ExpSignature} and \ref{VortexLineYield}). This
accumulation record shows one oversize step in the absorption
amplitude, which coincides with a similar discontinuous jump in
the out-of-phase dispersion signal. As explained in the context of
Fig.~\ref{VortexLineYield}, the total number of vortex lines
extracted from an irradiation session can be determined in several
different ways. In the case of Fig.~\ref{SpinMassStepSize} a
comparison of the different line counts shows that, within the
uncertainty limits, the large jump cannot represent more than a
few vortex lines. As will be seen below, this is the signature
from a spin-mass vortex, which in Fig.~\ref{SpinMassStepSize}
should be counted as one line.

\begin{figure}[!!!tb]
  \centerline{\includegraphics[width=1.0\columnwidth]{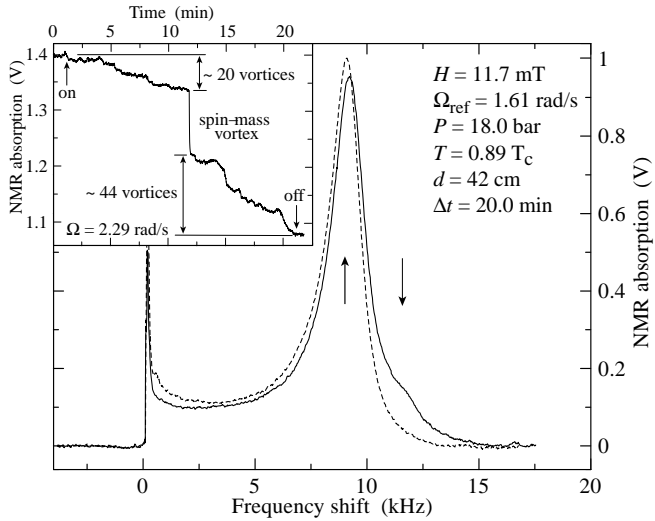}}
  \bigskip
\caption[SpinMassSignals] {NMR signal characteristics of the spin-mass vortex. {\it
(Insert)} This 20 min irradiation session includes one large jump, in addition to
vortex-line formation. The latter amounts to 64 lines, as determined from the
measured annihilation threshold $\Omega_{\rm v} = 0.19$ rad/s. The rate
Eq.~(\ref{Ndot}) gives  ${\dot N} = 3.35$ lines/min or a total of 67 lines (with the
measured $\Omega_{\rm cn} = 1.10$ rad/s). The combined drop in signal amplitude from
these lines is 206 mV, which corresponds to 3.2 mV/line.  Here at a lower
temperature the step size per vortex is much reduced from that in
Fig.~\protect\ref{VortexLineYield}. The single large jump corresponds to 36
additional lines, but like in Fig.~\protect\ref{SpinMassStepSize} it is interpreted
to represent one spin-mass vortex. {\it (Main panel)} NMR spectra of the accumulated
vortex sample, recorded at a lower reference velocity $\Omega_{\rm ref}$: The
solid curve with the smaller counterflow peak is the original state after irradiation and deceleration from 2.29 to 1.61 rad/s. The dashed curve with the higher counterflow peak was traced later after cycling $\Omega$ from 1.61 to 0.20 rad/s and back, ie. after decelerating close to, but
still above the annihilation threshold of the regular vortex cluster. In the former spectrum (solid curve) the soliton sheet shifts a sizeable fraction of the absorption to the maximum possible
frequency shift (at right vertical arrow). After cycling $\Omega$ the outermost spin-mass
vortex (see Fig.~\protect\ref{SpinMassStructure} top panel) is annihilated (dashed curve) and the CF peak is shifted to the value (at left vertical arrow) without the soliton sheet. This shift in the frequency of the counterflow peak creates the jump in the dispersion signal in Fig.~\protect\ref{SpinMassStepSize}. Since the remaining regular cluster was not decelerated at any stage below its annihilation threshold of 0.19 rad/s, no usual vortex lines were annihilated from the dashed-curve spectrum. } \label{SpinMassSignals}
\end{figure}

The identification of the spin-mass vortex is based on its peculiar NMR spectrum
(Kopu and Krusius 2001). In an axially oriented magnetic field the NMR spectrum has a textural
cut off frequency, beyond which no absorption is allowed (excepting line broadening
effects). The distinguishing feature of the spin-mass vortex is the absorption which
has been pushed to the maximum possible frequency shift, beyond that of the
counterflow (CF) peak (which corresponds to 80\,\% of the maximum value, see
Fig.~\ref{SpinMassSignals}).

Because of this shifted absorption, the CF peak is slightly shifted both in
frequency and in height in the presence of a spin-mass vortex, compared to its
location without the spin-mass vortex. This difference is displayed in
Fig.~\ref{SpinMassSignals}, where the spectrum plotted with the solid curve includes
one spin-mass vortex while the dashed spectrum is without it. The small frequency
shift between these two peaks is the reason why also the NMR dispersion signal is
discontinuous when the spin-mass vortex is formed during the neutron irradiation in
Fig.~\ref{SpinMassStepSize}. Since both the NMR absorption and dispersion signals
are continuously monitored at the lock-in-amplifier output, a jump in the dispersion
channel provides the immediate alert for a spin-mass vortex. In contrast, no
discontinuity in the dispersion signal is observed in first order when a small
number of vortex lines is added to the cluster at high bias velocity (see spectra in
Fig.~\ref{VorNMR}). In this case some absorption intensity is shifted towards the
Larmor edge to small frequency shifts, and the change in the dispersion signal at
the site of the CF maximum is negligible.


\begin{figure}[!!!tb]
\centerline{\includegraphics[width=0.7\columnwidth]{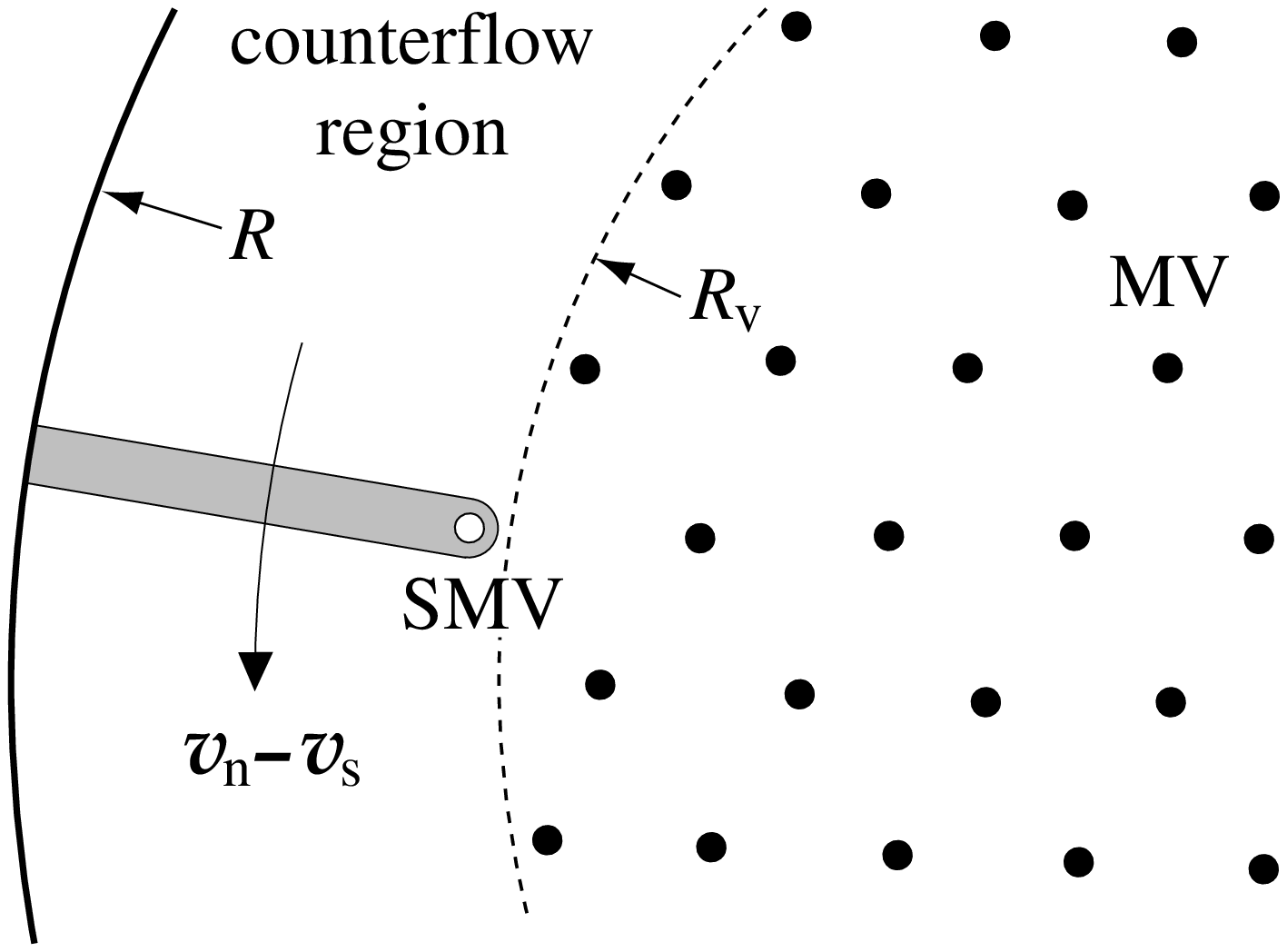}}
  \medskip
  \centerline{\includegraphics[width=0.9\columnwidth]{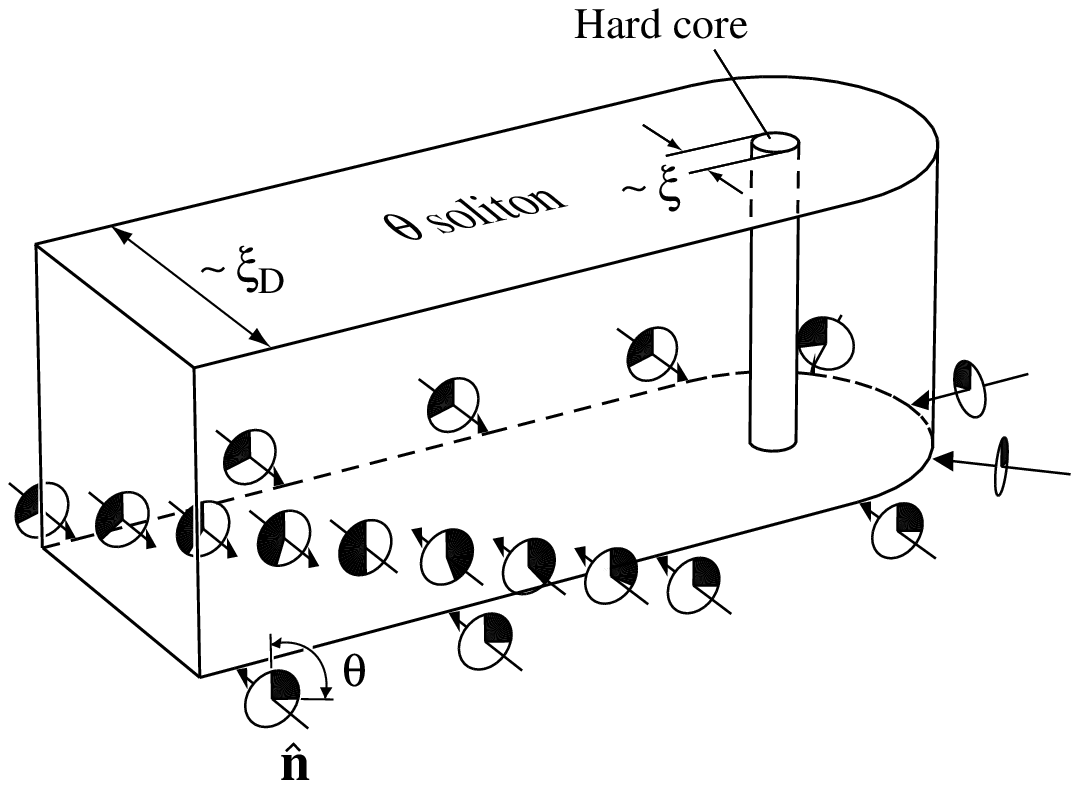}}
\caption[SpinMassStructure] {{\it (Top)} The spin-mass vortex (SMV) is a composite
object, which carries a trapped spin current as well as a mass current. In the
rotating container it is confined by the Magnus force to the edge of the vortex
cluster, which also includes usual mass-current vortex lines (MV). The extension of
the spin vortex into the bulk liquid is a soliton tail (grey strip). The spin-mass
vortex itself provides one termination line of the soliton while its second end is
anchored on the cylindrical wall. {\it (Bottom)} Inside the soliton the angle
$\theta$ changes through $\pi$ while outside it is homogeneously at $104^{\circ}$.
At the soliton surface the rotation axis $\hat{\bf {n}}$ is oriented perpendicular
to the surface while further away it goes smoothly over into the flare-out texture
of an ideal cylinder in an axially oriented magnetic field. The distortion of the
$\hat{\bf {n}}$ texture from the flare-out configuration in the region around the
soliton (Kopu and Krusius 2001) gives rise to the NMR signature of the spin-mass vortex in
Fig.~\protect\ref{SpinMassSignals}. } \label{SpinMassStructure}
\end{figure}


The large-scale configuration, in which the spin-mass vortex appears, is shown in
the top part of Fig.~\ref{SpinMassStructure}. It consists of a linear object and a
planar domain-wall-like soliton sheet. The line defect provides one termination for
the sheet while its second end is anchored on the cylindrical wall of the container.
The line defect consists of a singular core, which supports a circulating superfluid
mass current, but has additionally trapped a disclination line in spin-orbit
coupling. The latter feature can also be described in terms of a spin vortex which
supports a persistent spin current. Thus the composite object, the spin-mass vortex,
has the combined properties of a usual mass-current vortex and a spin-current
vortex. A gain in the core energy of the combined structure provides the energy
barrier against their dissociation. The soliton sheet is the continuation into the
bulk superfluid from the breaking of the spin-orbit coupling: It separates two
degenerate domains with anti-parallel orientations of the B-phase anisotropy axis
$\hat {\bf {n}}$ (lower part of Fig.~\ref{SpinMassStructure}). The spin-mass vortex
feels the Magnus force from the applied bias flow, but this is partly compensated by
the surface tension of the soliton sheet. The equilibrium position of the spin-mass
vortex is therefore at the edge of the vortex cluster (top part of
Fig.~\ref{SpinMassStructure}) from where it can be selectively removed by
annihilation, as was done in the experiment of Fig.~\ref{SpinMassSignals}.

The origin of the two defects in the spin-mass vortex can be seen from the form
which the B-phase order parameter takes (Vollhardt and W{\"o}lfle 1990):
\begin{equation}
A_{\alpha j} = \Delta_{\rm B}(T,P) \; e^{i \phi} \; R_{\alpha j} (\hat {\bf {n}},
\theta) \;. \label{BphaseOP}
\end{equation}
It includes the isotropic energy gap $\Delta_{\rm B} (T,P)$, the phase factor $e^{i
\phi}$, and the rotation matrix $R_{\alpha j}(\hat {\bf {n}},\theta)$. The latter
defines the rotation by which the $SO(3)_{\rm L}$ orbital and $SO(3)_{\rm S}$ spin
spaces are rotated with respect to each other around the axis $\hat{\bf {n}}$ by
the angle $\theta$. To minimize the weak spin-orbit interaction, it is required that
$\theta = \arccos{(-{1 \over 4})} \approx 104^{\circ}$ homogeneously everywhere
within the bulk superfluid. The mass-current vortex is associated with a $2\pi$
circulation in the phase factor while the spin current involves a disclination line
in $R_{\alpha j} (\hat {\bf {n}}, \theta)$ field. The minimum energy configuration
becomes then the structure depicted in the lower part of
Fig.~\ref{SpinMassStructure}. Here the volume in the bulk, where the spin-orbit
interaction is not minimized and $\theta$ traverses through $\pi$, is concentrated
within a soliton tail.

In $^3$He-B weak anisotropy energies arise in an external magnetic field. They
produce an extended orientational distribution, or texture, of the anisotropy axis
$\hat {\bf {n}}$. The characteristic length scale of this texture in the conditions
of the rotating NMR measurements is $\xi_H \propto 1/H \sim 1$\,mm. Since the NMR
frequency is controlled locally by the orientation of $\hat {\bf {n}}$ with respect
to $\bf {H}$, the NMR spectrum is an image of the $\hat {\bf {n}}$ texture in the
cylindrical container. In an axially oriented magnetic field and an infinitely long
cylinder, the texture is known to be of the ``flare-out'' form (Vollhardt and W{\"o}lfle 1990). In
this texture there are no regions where $\hat {\bf {n}}$ would lie in the
transverse plane perpendicular to the field $\bf {H}$. In contrast, at the soliton
sheet $\hat {\bf {n}}$ is oriented strictly perpendicular to the sheet and thus to
$\bf {H}$. This part of the texture is responsible for shifting NMR absorption into
the region of maximum possible frequency shift and becomes thus the signature of the
spin-mass vortex in the NMR spectrum (Fig.~\ref{SpinMassSignals}).

The presence of the soliton sheet has many important experimental implications: It
determines the location of the spin-mass vortex at the edge of the cluster and it
leaves a clear signature in the NMR spectrum at high bias velocities when the sheet
becomes stretched. It also affects the threshold velocity at which a spin-mass
vortex might be expected to escape from the neutron bubble. This was only observed
to happen well above $v_{\rm cn}$, the threshold velocity for usual mass-vortex
rings (Fig.~\ref{CritVel&NeutrThreshold}). It is explained by the fact that for
spontaneous expansion a spin-mass vortex loop has to exceed, in addition to the
usual barrier formed by the line tension of the small loop, also the surface tension
from the soliton sheet.

The spin-mass vortex was originally discovered as a defect which is formed when an
A$\rightarrow$B transition front moves slowly close to adiabatic conditions through
a rotating container (Kondo et~al.\ 1992; Korhonen et~al.\ 1993). In this experiment
the initial A-phase state is one with the equilibrium number of doubly quantized
singularity-free vortex lines while the final state is found to contain less than
the equilibrium number of singly quantized B-phase vortex lines plus some number of
spin-mass vortices. This means that A-phase vortex lines interact with the moving AB
interface, they are not easily converted to B-phase vorticity, a critical value of
bias flow velocity has to be exceeded before the conversion becomes possible, and
even then some fraction of the conversion leads to vorticity with the additional
defect in $R_{\alpha j} (\hat {\bbox{n}}, \theta)$
(Parts et~al.\ 1993; Krusius et~al.\ 1994). In neutron measurements the spin-mass vortex
was not observed at 2.0 bar, but at 18.0 bar. In the KZ scenario this is the
pressure regime where the probability for the formation of A phase blobs in the
neutron bubble increases (Fig.~\ref{AB_EnergyDifference}). Interestingly one might
therefore surmise that even in the neutron bubble the spin-mass vortex could result
from vorticity which originally was contained within A-phase blobs and which is
partly transferred to the B phase when the A phase blobs shrink away.

The two NMR spectra in Fig.~\ref{SpinMassSignals} illustrate the easy identification
of the spin-mass vortex after neutron irradiation. Although the spin-mass vortex is a rare event in neutron irradiation, compared to the yield of vortex lines, its presence demonstrates that in addition to usual mass-current vortices also other order-parameter defects are created.  This lends support to the discussion in Sec.~\ref{A-phase} that also the AB
interface could be among such defects. The presence of the spin-mass vortex also
limits the possible mechanisms of defect formation. An explanation in terms of the
superflow instability at the neutron bubble boundary (Sec.~\ref{MechVorForm}) is not a probable alternative: The spin-mass vortex is not normally created at the bulk critical velocity limit $v_{\rm cb}(T,P)$ as a response to superflow.

\subsection{Vortex formation in gamma radiation}  \label{GammaRadiation}

To study the dependence of the vortex formation rate ${\dot N}$ on the
dimensions of the heated neutron bubble in the $^3$He-B bath, it would be
useful to investigate alternative heating methods. In addition to thermal
neutrons, beta and gamma radiation have frequently been used as a source of
heat. In the rotating experiments a $^{60}$Co gamma source was experimented
with, which emits $\gamma$ rays at 1.17 and 1.33 MeV energy.

A gamma ray, which scatters in the container wall or in the liquid $^3$He bath,
knocks off a secondary electron which gives rise to a heat release in the liquid
$^3$He bath. The interaction probability is roughly proportional to the density of
the material and thus a large fraction of the secondary electrons originate from the
quartz glass wall. Although the initial energies of the secondary electrons cover a
broad spectrum, the variation in the effective heat release is narrower. Energetic
electrons knock off additional electrons which produce their own ionization tracks.
However, once the energy has dropped in the regime of several keV, the remaining
ionization track is short: roughly the rate of energy loss per unit distance
travelled increases inversely proportional to the electron's decreasing energy.
Therefore the heating becomes concentrated within a distance $\lesssim 10\;\mu$m of
the final stopping point.

As a source for localized heating in liquid $^3$He experiments, gamma rays suffer
from several shortcomings, compared to thermal neutrons: 1) It is not possible to
concentrate on single heat-release events, since many secondary electrons can be
produced by a single $\gamma$ ray. Thus there is more variation in these events. For
instance, the endpoints of different ionization tracks may fall close to each other
and may merge to produce a large hot bubble of variable size. 2) The events are not
localized in a well-determined location within the container. 3) It is not only the
liquid $^3$He bath which preferentially absorbs the heat, but even more so the
structural materials of the refrigerator. This interferes with the temperature
control of the measurements: The heating produced by the $\gamma$ radiation in the
metal parts of the cryogenic equipment may rise to excessive levels, and radiation
shielding and collimation need to be built into the experiment.

It was found that qualitatively vortex formation in gamma and neutron radiation are
similar. But the measurements also showed that quantitative differences are large:
The threshold velocity $v_{\rm cn}$ is substantially increased and thus the yield
${\dot N}(\Omega)$ at a given rotation velocity $\Omega$ in the vortex-free state is
smaller. So far no careful quantitative measurements have been carried out with
$\gamma$ radiation.

\subsection{Bias dependence of loop extraction}  \label{BiasDepSec}

The dependence of the vortex formation rate ${\dot N}$ on the externally
applied bias $v$ is the most tangible quantitative result from the rotating
measurements. It will be described and analyzed below (Ruutu et~al.\ 1998a). The
empirical rate equation (\ref{Ndot}) is shown to be consistent with the KZ
model, including the measured value of the prefactor $\gamma$.

\subsubsection{Experimental velocity dependence} \label{ExpVelDep}


\begin{figure}[!!!tb]
\centerline{\includegraphics[width=0.90\columnwidth]{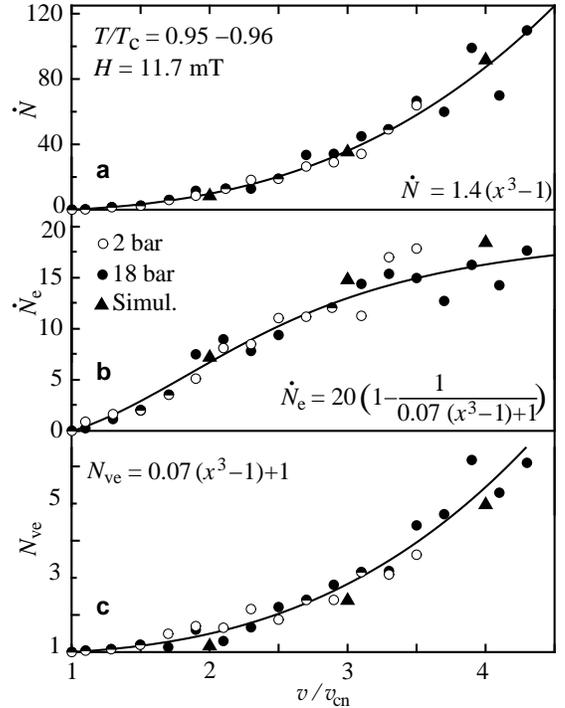}}
\caption[BiasDependence] {Rates of vortex line formation plotted
  {\it vs} normalized bias velocity $x=v/v_{\rm cn}$: (a)
  total number of lines formed per minute ($\dot N$), (b)
  number of observed neutron absorption events per minute ($\dot N_{\rm e}$),
  and (c) average number of lines per observed event ($N_{\rm ve}$).
  All three quantities have been determined {\it independently}
  from discontinuities in NMR absorption, ie. each data point is based on
  one or several accumulation plots like those in
  Fig.~\protect\ref{ExpSignature}. The data in the two upper plots are
  proportional to the neutron flux, while the data in the bottom plot are
  neutron-flux independent. Solid curves are fits to the data,
  given by the expressions in each panel. Triangles are results from numerical
  simulations described in Sec.~\protect\ref{Simulation}. (From
  Ruutu et~al.\ 1998a). }
\label{BiasDependence}
\end{figure}


The frequency and number of vortex lines can be counted from the
accumulation record (Figs.~\ref{ExpSignature}, \ref{VortexLineYield}), when
the NMR absorption is monitored as a function of time during neutron
irradiation.  Fig.~\ref{BiasDependence} shows the vortex formation rate
$\dot N$ (a), the rate of those neutron absorption events $\dot N_{\rm e}$
which produce at least one line and thus become observable (b), and the
average number of lines produced by each
absorption event (c). All three quantities are determined
independently and directly from the absorption records. In
Fig.~\ref{BiasDependence} the results are plotted as a function of the
normalized bias $v/v_{\rm cn}$. To construct the plot, the horizontal axis
was divided into equal bins from which all individual measurements were
averaged to yield the evenly distributed data points displayed in the
graphs.

The rates in Fig.~\ref{BiasDependence}a,b are proportional to the intensity
of the neutron flux, which is contained in the prefactors of the
expressions given in the different panels of the figure. For counting the
rates from the NMR absorption records it is vital that absorption events do
not overlap, when the applied counterflow velocity is increased. Therefore
three different source positions were used which were scaled to the same
distance using the measured graph in Fig.~\ref{SourceDistance}.

The rates in Fig.~\ref{BiasDependence} increase rapidly with the applied
bias velocity $v$: At $v/v_{\rm cn} \approx 4.5$, close to the maximum
velocity limit imposed by the spontaneous instability limit
(Fig.~\ref{CritVel&NeutrThreshold}), there are almost no unsuccessful (or
unobserved) absorption events left: $\dot N_{\rm e}(\infty) \approx \dot N_{\rm
  e}(4.5 \, v_{\rm cn})$. The panel in the middle shows that this
saturation corresponds to a flux of 20 neutrons/min. The same flux of
neutrons was estimated to be absorbed in the sample based on independent
measurements with commercial $^3$He counters. Also we note that in
agreement with the scaling properties expressed by the rate equation
(\ref{Ndot}), measurements at the two pressures of 2 and 18 bar fall on the
same universal curves.

Fig.~\ref{BiasDependence} demonstrates that the neutron-induced vortex formation
process involves a strong stochastic element: Close to the critical threshold at $v
= 1.1 v_{\rm cn}$ only one neutron capture event from 40 manages to produce a
sufficiently large vortex loop for spontaneous expansion. On increasing the bias
flow by a factor of four, almost all neutron capture events give rise to at least
one escaping vortex loop. Secondly, the nonlinear increase of $\dot N$ as a function
of $v/v_{\rm cn}$  in the top panel has to be attributed to two factors: (i) the
increase in the event rate $\dot N_{\rm e}$ in the middle panel and (ii) the rapid
rise in the number of lines produced per event in the bottom panel. These
conclusions fit with the KZ predictions, as we shall see in the next section.

\begin{figure}[!!!tb]
\centerline{\includegraphics[width=0.95\columnwidth]{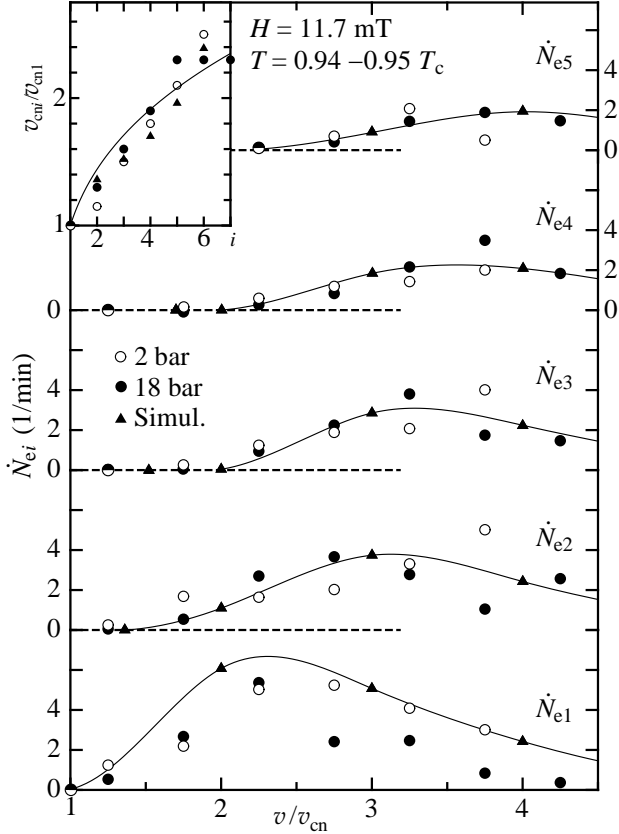}}
\caption[VorRingCount] { Rates $\dot N_{{\rm e}i}$ of vortex
  formation events, grouped according to the number of lines $i$ formed per
  absorption event, plotted vs $v/v_{\rm cn}$. Triangles are
  result of numerical simulations described in
  Sec.~\protect\ref{Simulation}. Solid curves are spline fits to
  simulation data. {\it (Inset)} Threshold velocity $v_{{\rm cn}i}/v_{\rm
    cn1}$ for the onset of an event with $i$ lines, plotted vs the number
  of lines $i$. The solid curve represents the fit $v_{{\rm cn}i}/v_{\rm
    cn1}= [2.0 \; (i-1)+1]^{1 \over 3}$.  (From Ruutu et~al.\ 1998a).}
\label{VorRingCount}
\end{figure}

The most detailed information from the rate measurements is the dispersion
into events in which a given number of lines is formed.
Fig.~\ref{VorRingCount} displays the rates $\dot N_{{\rm e}i}$ of events
which produce $i=1$\,--\,5 lines. This data exhibits larger scatter, owing
to its statistical variation, but again after averaging one gets for each
value of $i$ a curve, which peaks at a maximum, and then trails off. With
increasing value of $i$ the curves shift to successively higher velocities.
The starting points of each curve, the threshold velocities $v_{{\rm
    cn}i}$, are plotted in the inset.  Their values increase monotonically
with each consecutive value of $i$.  This means that at and immediately
above the first threshold, $v_{\rm cn} = v_{{\rm cn}1}$, only single-vortex
events occur.

A surprising finding from the measurements is the total absence of a background
contribution to the measured rates. A number of tests were performed to look for
spontaneous events in the absence of the neutron source, to check whether a
contribution from the background radiation level should be subtracted from the
measured rates. For instance, a vortex-free sample was rotated for 90 min at
different velocities (0.9, 1.3, and 2.1 rad/s at 2.0 bar and 0.94 $T_{\rm c}$), but
not a single event was noted.

\subsubsection{Analytic model of vortex loop escape} \label{AnalVorEsc}

An analytic calculation can be constructed for independent vortex loops in
the applied flow, which explains the cubic velocity dependence of the
escape rate in Fig.~\ref{BiasDependence}. Vortex rings, which are part of
the random vortex network after the quench, have a critical radius
(Eq.~(\ref{VorRing})) below which they contract and above which they expand
spontaneously in homogeneous superflow.  By comparing this radius to the
typical curvature in the random network, we arrive at the desired result
(Ruutu et~al.\ 1996a).

During the quench through the superfluid transition a random vortex network
is formed with a characteristic length scale of the order of $\xi_{\rm v}$.
The later evolution of the network leads to a gradual increase in this
length. The average inter-vortex distance or, equivalently, the typical
radius of curvature of the loops increases with time.  We shall call this
length $\tilde \xi(t)$. This ``coarse-graining'' process preserves the
random character of the network, in other words the network remains self
similar or scale invariant.  Only later a change occurs in this respect,
when the loops become sufficiently large to interact with the externally
applied bias flow: This causes large loops to expand, if they are oriented
transverse to the flow with the correct winding direction, while small
loops contract, and loops with the wrong sign of circulation tend to both
contract and to change orientation. The time scale of these processes is
determined by the magnitude of mutual friction between the normal and
superfluid components. In the Ginzburg-Landau temperature range the
evolution of the network occurs in milliseconds and the expansion of the
extracted loops into rectilinear lines in a fraction of a second.

In superfluid $^3$He the large viscosity of the normal component
clamps it to corotation with the container. In the rotating frame
of reference we may write $v_{\rm n} =0$ and $v_{\rm s}=v$.  The
energy of a vortex loop, which is stationary with respect to the
walls, is given by (Donnelly 1991)
\begin{equation}
{\cal E}=E_{\rm kin}+{\bf p} {\bf v}\,, \label{eq10}
\end{equation}
where ${\bf v}$ is the velocity of the applied superflow. In simple configurations,
the hydrodynamic kinetic energy or self-energy of the loop arises from the trapped
superfluid circulation with the velocity ${\bf v}_{\rm s,vort} = ({\kappa / 2\pi})
\,{\bf \nabla} \Phi$,
\begin{equation}
E_{\rm kin}={1\over 2}\int \rho _{\rm s} v^2_{\rm s,vort} \, dV=
\varepsilon L~, \label{eq11}
\end{equation}
and is proportional to the length $L$ of the loop and its line
tension,
\begin{equation}
\varepsilon  = {{\rho_{\rm s} \kappa^2} \over {4\pi}} \; \ln
 {\tilde\xi(t) \over {\xi}}~~.\label{eq12}
\end{equation}
Here we neglect the small contribution from the core energy, use
intervortex spacing
$\tilde\xi(t)$ as the upper cutoff, and the superfluid coherence
length $\xi(T,P)$ as the lower cutoff for the integration in
Eq.~(\ref{eq11}). This equation is valid in
the logarithmic approximation, when $\tilde\xi(t) \gg \xi(T,P)$.  While the
first term in Eq.~(\ref{eq10}) is proportional to the length $L$ of the
loop, the second term involves its linear momentum,
\begin{equation}{\bf p} = \int \rho_{\rm s} {\bf v}_{\rm s,vort} \,dV= {1
\over 2\pi} \, \rho_{\rm s} \kappa \,  \int {\bf \nabla} \Phi\,
dV= \rho_{\rm s} \kappa {\bf S}~~,\label{eq13}
\end{equation}
where the  last step follows from Gauss's theorem and involves the area  $S$ of the
loop in the direction of the normal ${\bf S}/S$ to the plane of the loop. Thus we
write for the energy of a loop
\begin{equation}
  {\cal E}(L,S,t) =\rho_{\rm s} \kappa ~\left[ ~L~{\kappa \over 4\pi} \; \ln
 {\tilde\xi(t) \over {\xi}}~-~ v S ~\right]~~,
\label{eq14}
\end{equation}
where $S$ is the algebraic area, perpendicular to the flow and of proper winding
direction. This equation expresses the balance between a contracting loop due to its
own line tension, which dominates at small applied velocities, and expansion by the
Magnus force from the external superflow, which dominates at high applied
velocities. The divide is the equilibrium condition, which was expressed in
Eq.~(\ref{VorRing}) and corresponds to the situation when the height of the energy
barrier, which resists loop expansion, vanishes. In this extremal configuration
${\bf p}$ is antiparallel to ${\bf v}$, the loop moves with the velocity $-{\bf v}$
in a frame of the superfluid component, but is stationary in the rotating frame. In
a more general sense, if we consider loops in the random network which still deviate
from circular shape, the extremal case degenerates to a saddle point. This is
because the extremum requires also minimization with respect to deviations from
circular shape, ie. the total energy is invariant under small variations of the
radius of the ring, or $\delta {\cal E} = \delta E_{\rm kin} + {\bf v} \cdot \delta
{\bf p} =0$.

The expansion of the vortex loop should be calculated by including
the mutual friction forces. In our analytic description of vortex
loop escape we shall neglect such complexity. Instead we shall
make use of three scaling relations which apply to Brownian
networks and are described in more detail in
Sec.~\ref{Simulation}. These expressions relate the mean values in
the statistical distributions of the loop diameter ${\cal D}$,
area $S$, and density $n$ to the length $L$ of the loop:
\begin{equation}
  {\cal D} = A L^\delta\,\tilde\xi^{1-\delta}~,
  \quad (A \approx 0.93,\;\;\delta \approx
  0.47)~,
\label{DLx} \end{equation}
\begin{equation}
  |S| = B {\cal D}^{2-\zeta}\,\tilde\xi^\zeta,
  \quad (B \approx 0.14,\;\;\zeta \approx 0)~,
\label{SDx} \end{equation}
\begin{equation}
  n = CL^{-\beta}\,\tilde\xi^{\beta-3}~,
  \quad (C \approx 0.29,\;\;\beta \approx 2.3)~.
\label{NLx}
\end{equation}
For a Brownian random walk in
infinite space the values of $\delta$, $\beta$ and $\zeta$ are
$1/2$, $5/2$ and 0.

The important simplification here is that these relations are assumed valid during
the entire evolution of the network, until sufficiently large rings are extracted by
the  bias flow into the bulk. Using Eqs.~(\ref{DLx}) and (\ref{SDx}), we may write
Eq.~(\ref{eq14}) for the energy of a loop in the form
\begin{equation}
  {\cal E} ({\cal D},t) =
  \rho_{\rm s} \kappa {\cal D}^2~\left[ ~{\kappa\over 4\pi
      \tilde\xi(t)A^2} \; \ln {\tilde\xi(t) \over {\xi}}~-~
    v B ~\right]~~.
\label{eq15}
\end{equation}

When the scale $\tilde \xi(t) $ exceeds a critical size
$\tilde \xi_{\rm c} (v)$, which depends on the particular value of
the applied superflow velocity $v$,
\begin{equation}
\tilde \xi_{\rm c} (v)= { 1 \over {A^2B}} \; {\kappa \over {4\pi
v}} \; \ln {\tilde \xi_{\rm c} \over {\xi}}~~~,\label{eq16}
\end{equation}
the energy in Eq.~(\ref{eq15}) becomes negative and the loops
in the network
start expanding spontaneously.  The total number of loops $N_{\rm b}$, which will be extracted from one
neutron bubble, can then be obtained by integrating their density from the
smallest size $\tilde \xi_{\rm c}$ to the
upper cutoff, which is
provided by the diameter of the entire network, or that of the
heated bubble, $2R_{\rm
  b}$:
\begin{equation}
N_{\rm b} = V_{\rm b} \, \int_{\tilde\xi_{\rm c}}^{2R_{\rm b}}d{\cal
D}~n({\cal D})~. \label{eq17}
\end{equation}
Here the density distribution $n(L)=C~\tilde \xi^{-3/2}~L^{-5/2}$,
combined with that for the average diameter ${\cal D}(L) =A~(L~
\tilde \xi ~)^{1/2}$, gives $n({\cal D}) \, d{\cal D} = 2A^3C{\cal
D}^{-4} \; d{\cal D}$. On inserting this into the
integral~(\ref{eq17}) we obtain
\begin{equation}
N_{\rm b} = {1 \over 9} \pi A^3 C \left[\left({{2R_{\rm b}} \over
\tilde\xi_{\rm c}}\right)^3 -1\right] \; . \label{eq18}
\end{equation}
From this equation we see that the requirement $ N_{\rm b}(v_{\rm cn})=0$
returns us the definition of the threshold velocity $ v_{\rm cn}$:
$\tilde \xi_{\rm c} (v=v_{\rm cn}) = 2 R_{\rm b}$. This in turn
gives us from Eq.~(\ref{eq16}) for the radius of the heated bubble
\begin{equation}
R_{\rm b} = { 1 \over {A^2B}} \; {\kappa \over {8\pi v_{\rm cn}}}
\; \ln {2R_{\rm b} \over {\xi(T,P)}}~~~,\label{eq19}
\end{equation}
which we used in Sec.~\ref{ThreshVelocity} to derive the
temperature dependence of the threshold velocity: $ v_{\rm cn}
\propto (1-T/T_{\rm
  c})^{1/3}$.

Eqs.~(\ref{eq16}) and (\ref{eq19}) thus show that $\tilde \xi_{\rm
c} \propto 1/v$ and $R_{\rm b} \propto 1/ v_{\rm cn}$, so that we
may write for the vortex-formation rate $\dot N =\phi_{\rm n} N_{\rm b}$
from Eq.~(\ref{eq18})
\begin{equation}
{\dot N} = {1 \over 9} \pi A^3 C \phi_{\rm n} \, \left[\left({v
\over v_{\rm cn}}\right)^3 -1\right] \; , \label{eq20}
\end{equation}
where $\phi_{\rm n}$ is the neutron flux. This is the form of the measured
cubic velocity dependence in the empiric Eq.~(\ref{Ndot}). By inserting $A
\approx 0.93$, $C \approx 0.29$ from Eqs.~(\ref{DLx}) and (\ref{NLx})
respectively and $\phi_{\rm n} \approx 20$ neutrons/min, as determined from
the saturation of the event rate $\dot N_{\rm e}$ in
Fig.~\ref{BiasDependence}b, we obtain for the rate factor in
Eq.~(\ref{eq20}) $\gamma \approx 1.6$ min$^{-1}$, which agrees with the
experimental value in Figs.~\ref{CritVel} and \ref{BiasDependence}a. We note
that the cubic dependence on the applied bias flow in Eq.~(\ref{eq20})
comes only from the assumption that the whole volume of the heated bubble
contributes equally to the production of vortices, while the values of the
prefactor $\gamma$ and of the constant term in Eq.~(\ref{eq20}) depend on
the scaling relations (\ref{DLx}\,--\,\ref{NLx}).

The definition of the threshold velocity
$v_{\rm cni}$, which applies for an event in which $i$ rings are
formed simultaneously, is roughly consistent with the requirement
$N_{\rm b} (v=v_{\rm cni}) \approx i$. This gives $v_{{\rm cn}i}/v_{\rm
cn}\sim i^{1/3}$, which agrees with the measured result in
Fig.~\ref{VorRingCount}.

To summarize, we note that the KZ model, combined with the simplest possible
interpretation for the loop escape from a random vortex network, reproduces both the
measured cubic dependence on the normalized velocity, $(v / v_{\rm cn})^3$, and the
magnitude of the extraction rate.

\subsection{Neutron-induced vortex formation at low temperatures} \label{LowTempVorForm}

The low temperature properties of superfluid $^3$He could not have been extrapolated
from measurements restricted to temperatures above $0.80\,T_{\rm c}$, even within
the framework of the general theory. Similarly, to understand the neutron-induced
vortex formation process, it is important to extend the measurements to lower
temperatures. One such study has been performed with NMR down to $0.4\,T_{\rm c}$ in
a rotating sample (Finne et~al.\ 2004a). This measurement is a continuation of the NMR
work, which we have discussed so far above $0.80\,T_{\rm c}$. Another series of
measurements has been performed calorimetrically below $0.2\,T_{\rm c}$ in a
quiescent bath (B{\"a}uerle et~al.\ 1996, 1998a). At these very low temperatures quantized
vortex lines have a long life time in zero flow. Thus there is sufficient time after
the neutron absorption event to detect the vortices.

NMR measurement in $^3$He-B loses rapidly in amplitude resolution on cooling to
lower temperatures: the susceptibility drops with decreasing temperature and the
width of the NMR spectrum increases. Thus the change of NMR absorption per one
rectilinear vortex line quickly diminishes and single-vortex resolution is lost.
Also below $0.6\,T_{\rm c}$ a new phenomenon enters, namely superfluid turbulence
(Finne et~al.\ 2003). If here a few seed vortex loops are injected into rotating
vortex-free superflow, they become easily unstable and form a turbulent vortex
tangle. In the turbulent state the number of vortices rapidly multiplies and the
final state becomes an equilibrium array of rectilinear lines.

In neutron-irradiated superflow this is exactly what happens: First vortex loops are
extracted from the neutron bubble and injected into the bias flow. Their number can
be tuned to some extent (in the range 1--5), by choosing
the value of the bias velocity (Fig.~\ref{BiasDependence}c). Then with some
probability, which increases towards lower temperatures and higher bias velocities,
turbulence starts. As a result the sample is suddenly filled with the equilibrium
number of vortex lines, which means that instead of a few rectilinear lines the
final state contains of order $\sim 10^3$ lines. In these conditions the final state
has little connection with the loop extraction process and the details of the
neutron-induced vortex formation become lost. But we can use this method to start
turbulence in a rotating superfluid column.

\subsubsection{Experimental techniques} \label{ExperimentalTechniques}


\begin{figure}[t]
\centerline{\includegraphics[width=0.75\linewidth]{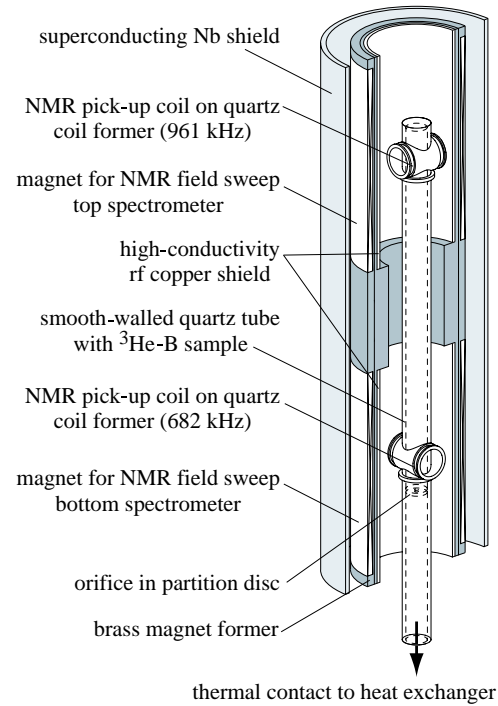}} \caption{$^3$He
sample with NMR measuring setup. The sample is contained in a fused quartz tube
which has a diameter of 6\,mm and length 110\,mm. This space is separated from the
rest of the liquid $^3$He volume with a partition disc. In the center of the disc an
orifice of 0.75\,mm diameter provides the thermal contact to the liquid column which
connects to the sintered heat exchanger on the nuclear cooling stage. Two
superconducting solenoidal coil systems with end-compensation sections produce two
independent homogeneous field regions with axially oriented magnetic fields. An
exterior niobium cylinder provides shielding from external fields and additional
homogenization of the NMR fields. The NMR magnets and the Nb shield are thermally
connected to the mixing chamber of the pre-cooling dilution refrigerator and have no
solid connection to the sample container in the center. The two split-half detection
coils are fixed directly on the sample container.  (From
Finne et~al.\ 2004a).} \label{Experiment}
\end{figure}


The rotating NMR measurements at low temperatures are performed in the setup of
Fig.~\ref{Experiment}. For the study of turbulence it proved fortunate that this
arrangement includes two NMR detection coils at both ends of the long sample
cylinder. A second bonus is the relatively high critical velocity of the container
so that vortex-free superflow could be maintained to above 3.5\,rad/s below
$0.8\,T_{\rm c}$ at 29.0\,bar.

On cooling to lower temperatures vortex-free superflow generally becomes more and
more difficult to generate and maintain. The lower section of the $^3$He volume
below the orifice is directly connected with the sintered heat exchanger. This
region is flooded with remanent vortices from the porous sinter already at low
rotation. Generally one finds that these vortices start to leak through the orifice
into the sample volume on cooling below $0.6\,T_{\rm c}$. Surprisingly the sample
tube, with which the neutron measurements were performed, seemed immune to this
problem in spite of the fact that the orifice had a relatively large diameter of
0.75\,mm.

Below $0.5\,T_{\rm c}$ remanent vortices become a problem. Here successive
accelerations to rotation have to be separated by extensive waiting periods at stand
still, to allow the slow annihilation of remanent vortices. If rotation is started
too soon, remanent vortices are still present and become easily unstable in
rotation. As a result vortex-free rotation at velocities above 1\,rad/s is then not
possible to achieve. Just below $0.60\,T_{\rm c}$ the waiting time is around 5\,min
in the container of Fig.~\ref{Experiment}, but at $0.40\,T_{\rm c}$ it is found to
be 30\,min or more. No procedure using rotation in different directions or with
different deceleration rates seemed to help in cutting down on the waiting period.

\begin{figure}[t]
\begin{center}
\leavevmode
\includegraphics[width=0.9\linewidth]{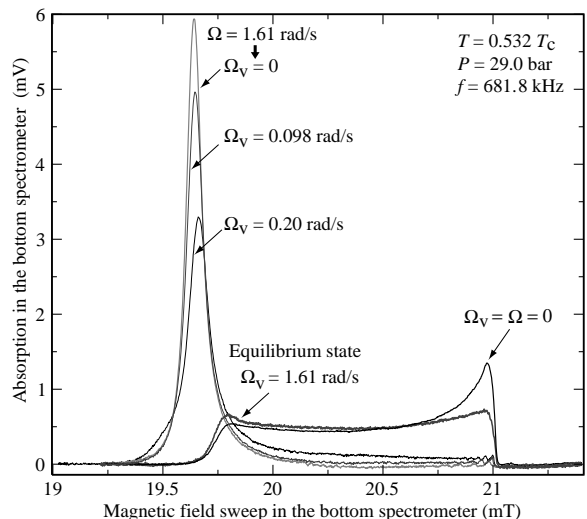}
\caption{NMR absorption spectra of $^3$He-B in rotation at low temperatures. As in
the left panel of Fig.~\protect\ref{VorNMR}, changes in the line shape can be
calibrated to give the number of vortex lines $N$. Here $N$ is characterized
  by the rotation velocity $\Omega_{\rm v}(N)$ at which a given number of lines
  $N$ is in the equilibrium state. The different spectra have been measured
  with the RF excitation at constant frequency $f$, using a linear sweep of the
  axially oriented polarization field $H$. The spectra have been recorded at
  constant temperature and thus all have the same integrated total
  absorption. The sharp absorption maximum at low field is the
  counterflow peak (CF). Its shift from the Larmor field (at 21.02\,mT)
  is used for temperature measurement. When a central cluster of
  rectilinear vortex lines is formed, the height of the CF peak is reduced.
  In the equilibrium vortex state $(\Omega = \Omega_{\rm v})$,
  where the number of vortex lines reaches
  its maximum, the spectrum looks very different: it has appreciable
  absorption at high fields and borders prominently to the Larmor edge.
  This spectrum is more similar to that of the  non-rotating state
  $(\Omega = 0)$. As shown in Fig.~\protect\ref{ClusterCalibration}, when
  the vortex number is small, $\Omega_{\rm v} \ll \Omega$, the reduction in
  the CF peak height can be calibrated to give $\Omega_{\rm
  v}$ and thus $N$. (From Finne et~al.\ 2004a). }
\label{NMR-Spectra}
\end{center}
\vspace{-6mm}
\end{figure}

The experimental setup in Fig.~\ref{Experiment} is equipped with two
independent continuous-wave NMR spectrometers. Each spectrometer includes a
split-half excitation/detection coil, wound from thin superconducting wire.
The two coils are installed at both ends of the sample cylinder. Each coil
is part of a high-Q tank circuit, with a resonance frequency approaching
1\,MHz and Q$\sim 10^4$.  Fig.~\ref{NMR-Spectra} shows a number of
absorption spectra measured with the bottom spectrometer, similar to the
high-temperature examples in the left panel of Fig.~\ref{VorNMR}. The
spectra illustrate the relevant NMR line shapes and their large width at
low temperatures, with varying numbers of rectilinear vortex lines in the
sample. Again the peak height of the CF maximum can be calibrated to yield
the number of lines $N$ in the linear limit, when $N \ll N_{\rm eq}$. Often
enough, to determine $N$, one has to increase $\Omega$ to some higher
reference value where the sensitivity of the CF peak is restored.


\begin{figure}[t]
\centerline{\includegraphics[width=0.9\linewidth]{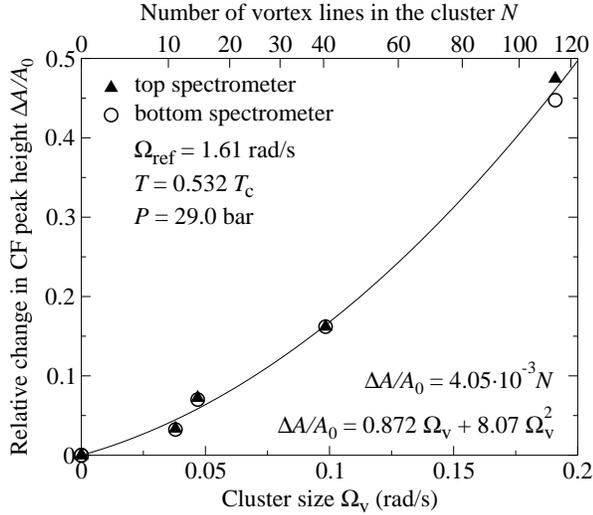}}
\caption{Calibration of CF peak height $A$ versus vortex line number $N$.
  The CF peak height in the vortex-free state, $A_0(\Omega_{\rm ref})$, is
  compared to the peak height $A(\Omega_{\rm ref},\Omega_{\rm v}(N))$,
  which is measured at the same rotation velocity $\Omega_{\rm ref}$ for a
  small vortex cluster of size $\Omega_{\rm v}$, which is prepared as
  described in the text. The quantity plotted on the vertical scale is the
  relative reduction in peak heights, $\Delta A/A_0 = [A_0(\Omega_{\rm
    ref}) - A(\Omega_{\rm ref},\Omega_{\rm v}(N))]/A_0(\Omega_{\rm ref})$,
  measured in constant conditions. The solid line is a fit, given by the expression
  in the figure. The conversion from $\Omega_{\rm v}$ (bottom axis) to
  $N$ (top axis) in the continuum picture is (Ruutu et~al.\ 1998b) $N = N_0
  (1-d_{\rm eq}/R)^2$, where $N_0 = \pi R^2 (2 \Omega_{\rm v}/\kappa)$ and
  $d_{\rm eq} = [\kappa/(8\pi\Omega_{\rm v}) \, \ln(\kappa/2\pi\Omega_{\rm
    v} r_{\rm c}^2)]^{1/2}$. Here $r_{\rm c} \sim \xi(T,P) $ is the radius
  of the vortex core. Using this conversion it is found that $\Delta A/A_0$
  is a linear function of $N$, similar to the right panel in Fig.~\protect\ref{VorNMR}.
  (From Finne et~al.\ 2004a). }
\label{ClusterCalibration}
\end{figure}


Fig.~\ref{ClusterCalibration} shows an example of a calibration where the
CF peak height is measured as a function of $N$ (or equivalently
$\Omega_{\rm v}$). The measurement has been performed for small vortex
clusters ($N \lesssim 100)$, by measuring each calibration point in a four
step process: 1) First the spectrum is recorded in the vortex-free state in
the reference conditions and the CF peak height $A_0(\Omega_{\rm ref})$ is
obtained. 2) Next a large number of vortex lines is created by irradiating
with neutrons (with the rotation at $\Omega_{\rm ref}$ or higher).  3) The
sample is then decelerated to the low rotation velocity $\Omega_{\rm low}
\ll \Omega_{\rm ref}$ so that part of the vortex lines are observed to
annihilate. Since the long cylinder is oriented along the rotation axis
only within a precision of $\sim 0.5^{\circ}$, this means that any
annihilation barrier must be negligible or that at $\Omega_{\rm low}$ the
sample is in the equilibrium vortex state $\Omega_{\rm low} = \Omega_{\rm
  v}(N)$, with a known number of vortex lines $N$.~ 4) Finally the rotation
is increased back to the reference value $\Omega_{\rm ref}$ and the
spectrum is recorded in order to measure the CF peak height $A(\Omega_{\rm
  ref}, \Omega_{\rm v}(N))$.  In Fig.~\ref{ClusterCalibration} the
reduction $\Delta A = A_0(\Omega_{\rm ref}) - A(\Omega_{\rm ref},
\Omega_{\rm v})$ in the CF peak height of the two spectra is plotted as a
function of $\Omega_{\rm v}$. To reduce the dependence on drift and other
irregularities it is normalized to the CF peak height $A_0(\Omega_{\rm
  ref})$ of the vortex-free reference state.  In contrast to the
calibration in the right panel of Fig.~\ref{VorNMR}, this procedure
does not rely on single-vortex resolution. However like always, it does
require that the order parameter texture is stable and reproducible and
that new vortices are not formed during rotational acceleration.

The result in Fig.~\ref{ClusterCalibration} is a smooth parabola, which becomes a
straight line if $\Delta A/A_0$ is plotted as a function of the number of
rectilinear lines $N$. Any sample with an unknown number of lines, which is less
than the maximum calibrated number, can then be measured in the same reference
conditions ($\Omega_{\rm ref}$, $T$, and $P$) and compared to this plot, to
determine $N$.  In Fig.~\ref{VortexProductionRate} a measurement of vortex formation
in neutron irradiation is shown for which the calibration was used. After each
irradiation session at different bias flow velocity $v= \Omega R$ the rotation is
changed to $\Omega_{\rm ref}$ and the NMR absorption spectrum is recorded. From this
spectrum the reduction in CF peak height is determined, by comparing to the spectra
of the vortex-free state which are measured regularly between neutron irradiation
sessions.

\subsubsection{Measurement of vortex formation rate} \label{MeasureRate}

\begin{figure}[t]
\begin{center}
\leavevmode
\includegraphics[width=0.9\linewidth]{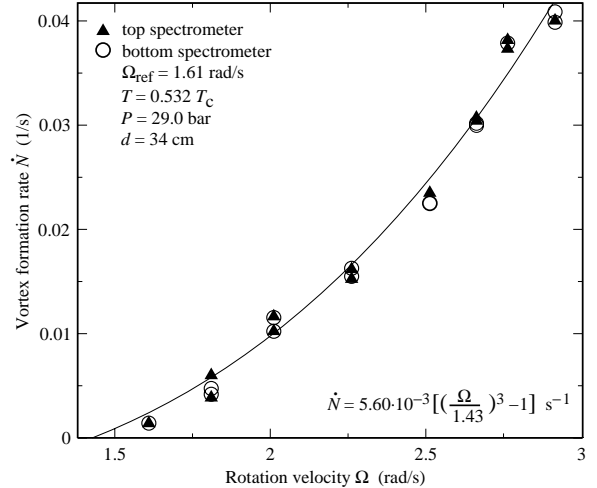}
\caption{Rate of vortex formation in neutron irradiation. The average number of
  rectilinear vortex lines created per unit time during the irradiation
  period is shown as a function of the rotation velocity $\Omega$. The data
  are fitted with the expression ${\dot N} = 0.336\,[(\Omega/1.43)^3 -
  1]\,$min$^{-1}$ (with $\Omega$ in rad/s). Depending on the rate ${\dot
    N}$, the irradiation time varies here from 30\,min to 4.5\,h, so that
  the number of accumulated vortices remains within the range of the calibration
  in Fig.~\protect\ref{ClusterCalibration}. This calibration is used to
  determine the number of vortices from the relative reduction in
  the CF peak height, $\Delta A/A_0$.  The distance of
  the neutron source from the sample was $d = 34\,$cm. The range of the
  bias flow is limited in these measurements between the critical velocity
  $\Omega_{\rm cn} = 1.43\,$rad/s and the upper limit $\Omega = 3.0\,$rad/s,
  where the turbulent events start to occur. (From Finne et~al.\ 2004a).}
  \label{VortexProductionRate}
  \end{center}
\vspace{-6mm}
\end{figure}

In Fig.~\ref{VortexProductionRate} the rate of vortex formation $\dot N$ is
measured at $0.53\,T_{\rm c}$ as a function of the bias velocity $v= \Omega
R$, during neutron irradiation. The result supports the cubic rate equation
(\ref{eq20}). The critical velocity $v_{\rm cn} = \Omega_{\rm cn} R =
4.3\,$mm/s is in the same range as the data in
Fig.~\ref{CritVel&NeutrThreshold}. This is plausible if the critical
velocity is determined only by the size of the neutron bubble which does
not change appreciably with decreasing temperature, Eq.~(\ref{e.2}).
However, the rate factor $\gamma$ is 40 times smaller than in the
measurements above $0.80\,T_{\rm c}$ and below 21.2\,bar (when scaled to
the same neutron flux). A second similar measurement at $10.2\,$bar and
$0.57\,T_{\rm c}$ gives a critical velocity $v_{\rm cn} = 3.6\,$mm/s and a
rate constant $\gamma$ which is 3 times larger than the measurement in
Fig.~\ref{VortexProductionRate}.

These results raise the question how do $v_{\rm cn}$ and $\gamma$ vary over
a wider temperature and pressure range. In particular this becomes a
problem at high pressures, where a wide range of stable A phase exists
between $T_{\rm c}$ and the ambient B-phase bath temperature $T_0$, where
the irradiation is performed. In the 10\,bar measurement A phase is not
stable.  In both of these measurements the available range of bias
velocities is limited (to approximately $v_{\rm cn} < v \lesssim 2\, v_{\rm
  cn}$), by the appearance of turbulent events at higher velocities. Both
of these low temperature measurements are consistent with the cubic form of
the rate equation, but do not conclusively prove it owing to the limited
velocity range (unlike the high-temperature result in
Fig.~\ref{BiasDependence}). Clearly more extensive measurements as a
function of pressure and temperature are called for.

\subsubsection{Superfluid turbulence in neutron irradiation} \label{Turbulence}

\begin{figure}[t]
\centerline{\includegraphics[width=0.9\linewidth]{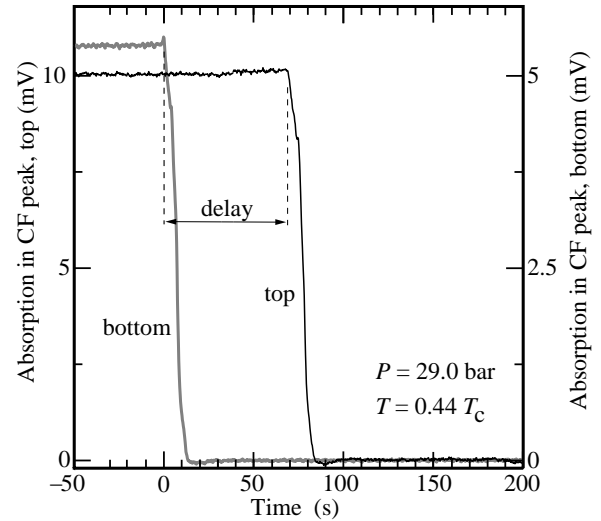}}
\caption{NMR signatures of turbulent vortex formation in neutron irradiation. The CF
  peak height (Fig.~\protect\ref{NMR-Spectra}) is monitored as a function of time in
  the two NMR detection coils. Here a vortex tangle suddenly starts
  to expand, first in the bottom coil (at $t=0$), when its CF peak
  rapidly drops to zero. After a delay
  of 70\,s, which the turbulent front needs to travel to the lower edge of the
  top coil 90\,mm higher along the column, a similar collapse is recorded
  by the top spectrometer. The flight time $\tau = z/(\alpha \Omega R)$
  corresponds to the situation where $z=95\,$mm and
  turbulence is first formed in the middle of the bottom
  coil. This sequence of events is
  schematically displayed here at $\Omega = 1.61\,$rad/s. The sudden collapse
  of the CF peak means that the turbulent tangle
  is rapidly polarized at these large flow velocities and that the global
  counterflow between the normal
  and superfluid components is thereby removed. Measurements of this type show
  that from the initial injection site, where the extracted vortex loops first start
  to intersect in the bias flow, turbulence expands by forming
  two fronts which move at constant velocity towards the top and
  bottom ends of the rotating column. (From Finne et~al.\ 2004a).}
\label{TurbulenceSignal}
\end{figure}

At high rotation and low temperatures below $0.60\,T_{\rm c}$ neutron
irradiation events may become turbulent, similar to vortex formation from
other sources (Finne et~al.\ 2003). This means that vortex loops, which
have been extracted from the neutron bubble and are injected into the bias
flow, may start to interact, to produce a vortex network of large scale.
This tangle then blows up and fills the rotating sample with the
equilibrium number of vortex lines. Some NMR characteristics of such a
neutron capture event are illustrated in Fig.~\ref{TurbulenceSignal}. The
basic experimental feature here is that the sample is suddenly filled with
the equilibrium number of rectilinear vortex lines, apparently as a result
of one neutron capture event: The NMR absorption spectrum jumps (via a
brief transitory period) from a line shape with a large CF peak to that of
the rotating equilibrium state of totally different form
(Fig.~\ref{NMR-Spectra}).

Turbulent events are observed in neutron irradiation only if (i) the rotation
velocity exceeds $\Omega_{\rm cn}$ and (ii) the temperature is sufficiently low so
that vortex motion is not heavily damped by mutual friction. Even then these
processes are stochastic such that the vortex loops extracted from the neutron
bubble only rarely achieve the proper initial conditions in which turbulent loop
expansion starts to evolve. With decreasing temperature and increasing rotation the
probability of turbulent events increases. This is understandable for the following
reasons: (i) With decreasing temperature mutual friction damping is reduced, Kelvin
wave excitations on the existing vortices grow in amplitude, more new loops are
formed, and via reconnection processes the intersecting loops multiply to a
turbulent cascade. (ii) With increasing bias velocity the number of vortex loops,
which are injected into the bias flow, rapidly increases (Fig.~\ref{BiasDependence},
bottom) and the probability of their intersections increases. Here we shall not
discuss general features of turbulent superfluid dynamics, only describe the main
signatures of the neutron-induced turbulent events.

Fig.~\ref{TurbulenceSignal} illustrates how the height of the CF peak suddenly drops
to zero in the NMR spectrum when a turbulent event starts to evolve.  In this
schematic example the turbulent tangle first appears inside the bottom coil. The
collapse of the CF peak height is the first feature in the NMR absorption spectrum
which signals the turbulence: Its rapid decay shows that the turbulent state becomes
polarized, to mimic on an average solid-body rotation. Simultaneously the NMR
absorption intensity from the CF peak is transferred close to the Larmor edge of the
spectrum where a new sharp peak rapidly grows in intensity. This new peak then
slowly decays to the line shape of the equilibrium state (Fig.~\ref{NMR-Spectra}).
The intensity in the Larmor region reflects how the vortex density evolves within
the coil: It first rapidly overshoots to a value which is about twice that in
equilibrium and then slowly rarefies to the equilibrium value.


\begin{figure}[!!!tb]
  \centerline{\includegraphics[width=0.9\columnwidth]{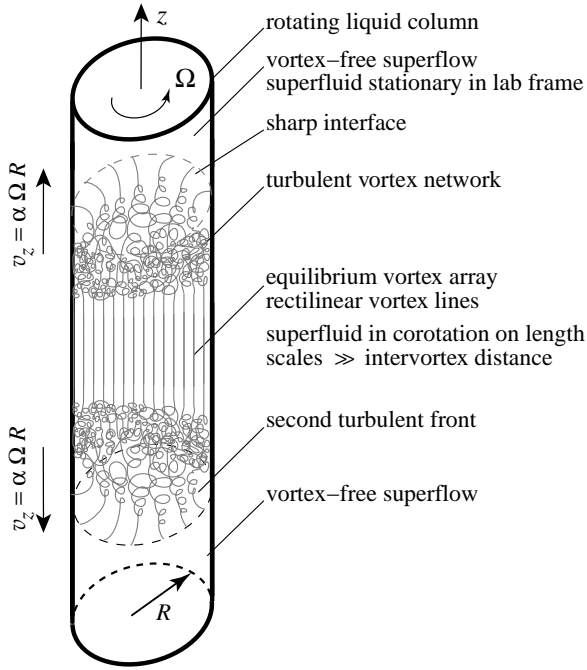}}
  \bigskip
\caption[RotColumnTurbulence] {Turbulent expansion of quantized vorticity in a
rotating superfluid column. Initially the column rotates in the vortex-free state.
If sufficiently many loops are extracted from a neutron bubble, they may interact to
develop a superfluid tangle. The tangle forms with a sharp advancing front and a
trailing tail which decays into the equilibrium rotating state with rectilinear
lines. Two turbulent fronts thus propagate along the column in opposite directions,
while in the middle there exists already an ordered array of rectilinear vortex
lines, if the column is sufficiently long. The axial velocity of the fronts is
controlled by the dissipative mutual friction $\alpha$. The ordered array in the
middle remains fixed in the rotating frame, while the two fronts rotate, as
determined by the reactive mutual friction $\alpha^{\prime}$. }
  \label{RotColumnTurbulence}
\end{figure}


The decaying CF signals from the two coils do not overlap in
Fig.~\ref{TurbulenceSignal}.  Even the more slowly relaxing overshoots in the Larmor
region do not overlap in time when $T<0.5 T_{\rm c}$ and turbulence is initiated at
one end of the sample tube. This means that turbulence propagates along the column
as a stratified layer: By the time the NMR absorption in the top coil gives the
first indication of the approaching turbulent front, the bottom coil has already
settled into its stable equilibrium line shape. The turbulent front is formed from a
relatively thin layer of disordered tangle. Here the polarization of the circulation
reaches its final equilibrium value almost immediately, even before the vortex
density has reached its peak, and well before the vortex configuration has
stabilized to its equilibrium state of rectilinear lines. These observations suggest
that vorticity propagates in the rotating column in the peculiar configuration shown
in Fig.~\ref{RotColumnTurbulence}. Here turbulence does not fill the entire column,
only short sections, before it straightens into rectilinear lines. Such a
configuration has not been described before -- simply because it has not been
possible to set up this situation in $^4$He-II, {\it ie.} one where vortices are
injected in high-velocity rotating superflow.

\begin{figure}[t]
\centerline{
  \includegraphics[width=0.9\linewidth]{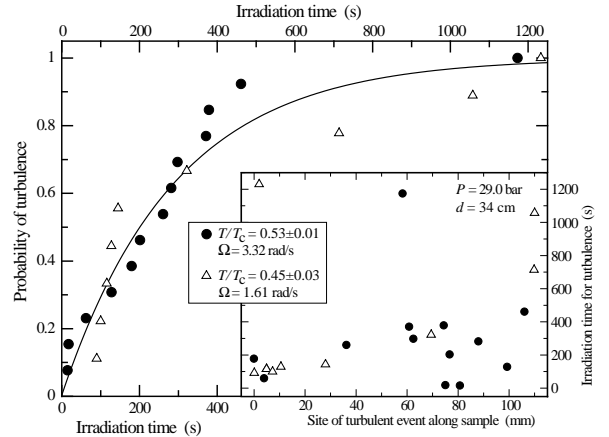}}
\caption{Turbulent vortex formation in neutron irradiation at 29.0\,bar
  pressure. The sample is irradiated in constant conditions until a
  turbulent vortex expansion event takes place. The irradiation time
  required to achieve the turbulent event is measured. Results from
  measurements at two different constant conditions are shown in this plot
  in the form of cumulative probability distributions. The solid curve is a fit of the data measured at $0.53\,T_{\rm c}$ to a distribution function of the form $P(t) = 1-(1-p_{\rm n})^{{\dot N}_{\rm e}t}$, where the probability of a single neutron absorption event to start turbulence is $p_{\rm n} = 0.092$. The rate ${\dot N}_{\rm e}$ of the neutron-induced vortex injection events is taken as $0.55\, \dot{N}$ from Fig.~\ref{VortexProductionRate} as appropriate for $v/v_{\rm cn} = 2.32$ (see Fig.~\ref{BiasDependence}). There were no cases among the two sets
  of measurements where a turbulent event would not have been observed
  after an irradiation period of 20\,min. The {\it inset} shows
  the axial location of the initial turbulent seed, the site of the neutron
  capture, measured from the orifice upward with a technique explained in
  Fig.~\protect\ref{TurbulenceSignal}. The corresponding irradiation time needed to
  achieve the event is shown on the vertical axis. As expected, the sites
  are randomly distributed along the sample.}
\label{NeutronTurbulence}
\end{figure}

The turbulent front moves along the column with a fixed velocity which was
measured by Finne et~al.\ (2004b). This velocity has the same value as that
of a single short vortex filament moving along the cylinder wall in the
initial vortex-free bias flow: $v_{\rm z} = \alpha \Omega R$. Here $\alpha$
is the dissipative mutual friction coefficient which was measured by
Bevan et~al.\ (1997a). This means that from the delay between the signals of
the two detector coils (as marked in Fig.~\ref{TurbulenceSignal}), we may
calculate the axial location $z$ where the turbulent event started. This
location has been plotted for the data in Fig.~\ref{NeutronTurbulence} in
the inset of this figure.

The probability of a neutron capture to trigger a turbulent process is studied in
Fig.~\ref{NeutronTurbulence}.  Two measurements are shown, at different temperatures
and rotation velocities. The cumulative probability distribution has been plotted
for the irradiation time needed to achieve the turbulent event. The sample is
irradiated at constant conditions until the CF peak is observed to collapse
suddenly. The irradiation time is plotted on the horizontal axis. On the vertical
axis the number of turbulent events observed within this time is shown, normalized
to the total number of events: 13 events in one case (at $0.53\,T_{\rm c}$) and 9 in
the second (at $0.45\,T_{\rm c}$). As seen from the plot, the number of events is
insufficient to produce smooth probability distributions, but the irradiation times
are observed to be distributed over the same range in the two cases, {\it ie.} their
distributions have similar average values and widths. This in spite of the fact that
the measurements at $0.45\,T_{\rm c}$ were performed at half the rotation of those
at $0.53\,T_{\rm c}$. At high temperatures a reduction by two in velocity results in
a significant decrease in the yield of vortex loops from one neutron absorption
event (Fig.~\ref{BiasDependence}). The fact that the two distributions in
Fig.~\ref{NeutronTurbulence} do not differ significantly indicates that with
decreasing temperature the transition to turbulence becomes more probable and less
sensitive to the initial configuration of the injected loops.

Let us consider the measurement at $0.53\,T_{\rm c}$ in more detail. The state of
the sample changes during the irradiation at constant $\Omega$: (i) the initial
state is vortex-free, (ii) during the irradiation rectilinear vortex lines are
formed at a rate which can be extrapolated from
Fig.~\protect\ref{VortexProductionRate}, (iii) until finally all vortex-free CF is
completely terminated in a turbulent event. In the final step the state of the
sample changes from one with only a small central vortex cluster to one with the
equilibrium number of vortex lines $(N_{\rm
  eq} \sim 2600)$. Note that to observe a new turbulent event the existing
vortices have to be annihilated, by stopping rotation. Then the vortex-free state
can be prepared again and a new irradiation session can be started. The longest
irradiation time is here $\sim 20\,$min. During this period the cluster grows at the
rate ${\dot N} = 3.9\,$vortices/min, so that it contains about 80 vortices when the
turbulent event finally starts. At this point the CF velocity at the sample boundary
$v = v_{\rm n} - v_{\rm s} = \Omega R - \kappa N/(2\pi R)$ has been reduced by
2.7\,\% from the initial state. Since the mean irradiation time in the measured
distribution is only $\sim 250\,$s, the reduction in CF velocity by vortices formed
before a turbulent event is minor. We may thus view the result in
Fig.~\ref{NeutronTurbulence} as representative of these particular values of
rotation and temperature.

One may wonder whether a turbulent process results from a single neutron capture
event or from the coincidence of two or more events. In the latter case the
simultaneous events need to be sufficiently close not only in time, but also in
space, so that expanding loops, which have been extracted from the two neutron
bubbles, have a possibility to intersect. (The probability of the bubbles themselves
to intersect is very small.) The intersection of the loops expanding from the two
random positions along the height of the cylindrical sample is more likely to occur
closer to the middle than at the top or bottom end. However, the events listed in
the inset of Fig.~\ref{NeutronTurbulence} occur randomly along the sample. In all
cases the NMR signatures from the turbulent events are similar, there is no
prominent variation in their appearance depending on where the event starts.
Additionally, at lower temperatures, the turbulent events occur close above
$\Omega_{\rm cn}$.  Here successful neutron absorption events, which lead to the
extraction of loops to the bulk, are rare but, nevertheless, turbulence becomes more
probable. From this we presume that a single neutron capture event in suitable
conditions must be able to start the turbulence. For a more careful proof of this
point the measurements should be repeated as a function of the neutron flux.

To conclude, with deceasing temperature turbulence moves closer to the critical
threshold $v_{\rm cn}$. This means that the range of bias velocities, in which the
neutron-induced vortex formation process can be studied, decreases and ultimately
below $0.45\,T_{\rm c}$ a measurement of the bias dependence becomes obsolete.
Nevertheless, here and at lower temperatures it is a practical technique to start
turbulence.


\begin{figure}[!!!tb]
  \centerline{\includegraphics[width=1.0\columnwidth]{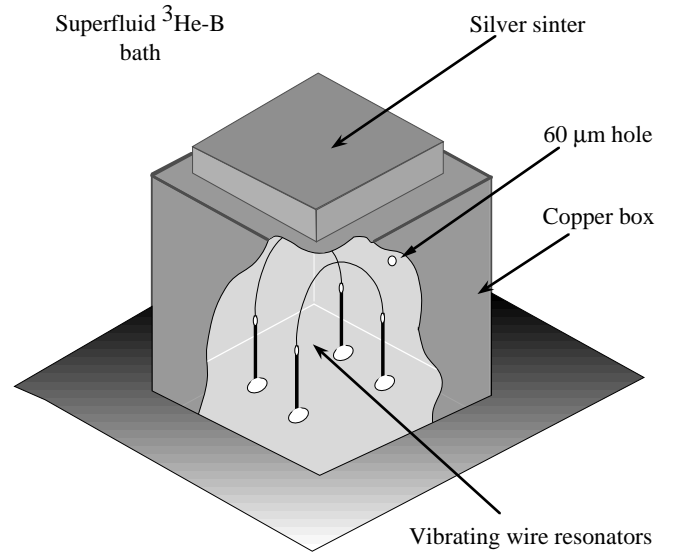}}
  \bigskip
\caption[QPBolo] {Bolometer box, which is used to measure calorimetrically the
  energy balance of the neutron capture reaction
  in the low temperature limit of ballistic quasiparticle
  motion. Two vibrating superconducting wire loops are included, one
  for measuring the temperature of the liquid within the box and the
  other as a heater to calibrate the calorimetric measurement.
  (From B{\"a}uerle et~al.\ 1998b).}
   \label{QPBolo}
\end{figure}


\subsubsection{Calorimetry of vortex network} \label{GrenobleExp}

Another type of low temperature experiments are the calorimetric
measurements of B{\"a}uerle et~al.\ (1996, 1998a) which determine the amount of
heat which is dumped into the liquid $^3$He-B bath in each individual
neutron absorption event. They allow quantitative comparison to the
KZ model. These measurements are performed in the very low temperature
limit of ballistic quasiparticle motion in a quiescent bath, with zero bias
flow.

\begin{table*}[t]
  \renewcommand{\arraystretch}{1.4}
  \begin{tabular}{lcccl}
    &\multicolumn{3}{c}{ $P$\,(bar)}\\
    \cline{2-4}
    & 0 & 6 & 19.4 & \multicolumn{1}{c}{Source} \\
    \hline
    $\Delta E$ exper. (keV) & 85 & 95 & 150 & B{\"a}uerle et~al.\ (1996, 1998a) \\
    $T_{\rm c}$\,(mK) & 0.93 & 1.56 & 2.22 & Greywall (1986)\\
    $v_{\rm F}\,(10^3\mbox{cm}/\mbox{s})$ &
       5.95 & 5.04 & 3.94 & Greywall (1986)\\
    $C_{\rm v}\,[10^3\mbox{erg}/(\mbox{cm}^3\,\mbox{K})]$ &
       5.83 & 12.8 & 25.4 & Greywall (1986)\\
    $\xi_0\,$(nm) & 65 & 33 & 18 &
       $\xi_0 = \sqrt{7\zeta(3)/(48\pi^2)} \;
       \hbar v_{\rm F}/(k_{\rm B} T_{\rm c})$\\
    $\tau_0\,$(ns) & 1.1 & 0.65 & 0.46 & $\tau_0 = \xi_0/v_{\rm F}$ \\
    $R_{\rm b}\,(\mu\mbox{m})$ & 27 & 17 & 12 &
       $\displaystyle R_{\rm b} = \sqrt{\frac{3}{2\pi e}} \Bigl(
       \frac{E_0}{C_{\rm v} (T_{\rm c} - T_0)}\Bigr)^{1/3}$ \\
    $\tau_{\rm T} \,(\mu\mbox{s})$ & 0.59 & 0.17 & 0.12 &
       Wheatley (1975)\\
    $D\,(\mbox{cm}^2/\mbox{s})$ & 21 & 4.3 & 0.94 &
       $D = v_{\rm F}^2 \tau_{\rm T}$ \\
    $\tau_{\rm Q}\,(\mu\mbox{s})$ & 0.16 & 0.31 & 0.69 &
       $\tau_{\rm Q} = (e/6) R_{\rm b}^2/D$ \\
    $\xi_{\rm v}\,(\mu\mbox{m})$ & 0.23 & 0.15 & 0.11 &
       $\xi_{\rm v} = \xi_0 (\tau_{\rm Q}/\tau_0)^{1/4}$ \\
    $\tilde D_{\rm b}$ & 210 & 200 & 190 &
       $\tilde D_{\rm b}=\bigl(\frac43\pi R_{\rm b}^3/\xi_{\rm
v}^3\bigr)^{1/3}$ \\
    $L_{\rm v}\,$(cm) & 79 & 43 & 28 &
       $L_{\rm v} = (4\pi/3a_l) R_{\rm b}^3/\xi_{\rm v}^2$ \\
    $\rho_{\rm s}\,(\mbox{g}/\mbox{cm}^3)$ & 0.081 & 0.094 & 0.108 &
       $\rho_{\rm s}\approx\rho$, Greywall (1986) \\
    $\Delta E$ theor. (keV) & 170 & 140 & 120 &
       $\Delta E = (\rho_{\rm s} \kappa^2/4\pi)
       L_{\rm v} \ln(\xi_{\rm v}/\xi_0)$ \\[3pt]
  \end{tabular}
\vskip3mm
\caption{ Comparison of measured and estimated total vortex line
  energies at different pressures in a random vortex network, as generated
  by  a neutron absorption event according to the KZ model. The ambient
  temperature of the $^3$He-B bath is taken to be $T_0 = 0.16$ mK. For the
  value of $\tau_{\rm Q}$ we use the time it takes for the normal phase
  bubble to disappear. The coefficient $a_l=2.1$, for estimating the total
  vortex-line length $L_{\rm v}$, is taken from the simulation results
  in Sec.~\protect\ref{IniDistrib}, when the value of the dimensionless
  bubble diameter is $\tilde D_{\rm b} = 200$.} \label{TableBunkov}
\end{table*}

The measuring probe is a superconducting wire loop. It is oscillated in the liquid
with a frequency of a few hundred Hz and a high Q value, by driving the loop at
resonance with an ac current in a dc magnetic field oriented perpendicular to the
wire. The damping of the wire oscillations measures the density of quasiparticle
excitations in the liquid and can be calibrated to give the temperature. For this a
second vibrating wire loop is used as a heater (Fig.~\ref{QPBolo}). The latter is
excited with a known current pulse to heat up the liquid, by driving the wire at
supercritical velocities where the break-down of Cooper pairs gives rise to a shower
of quasiparticles. To contain the ballistically moving quasiparticles, both wire
loops are placed inside a box from which the particles leak out through a pin hole
at a well-known rate ~(B{\"a}uerle et~al.\ 1998b).

The thermal connection between the liquid in the calorimeter box and that in the
surrounding $^3$He-B bath is only via the small pin hole. A neutron absorption event
inside the box heats up the liquid and a thermal pulse is recorded with the wire
resonator. The rise time is determined by the resonator properties, while the
trailing edge (with a time constant of about 1 min) monitors the much slower leakage
of quasiparticle excitations from the box through the orifice. If energy is released
into the box on a still longer time scale it is not recorded in the form of pulses.

It is found that neutron absorption events amount to roughly 100 keV
smaller thermal pulses than the 764 keV, which a slow neutron is expected
to yield for the absorption reaction with a $^3$He nucleus. Since the
various recombination channels of the ionized charge and the subsequent
thermalization processes in liquid $^3$He are poorly known, it is not quite
clear how large an energy deficit one should expect. Obvious losses include
the ultraviolet radiation absorbed in the walls of the bolometer box and
the retarded relaxation of excited electronic states of helium atoms and
molecular complexes. However, B{\"a}uerle et~al.\ (1996, 1998a) expect that these
contributions are not strongly pressure dependent and on the order of 7\%
of the reaction energy.  They then ascribe the remaining energy deficit to
the random vortex network which is created in the neutron bubble and which
in the low temperature limit should have a long life time, when mutual friction
approaches zero and no flow is applied.

In Table~\ref{TableBunkov} the measured missing energy at three different liquid
pressures has been compared to that estimated from the KZ model. The measured result
$\Delta E_{\rm exp}$ is recorded on the top most line while the calculated
comparison proceeds stepwise from one line to the next in the downward direction of
the table. The final result $\Delta E_{\rm theor}$ can be found on the lower most
line.  The thermal diffusion constant, $D = 3 k_{\rm T}/C_{\rm v} = v_{\rm F}^2
\tau_{\rm T}\,$, is derived from the conductivity $k_{\rm
  T}$, which has been tabulated by Wheatley (1975) while all
other liquid $^3$He values are taken from Greywall (1986). The agreement between
the top and bottom lines is within a factor of 2, which is surprising, if we
remember the uncertainties which are built into this comparison.
Table~\ref{TableBunkov} has been included in this context to draw attention on the
magnitudes of the different quantities.

The significance of the calorimetric measurement rests entirely on the quantitative
analysis of the measured data -- whether an energy deficit is present and can be
ascribed to vortices. The quantitative analysis is hampered by three kinds of
uncertainties:

(1) The proportions of the radiative and retarded contributions in the
thermalization after neutron absorption are not known sufficiently well
(Leggett 2002).

(2) To calibrate the bolometer box, the heater wire is vibrated at a
velocity which exceeds the pair-breaking limit. It is now known that when
the vibrating wire loop reaches this limit, then a beam of quasiparticles and vortex loops is created. The fraction of energy spent on generating vortices is temperature dependent and appears to decrease rapidly with increasing temperature. With increasing drive level more and more vortices are formed at different points along the wire loop (Fisher et~al.\ 2001; Bradley et~al.\ 2004). The
calibration of the bolometer box then becomes uncertain, if a sizeable fraction
of the energy of the calibrating electrical pulse ends up in a long lived turbulent vortex network.

(3) The decay time of a vortex tangle is unknown in the zero temperature
limit. The calorimetric measurement of the energy deficit in the neutron
absorption event requires a time constant of tens of seconds so that the
decay of the tangle would not contribute to the measured thermal pulse. On
the other hand, the measurements of Fisher et~al.\ (2001) on
vibrating-wire-generated tangles suggest that their time constant of decay
is seconds. The simulation calculations of Barenghi and  Samuels (2002) suggest
even faster decay.  However, the decay time is expected to depend on the
topology and line density of the vortex tangle since it proceeds via
reconnection into loops which fly away in the quiescent bath and finally
annihilate on solid surfaces (where part of the energy may go directly to
the wall, escaping the liquid). In this respect the random vortex network
created in the neutron absorption event (Sec.~\ref{IniDistrib}) is expected
to be quite different from the vortex tangle
generated by the vibrating wire: According to the
measurements of Bradley et~al.\ (2004), during stationary state vortex generation
and decay by a vibrating wire the line density reaches only a very low
value of 20\,mm$^{-2}$ (an inter-vortex distance $\sim 0.2\,$mm). In
comparison, in Table~\ref{TableBunkov} the initial line density expected in
the neutron bubble, as precipitated by the KZ process, is of order
$10^7\,$mm$^{-2}$.

The original interpretation of the calorimetric measurements supports
the KZ model of vortex formation. However, to keep this statement valid in
view of recent developments on vortex dynamics at ultra low temperatures,
more work is required. In principle a calorimetric measurement of the energy in the random vortex network, which is generated in a neutron absorption event, provides more direct proof of the KZ mechanism, than rotating measurements: Here in the absence of applied flow, the discussion about the superflow instability at the neutron bubble boundary is irrelevant. Also the calorimetric measurement gives directly the domain size $\xi_{\rm v}$ of the inhomogeneity in the order parameter distribution (Eq.~(\ref{eq:xi-initial})) right after the quench, which can otherwise be inferred only indirectly from other types of measurement.

It is interesting to note that the understanding gained from the calorimetric work has turned $^3$He-B into an attractive absorber material in the search for dark matter particles. In such a dark matter detector the collision energy will be measured with an array of bolometer boxes equipped with micro-mechanically fabricated vibrating resonators (Collin et~al.\ 2004). In the next sections we shall analyze further the initial loop formation in the neutron bubble and its later evolution,
which we so far have omitted.

\subsection{Simulation of loop extraction} \label{Simulation}

In spite of the agreement, which we have listed so far between the measured
characteristics of neutron-induced vortex formation and the KZ model, a deeper
understanding of the processes involved would be important. Owing to its
phenomenological content, the KZ model is based on general concepts and contains few
media-dependent parameters (such as the superfluid coherence length $\xi$ or the
order-parameter relaxation time $\tau$). These are known in the case of superfluid
$^3$He, hence the predictions of the model for the initial state of the vortex
network can be calculated. In the rotating experiments the initial state is
connected with observable quantities only through the complex evolution of the
network, governed by superfluid hydrodynamics. This means that a more rigorous
comparison requires numerical simulation.

A number of numerical simulations exist on the evolution of a network of linear
defects. These apply to cosmic strings, liquid crystals, and vortices in superfluid
$^4$He. Such results cannot be directly transferred to neutron-induced vortex
formation in $^3$He-B. The differences with the cosmic string and liquid crystal
calculations arise from the different boundary conditions and the equations which
govern the evolution of the network in the applied external bias fields.  In the
case of superfluid $^4$He, studies of random vortex networks often concentrate on
vortex flow driven by thermal counterflow between the normal and superfluid
components in a stationary situation (Tough 1982). In contrast, the vortex network,
which is produced in $^3$He-B in a neutron absorption event, is in a state of rapid
non-stationary evolution. Nevertheless, standard techniques exist (see eg.
Schwarz 1978, 1985, 1988) and can be applied for solving also the
transient problem in $^3$He-B. Here we describe calculations which address the
dependence of the vortex-formation rate on the normalized bias velocity $x =
v/v_{\rm cn}$ (Ruutu et~al.\ 1998a).

\subsubsection{Initial loop distribution}
\label{IniDistrib}

Vachaspati and Vilenkin (1984) developed an approach for the simulation of the initial network
of linear defects after a rapid phase transition, adapted to the case of cosmic
strings. This technique has been used in most studies since then (see e.g. the
reviews by Hindmarsh and Kibble 1995 and Bray 1994).

We approximate the neutron bubble with a cubic volume which is
subdivided into smaller cubes, such that the size of these volume
elements equals the length scale of the initial inhomogeneity in
the order-parameter distribution at the moment of defect
formation.  This length is on the order of the coherence length
and in the simulation it plays the role of a ``unit'' length: It
is a parameter of the model, which has to be derived by other
means. In the case of the superfluid transition this length is
given by Eq.~(\ref{eq:xi-initial}).

An arbitrary phase is assigned to each vertex of the grid, to
model the initial random inhomogeneity of the order parameter. It
is usually assumed that the distribution of the phase in each
segment of the grid between two vertices corresponds to the
shortest path in the phase circle. Thus it is possible to
determine whether a line defect pierces a given face. Then the
centers of the corresponding faces are connected to form closed or
open (depending upon specific boundary conditions) linear defects,
strings in the cosmological case and vortices in the superfluid.

Vachaspati and Vilenkin assigned to each vertex a value of the phase
from the following set: $\{0,2\pi/3,4\pi/3\}$ and then studied the
statistical properties of the resulting network of strings.  They
found that most (70\%) of the strings were in the form of open segments, which
extended from one boundary of the system to another.  For closed loops
they found two scaling relations to hold:
\begin{equation}
        n = C L^{-\beta},
\label{NL}
\end{equation}
and
\begin{equation}
        {\cal D} = A L^\delta,
\label{DL}
\end{equation}
where $\beta \approx \frac52$, $\delta \approx \frac12$, $n$ is
the density of loops with a given length $L$, and $\cal D$ is the
average spatial size of a loop. It is usually defined as an
average of straight-line dimensions in $x$, $y$ and $z$
directions. In this model both the characteristic inter-vortex
distance and the radius of curvature are on the order of the
length scale of the spatial inhomogeneity (i.e. the size of the
small cubes, which here has been set equal to unity). Later, other
variations of this model have been studied, including other types
of grids and other sets of allowed phases.  However, it has been
found that the scaling relations (\ref{NL}) and (\ref{DL}) hold
universally in each case.

\begin{figure}[!!!tb]
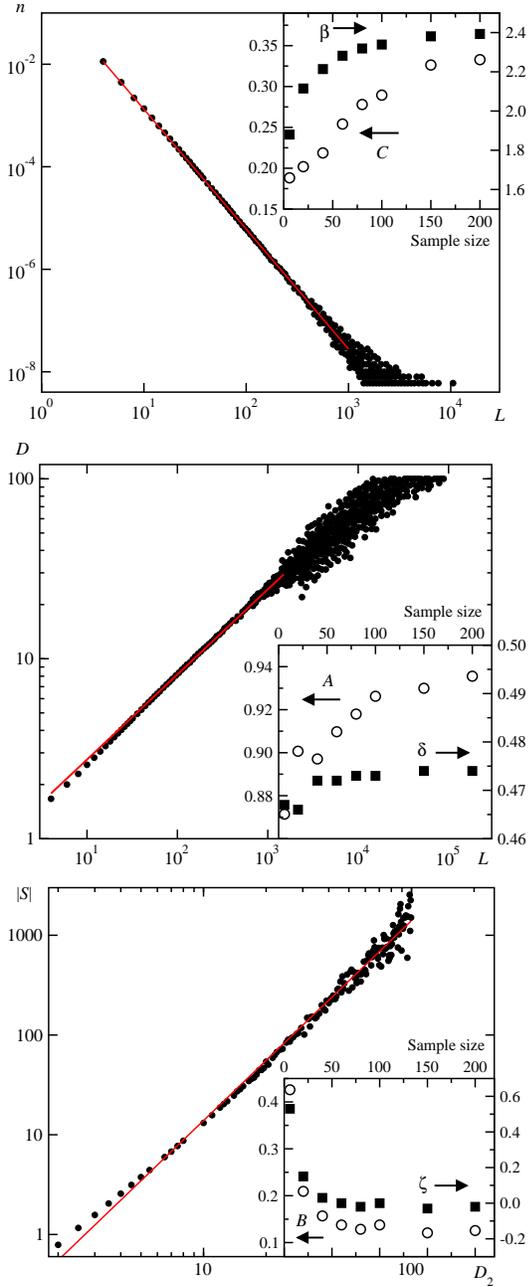

  \vbox{
    \centerline{\includegraphics[width=0.8\columnwidth]{rep-fig1x.eps}}
    \medskip
    \centerline{\includegraphics[width=0.8\columnwidth]{rep-fig2x.eps}}
    \medskip
    \centerline{\includegraphics[width=0.8\columnwidth]{rep-fig3x.eps}}}
  \bigskip
  \caption[fig:ini-distrib]{
    Scaling properties of vortex loops in the initial network:
    density $n$ {\em(top)} and average spatial size $\cal D$ {\em(middle)},
    plotted as a function of the length $L$ of the loops. Below their
    algebraic area $|S|$ is shown {\em(bottom)}, as a function of the
    2-dimensional spatial size ${\cal D}_2$ of the loop. The calculations
    are performed on a lattice of $100\times 100\times 100$ cells. The
    solid lines are fits to scaling laws obeying equations of the
    form of (\protect\ref{NL}),
    (\protect\ref{DL}), and (\protect\ref{SD}), respectively. The inset
    in each plot shows the dependence of the scaling parameters on the
    size of the cubic neutron ``bubble''.
    \label{fig:ini-distrib}
    }
\end{figure}

The direct applicability of these results to the case of vortex
formation in $^3$He-B is not evident. In cosmology the open lines
are the most significant ones: only these strings may survive
during the later evolution if no bias field exists to prevent
closed loops from contracting and annihilating. In the case of a
normal-liquid bubble within the bulk superfluid there exists an
obvious boundary condition: the phase is fixed at the boundary and
there are no open lines at all. Thus it is of interest to find out
the influence of the boundary conditions on the scaling relations
(\ref{NL}) and (\ref{DL}).

Another question, which arises in the case of superfluid vortices,
concerns the interaction of the loops with the external bias field
due to the normal-superfluid counterflow. The energy of a tangled
vortex loop in the counterflow is proportional to its algebraic
area $S = \oint\! y\, dz$ in the direction $x$ of the counterflow
velocity,  Eq.~(\ref{eq14}). Thus the dependence of $S$ on the
length of the loop is of interest as well.

The simulation is performed in a cubic lattice with a fixed (zero)
phase at the outer boundary. The set of allowed phases is not
limited. It is known that such limitations affect the number of
open lines in cosmic-string simulations, but this should be
irrelevant for the network in the neutron bubble. Several vortices
may pass through one cell in the grid. In this case the
corresponding faces of the cell are connected at random. For
calculating the length of a loop, it is assumed that both straight
and curved segments of the loop inside one cell have unit length.
The size of a loop in the direction of a specific coordinate axis
is measured as the number of cells in the projection of the loop
along this axis.

The size of the volume, which undergoes the superfluid transition
in the neutron absorption event, is about $50\,\mu$m, while the
characteristic inter-vortex distance in the initial network is of
the order of $1\,\mu$m. Calculations have been performed for
several ``bubble'' volumes: starting from $6\times 6\times 6$ up
to $200\times 200\times 200$. For each bubble size the loop
distributions obtained from a large number of (up to 1000) initial
distributions of random phases are averaged. The resulting
distributions of $n$, $\cal D$,  and $S$ are shown in
Fig.~\ref{fig:ini-distrib}. One can see that despite differences in the
boundary conditions the
Vachaspati-Vilenkin relations (\ref{NL}) and (\ref{DL}) hold in
the case of vortices in superfluid helium, but the exponents and
prefactors are slightly different and depend on the size of the
bubble.

For the algebraic area $S$ of a loop as a function of the
corresponding 2-dimensional diameter ${\cal D}_2$ the additional
scaling law is
\begin{equation}
       |S| = B {\cal D}_2^{2-\zeta},
\label{SD}
\end{equation}
where $\zeta \approx 0$. Here ${\cal D}_2$ is the average of the
straight-line dimensions of a loop in $y$ and $z$-directions. Thus
the oriented area of a tangled loop is proportional to the area of
a circle of the same straight-line size. The scaling relation
(\ref{SD}), as well as (\ref{NL}) and (\ref{DL}), are of the form
expected for a Brownian particle, for which the square of the
average displacement on the $i$-th step is proportional to $i$ and
therefore the mean value of the square of the oriented area is
given by
\begin{eqnarray*}
&\langle S^2\rangle =
\langle \bigl(\sum\limits_{1\le i \le L} y_i \Delta z_i \bigr)^2 \rangle =
\sum\limits_{1\le i,j\le L} \langle y_i
\Delta z_i y_j \Delta z_j\rangle=&\nonumber\\ &\sum\limits_{i,j}
\langle y_i y_j\rangle \langle \Delta z_i \Delta z_j\rangle=
\sum\limits_i \langle y_i^2\rangle \langle \Delta z_i^2\rangle
\propto \sum\limits_i i \propto L^2&\;.
\end{eqnarray*}
These scaling relations were used in the analytic model of vortex-loop extraction
from the cooling neutron bubble in the bias flow [{\it ie.} Eqs.~(\ref{DLx}) --
(\ref {NLx})].

\subsubsection{Network evolution under scaling assumptions}\label{scalsec}

After its formation, the vortex network evolves under the
influence of the inter-vortex interactions and the
normal-superfluid counterflow $v$. Its characteristic length scale
$\tilde\xi(t)$ increases with time.  Vortices start to escape from
the bubble when the energy gain due to the external counterflow
becomes larger than the energy of the superflow associated with
the vortex itself. This corresponds to the critical value of
$\tilde{\xi}$, expressed by Eqs.~(\ref{VorRing}) or (\ref{eq16}).

Tangled vortex flow in superfluid $^4$He has been studied numerically by
Schwarz (1978, 1985, 1988), Samuels (1992), Aarts and de~Waele (1994), Nemirovskii and Fiszdon (1994), Barenghi et~al.\ (1997), Tsubota and Yoneda (1995), and
others.  A large number of calculations has been devoted to the evolution of the
initial network of cosmic strings (see reviews by Hindmarsh and Kibble 1995; Bray 1994)
and also of linear defects in liquid crystals  (Toyoki 1994; Zapotocky et~al.\ 1995).
In all three cases the initial state is quite similar, but the equations controlling
the evolution are different. The common feature is that the interaction between the
loops leads to reconnections when the loops cross each other. It has been shown by
Kagan and Svistunov (1994) that the scaling relations remain valid in a random network if the
vortices are allowed to reconnect when they cross each other, but all other
interactions are neglected for simplicity. In most simulation work, both in the case
of cosmic strings and liquid crystals, it has been found that the scaling relations
are preserved during the evolution, even if inter-vortex interactions are included.

To calculate the escape rate from the network, two crude
assumptions are made, which are essentially the same as in the
analytic treatment in Sec.~\ref{AnalVorEsc}: 1) We assume that the
scaling laws remain valid until $\tilde{\xi}$ grows comparable in
size to the critical value in Eq.~(\ref{eq16}). 2) At this point
the influence of the external counterflow becomes suddenly so
significant that all sufficiently large loops immediately escape by
expanding to a rectilinear vortex line. In the numerical
simulations the state before escape is modelled by the same method
as was used to construct the initial state. It is assumed that not
only the scaling relations but also the statistical properties of
the vortex tangle remain the same during the evolution as in the
``initial'' state with the characteristic length $\tilde\xi$.

At the moment of escape, $\tilde{\xi} \sim r_\circ \propto v_{\rm
  cn}/v = 1/x$, and $\tilde\xi$ is used as the size of a cell in
the Vachaspati-Vilenkin method. For integer values of $x$ the
bubble is represented by a grid with $x\times x\times x$ vertices
and a random phase is assigned to each vertex. To satisfy the
boundary condition, this grid is surrounded by a shell of vertices
with fixed zero phase, representing the uniform bulk superfluid
outside the heated bubble.  Thus the whole grid contains
$(x+1)\times(x+1)\times(x+1)$ cells. Such a correspondence between
the size of the grid and the velocity is not too artificial even
at small values of $x$: for example, at the critical velocity
$v=v_{\rm cn}$ one gets a grid with $(x+1)^3=8$ cells and $3\times
3\times 3$ vertices, but the phase can be non-zero only at one
vertex in the middle of the grid. In this case no vortices can
appear in agreement with the definition of $v_{\rm cn}$.

\begin{figure}[!!!!tb]
  \centerline{\includegraphics[width=0.9\columnwidth]{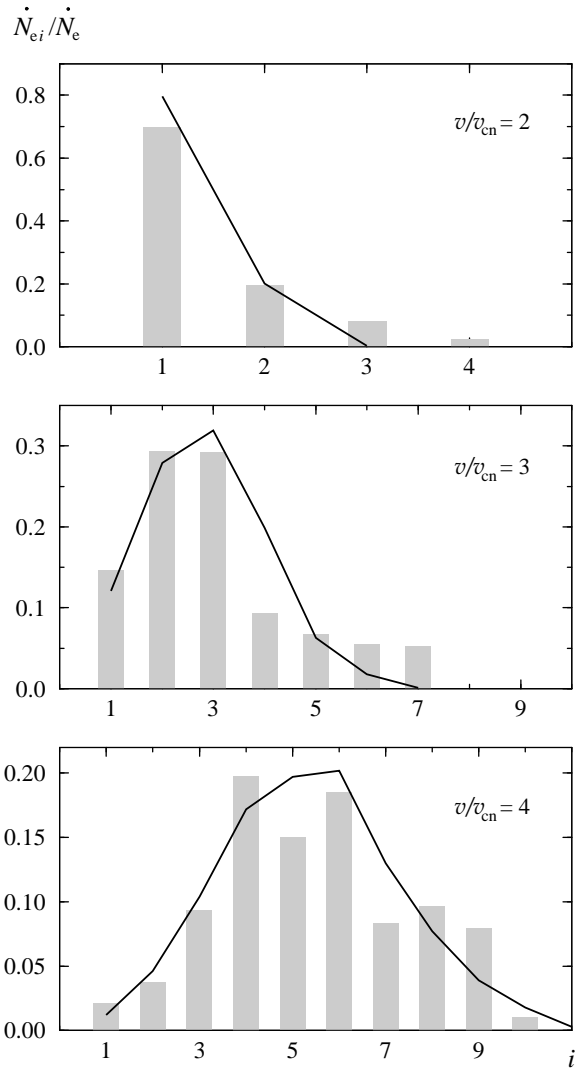}}
  \bigskip
  \caption[histo]{Distribution of the number of
    loops escaping per absorption event, normalized to the total
    number of absorption events and given at three different values
    of the external bias field, $v/v_{\rm cn}$.
    The solid lines represent the simulations while the bars denote
    the experiment.}
\label{histo}
\end{figure}

By counting the vortex lines produced from a number of
random-phase distributions it is possible to calculate the
probability distribution for the number of loops escaping per
absorption event. For these calculations it is assumed that each
loop, which survives until the moment when $\tilde{\xi} \sim
r_\circ $, will form an observable vortex line: In the case of a
tangled loop the probability is high that at least one arc is
oriented favorably with respect to the counterflow and will be
extracted.

In Fig.~\ref{histo} the calculated probability distribution is
plotted for observed neutron absorption events at different values
of $x$ ($x = 2$, $3$, and $4$) expressed as the fraction of those
neutron absorption events which produce a given number of
vortices, normalized by the number of all events which give rise
to at least one vortex. These results are in remarkably good
agreement with the experimental data (without any fitting
parameters). However, the agreement with the fraction of ``zero''
events, i.e. absorption events which produce no vortices at all,
is poorer, underestimating their number, especially at $x=2$.
Experimentally the ``zero'' events can be extracted from the
measured data by comparing the event rate at a given rotation
velocity with the saturated event rate at the highest velocity.

One reason for this discrepancy is that at low $R_{\rm b}/r_\circ$ ratio
(ie.  at low velocities) loops with a radius of curvature $\sim r_\circ$,
which are capable of escape into the bulk superfluid, have a rather large
probability to be oriented in such a way that they do not have sufficiently
large segments oriented favorably with respect to the counterflow, owing to
space constraints. Hence they contract and do not give a contribution to
the observed signal. Taking this into account one could develop more
elaborate techniques for counting vortices than the simple ``count all''
method described above. In this case better agreement with all experimental
data could be achieved.  Particularly, if only vortices with positive
algebraic area with respect to the counterflow ($S = \oint\! y\, dz$) are
counted, then good agreement with the number of ``zero'' events would be
obtained.  However, such refinements do not change the results essentially,
compared to uncertainties of the model.

\subsubsection{Direct simulation of network evolution} \label{EvolSimSec}

There is no direct evidence for the validity of the assumptions,
which were made in the previous section about the evolution of the
network, ie. whether the network remains self-similar (or scale
invariant) and the scaling laws Eqs.~(\ref{NL}) -- (\ref{SD}) can
be applied during its later evolution. Preliminary calculations
have been carried out using the techniques developed by Schwarz
for the simulation of superfluid turbulence in $^4$He. These
methods give good numerical agreement with experimental data in
$^4$He (Schwarz 1978, 1985, 1988) and have been used
extensively by Schwarz and many others
(Aarts and de~Waele 1994; Nemirovskii and Fiszdon 1994; Samuels 1992; Barenghi et~al.\ 1997; Tsubota and Yoneda 1995).

In such calculations vortices are generally considered to be
1-dimensional objects without internal structure. In $^4$He the
diameter of the core is much smaller than other characteristic
lengths, foremost the average radius of curvature for the loops or
the inter-vortex distance, and even $\ln(\tilde\xi/\xi)$ can be
treated as a large parameter. This is not the case in $^3$He-B due
to the large coherence length, especially in the early stages of
the evolution when the vortex density is largest. However, as
mentioned in Sec.~\ref{KW_mechanism}, the characteristic
inter-vortex distance and the radius of curvature may still be
several times larger than the coherence length and the diameter of
the vortex core. For now, we shall continue considering the
vortices as linear objects.

There are several forces acting on a vortex line in a
superfluid~(Donnelly 1991). The Magnus force appears in the
presence of superflow,
\begin{equation}
        \mbox{\bf f}_{\rm M} = \rho_{\rm s} \kappa \nu\, \s' \times (\v_{\rm L}-\v_{\rm sl}),
\label{magnus_force}
\end{equation}
where $\s$ is the radius vector of a point on the vortex line and
the prime denotes the derivative with respect to the length of the line (ie.
$\s'$ is a unit vector tangent to the vortex line at $\s$), $\v_{\rm L}=\dot{\s}$
is the local velocity of the vortex line, and $\v_{\rm sl}$ is the
local superfluid velocity at this point. For singular vortices in
$^3$He-B the number of circulation quanta is $\nu=1$. The local
superfluid velocity $\v_{\rm sl}$ is a sum of the superfluid velocity
$\v_{\rm s}$ far from the network and the velocity induced by all the
vortices in the tangle:
\begin{equation}
        \v_{\rm sl}(\r) = \v_{\rm s} + \frac{\kappa}{4\pi}
           \int_{\mbox{\small all}\atop\mbox{\small loops}}
             \frac{(\s-\r) \times d\s}{|\s-\r|^3}.
\label{vsl}
\end{equation}
The Iordanskii force arises from the Aharonov-Bohm scattering of
quasiparticles from the velocity field of the vortex,
\begin{equation}
        {\bf f}_{\rm Iordanskii}=  \rho_{\rm n} \kappa\nu\, \s' \times
(\v_{\rm L}-\v_{n})\,, \label{IordanskiiForceGeneral}
\end{equation}
where $\v_{\rm n}$ is the velocity of the normal component (ie. the heat
bath of the fermionic quasiparticles). The Kopnin or spectral flow
force has the same form, but originates from the spectral flow of
the quasiparticle levels in the vortex core:
\begin{equation}
        {\bf f}_{\rm sf}=  m_3 {\cal C}(T) \kappa \nu\, \s' \times
(\v_{\rm L}-\v_{n})\,, \label{SpFlForceGeneral}
\end{equation}
where the temperature dependent parameter ${\cal C}(T)$ determines
the spectral flow in the core. All these three forces are of
topological origin: they act in the transverse direction and are
thus nondissipative. They are discussed in more details in
Sec.~\ref{SecOtherAnalogs}.

In contrast, the nontopological friction
force ${\bf f}_{\rm fr}$ acts in the longitudinal direction,
\begin{equation}
        {\bf f}_{\rm fr} =  -d_\parallel\rho_{\rm s} \kappa \nu\, \s' \times
[\s'\times(\v_{\rm n}-\v_{L})]\,. \label{FrictionGeneral}
\end{equation}
Here the factor $\rho_{\rm s} \kappa \nu$ is the same as in the Magnus
force.

Neglecting the vortex mass, we may write the force balance
equation for the vortex element:
\begin{equation}
        {\bf f}_{\rm M} +{\bf f}_{\rm Iordanskii} +   {\bf f}_{\rm sf} + {\bf
f}_{\rm fr}=0\,.
\label{ForceBalance}
\end{equation}
It is convenient to rewrite the balance of forces in the following
form:
\begin{equation}
      \s' \times [(\v_{\rm L}-\v_{\rm sl})- \alpha'    (\v_{\rm n}-\v_{\rm sl})]  =\alpha \, \s'
\times [\s'\times(\v_{\rm n}-\v_{\rm sl})]\,,
           \label{balance}
\end{equation}
where $\alpha$ and $\alpha'$ are the dimensionless mutual friction
coefficients. The $\alpha$ parameters are actually the
experimentally determined mutual friction quantities (see eg.
Bevan et~al.\ 1997b). The inverse coefficients are the $d$
parameters:
\begin{equation}
\alpha+i(1-\alpha')=
\frac{1}{d_\parallel-i(1-d_\perp)}~~,
\label{alphaVsD}
\end{equation}
where $i=\sqrt{-1}$. The transverse mutual friction parameter
$d_\perp$ is expressed in terms of three temperature dependent
functions which determine the Magnus, Iordanskii and spectral-flow
forces:
\begin{equation}
      d_\perp(T)=(m_3 {\cal C}(T)-\rho_{\rm n}(T))/\rho_{\rm s}(T)\;.
      \label{dperp}
\end{equation}
Eq.~(\ref{balance}) is complicated because of the term (\ref{vsl})
which contains the integral over the vortex network. To solve
Eq.~(\ref{balance}), one may follow Schwarz and neglect the
influence of all other vortex segments on $\v_{\rm sl}$ except the one
containing the point of interest (the so called local self-induced
approximation). In this case
\begin{equation}
\v_{\rm sl} \approx \v_{\rm s} + \beta\s'\times\s''\,,\;\;{\rm where}\;\beta
= {\kappa\over{4\pi}} \ln{\tilde\xi\over\xi}\,.
\label{eqvsl_approx}
\end{equation}
The leading correction to this simplification comes from the
nearest neglected vortex segments and can be estimated to be of
order $\tilde\xi/[b\ln(\tilde\xi/\xi)]$, where $b$ is the average
inter-vortex distance. For a network in $^3$He-B, $\tilde\xi \sim
b$ and $\ln(\tilde\xi/\xi) \sim 1$ in the early stages of the
evolution and thus at this point the approximation is rather
crude. However, the main effect in the initial stages arises from
the reconnection of vortex lines which happen to cross each other
and this effect is taken into account below.

A second simplification close to $T_{\rm c}$ comes from $\alpha'
\ll \alpha$. We may rewrite the equation of vortex motion in the
form:
\begin{equation}
      \v_{\rm L} = \v_{\rm s} + \beta\s'\times\s'' +
                \alpha\s'\times[(\v_{\rm n}-\v_{\rm s}) - \beta\s'\times\s''].
\label{eqm1}
\end{equation}
In the rotating reference frame the normal component is at rest, $\v_{\rm n} =
0$, and one has $\v_{\rm s}=-\v$, where $\v =\v_{\rm n}-\v_{\rm s}$ is the counterflow
velocity. We shall be comparing to experimental results measured at $T =
0.96 \; T_{\rm c}$ and $P = 18$~bar, where the most detailed data were
collected. In these conditions $\alpha\approx 10$~(Bevan et~al.\ 1995) and
in the rotating frame Eq.~(\ref{eqm1}) can be simplified further to the
form
\begin{equation}
        \dot{\s} = \alpha\beta\s'' + \alpha\s'\times {\bf v}\,,
\label{eqm}
\end{equation}
The first term on the right hand side of Eq.~(\ref{eqm}) causes a loop to
shrink while the second represents growth or shrinking, depending upon the
orientation of the loop with respect to the counterflow.

In the numerical calculation vortices are considered as lines in
the same 3-dimensional lattice as before. The temporal and spatial
coordinates are discrete and therefore the network evolves in
discrete steps: During each step in time a vertex of a loop can
jump to one of the adjacent vertices of the lattice. The
probability of a jump in any direction is proportional to the
component of $\dot{\s}$ in Eq.~(\ref{eqm}) in this direction, so
that the average velocity equals $\dot{\s}$. Other more elaborate
representations of vortex loops, based on splines for instance,
have been used to model the dynamics of smooth vortex lines in the
continuous space. However, for the preliminary study of the
network evolution the simplest lattice representation was used. The
simple case of one circular loop in the counterflow, where the
result is known analytically, can be checked separately and it is
found to give correct results.

It is well known that the interactions between neighboring vortex
segments in the tangle play an important role in the evolution of
a dense network. These interactions lead to the reconnection of
loops which cross each other. The reconnections are the main
source for the rarefication of the network at the initial stages of
evolution, which process through formation and decay of small
loops. The reconnection process has been
studied numerically~(Schwarz 1985; Koplik and Levine 1993) and
was found to occur with a probability close to unity for vortex
lines approaching each other. In the simulations the vertices are
connected in such a way that closed loops are obtained after each
time step during the evolution. If a cell is pierced by two
vortices, the two segments coming in and the two segments going
out are connected randomly. This corresponds to a reconnection
probability of 0.5. It is reasonable to suppose that this
convention leads to a slowing down in the evolution, in particular
in the initial phase, but does not produce gross qualitative
changes.

\begin{figure}
  \centerline{\includegraphics[width=0.9\columnwidth]{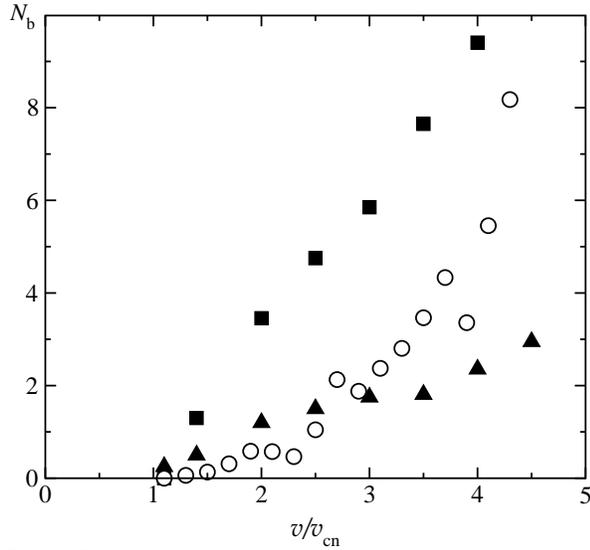}}
\caption[]{The number of vortices $N_{\rm b}$ escaped from the bubble
heated by one absorbed neutron
  as a function of $v/v_{\rm cn}$: ($\blacksquare$) simulation of the
  network evolution in the local self-induced approximation
  (Eq.~(\ref{eqvsl_approx})) and ($\blacktriangle$) by including in
  addition approximately the polarization of the tangle by the superflow
  (Eq.~(\ref{nvsl})).  Both calculations were performed on a $40\times
  40\times 40$ lattice, but the results remain unchanged if the lattice is
  reduced in size to $20\times 20\times 20$. The experimental data
  ($\circ$) represent measurements at $P = 18$ bar and $T = 0.96 \; T_{\rm
    c}$ (Ruutu et~al.\ 1998a). }
\label{evolres}
\end{figure}

The vortex tangle is initially produced by the procedure described
in Sec.~\ref{scalsec}, in a lattice $40\times 40\times 40$, and
its evolution is followed until all vortices have disappeared or
loops have expanded and formed circular planar rings far from
their location of origin. These large rings are counted, since it
is fair to assume that each one of them will produce a detectable
rectilinear vortex line in the rotating container. The calculation
is repeated for a large number of random initial distributions of
the phase at any given value of $v$. The results are averaged to
obtain the average number of vortices $N_{\rm b}$, which are produced
from one simulated bubble as a function of $v$, and are compared to
experimental data in Fig.~\ref{evolres}. It is also possible to
study the validity of the scaling law (\ref{NL}) at different
stages of the evolution in the course of these simulations. In the
absence of counterflow ($v=0$) it is found that the relation is
valid for large loops, $L > 4\tilde\xi$, at least during much of
the early evolution (in the late stages, when $n$ approaches zero,
the statistical noise exceeds $n$).

Fig.~\ref{evolres} illustrates the simulation results. No vortices
are obtained at low counterflow velocity $v<v_{\rm cn}$, but when
$v > v_{\rm cn}$, their number $N_{\rm b}$ starts to increase rapidly
with $v$. The value of the threshold velocity $v_{\rm cn}$
corresponds to the situation when the largest radius of curvature
$r_\circ(v_{\rm cn})$ becomes equal to the diameter of the initial
volume of the vortex network (ie. the diameter of the heated
bubble). It then becomes possible for a loop to escape into the
bulk superfluid if it consists of at least one arc with
sufficiently large radius of curvature $\geq r_\circ$. The same
calculations were also performed for a tangle with an initial
volume of $20\times 20\times 20$, but no differences were found in
the dependence $N_{\rm b}(v/v_{\rm cn})$. This is additional
circumstantial evidence for the fact that the scaling relations
approximately survive during the evolution of the network also in
the counterflow until the moment of escape and that the network
can be approximated as scale invariant.

The experimental data (denoted with ($\circ$) in Fig.~\ref{evolres}) fit
the universal dependence $N_{\rm b}(v/v_{\rm cn})= \tilde\gamma[(v/v_{\rm
  cn})^3-1]$, where $\tilde\gamma$ equals $\gamma$ in Eq.~(\ref{Ndot})
divided by the neutron flux. The numerical results ($\blacksquare$) lie
higher (note that no fitting parameters are involved) and do not display a
cubic dependence in the experimental range $1< v/v_{\rm cn} <4$.  The
differences could be explained by the approximations in the expression for
$\v_{\rm sl}$, which neglect significant contributions from the inter-vortex
interactions. The polarization by the external counterflow causes the loops
with unfavourable orientation to contract and loops with the proper winding
direction and orientation to grow. However, as the polarization evolves, it
also screens the vortex tangle from the external counterflow. To check
whether the polarization has a significant effect on the results, the
calculations were repeated by taking the screening approximately into
account in the expression for $\v_{\rm sl}$:
\begin{equation}
        \v_{\rm sl} = \v_{\rm s} + \beta\s'\times\s'' +
        \frac{\kappa}{4\pi}
        \int_{\mbox{\small other}\atop\mbox{\small loops}}
             \frac{1}{V} \int d\r\,
                        \frac{(\s-\r)\times d\s}{|\s-\r|^3}.
\label{nvsl}
\end{equation}
Here the contribution from each loop to the superflow is averaged
over the volume. The results ($\blacktriangle$) show that this
effect is significant and should be taken into account more
accurately, to reproduce the network evolution correctly.

To summarize we note that the preliminary simulation work shows
that it is possible to obtain numerical agreement between the KZ
mechanism and the rotating experiments, if one assumes that the
scaling relations of the initial vortex tangle are roughly obeyed
also during its later evolution in the presence of the external
superflow. This assumption should still be checked with
sufficiently accurate simulation, with full account of the
non-local inter-vortex interactions. Future simulation of
transient networks should then answer the question how much
information about the initial state of the vortex tangle can be
retrieved from the rotating experiment, where only the final
stationary state result is measured.

\subsection{Superfluid transition as a moving phase front}
\label{TransMoveFront}

In any real laboratory system a rapid phase transition becomes
inhomogeneous: The transition will be driven by a strong gradient
in one or several of the externally controlled variables. In the
case of the superfluid transition in the aftermath of a neutron
absorption event it is a steep and rapidly relaxing thermal
gradient. In this situation even a second order transition becomes
spatially inhomogeneous and is localized into a phase front. In
the limit of very fast transitions the order-parameter relaxation
slows down the propagation of the superfluid phase and the thermal
front, where the temperature drops below $T_{\rm c}$, may escape
ahead. This means that the moving phase boundary breaks down into
a leading thermal front and a trailing superfluid interface. The
width of the region between these two zones is determined by the
relative velocities of thermal and superfluid relaxations. Thermal
fluctuations in this region are amplified in the transition
process and are carried over as order-parameter inhomogeneity into
the new symmetry-broken phase. This is the central claim of the KZ
model. Below we shall briefly discuss the influence of the thermal
gradient on defect formation, as analyzed by Kibble and Volovik (1997) and
Kopnin and Thuneberg (1999).

\subsubsection{Neutron absorption and heating}

The decay products from the neutron absorption reaction generate
ionization tracks which can be estimated from a standard
calculation of stopping power (Meyer and Sloan 1997). This leads to a
cigar-shaped volume of ionized particles, with the largest
concentration at the end points of the two tracks. The
probabilities and relaxation times of the different recombination
channels for the ionized charge are not well known in liquid
$^3$He. Also the thermalization of the reaction energy may not
produce a heated region which preserves the shape of the original
volume with the ionized charge.

Initially the recombination processes are expected to lead to
particles with large kinetic energies in the eV range, which are
well outside the thermal distribution and for which the recoil
velocities become more and more randomly oriented. Energetic
particles suffer collisions with their nearest neighbors and the
mean free path increases only slowly for atoms participating in
these collisions, until all particles are slowed down and become
thermalized to the ambient conditions (Bunkov and Timofeevskaya 1998a,b).
This means that the reaction energy remains initially localized.
In the calorimetric experiments at the lowest temperatures
(B{\"a}uerle et~al.\ 1996, 1998a) the thermal mean free path exceeds the
container dimensions. Nevertheless, the energy is probably not immediately
dispersed into the entire container volume, but remains
centralized within a bubble of limited size during cooling through
$T_{\rm c}$, when the vortex network is formed. This is the
conclusion to be drawn from the comparison between experiment and
the KZ mechanism in Table~\ref{TableBunkov}, where it is assumed
that the thermal diffusion mean free path is the same as in the
normal fluid at $T_{\rm c}$.

In an inhomogeneous initial state with large thermal gradients the
second order phase transition is turned into one where a
normal-to-superfluid phase front with finite width sweeps through
the heated bubble. If the velocity of the phase front, $v_{\rm T}
\sim R_{\rm b}/\tau_{\rm Q} \sim 6$ m/s, is sufficiently high,
comparable to a critical value $v_{\rm Tc} \sim v_{\rm F} \,
(\tau_0 / \tau_{\rm Q})^{1 \over 4}$, then the KZ mechanism is
again expected to dominate, similarly as in the homogeneous case
(Kibble and Volovik 1997).

In contrast, if the majority of the kinetic energy is assumed to
be thermalized by quasiparticles with energies comparable to the
high-energy tail of the thermal Maxwellian velocity distribution,
then the mean free paths are long, the volume heated above the
ambient becomes large, and its temperature distribution may even
become nonmonotonic like in a ``Baked Alaska'', as has been
described by Leggett (1992). The Baked Alaska scenario is also
popular in high energy physics, where the formation of the false
vacuum with a chiral condensate after a hadron-hadron collision is
considered (Bjorken 1997; Amelino-Camelia et~al.\ 1997). In both cases a rather thin
shell of radiated high energy particles expands, with the speed of
light in a relativistic system and with the Fermi velocity $v_{\rm F}$
in $^3$He, leaving behind a region at reduced temperature. Since
this interior region is separated from the exterior vacuum by the
hot shell, the cooldown into the broken symmetry state in the
center is not biased by the external order parameter state. The
Baked Alaska mechanism thus can solve the problem of the
neutron-mediated formation of B phase from supercooled bulk A
liquid (Leggett 1992), while in high energy physics it opens the
possibility for the formation of a bubble of chiral condensate in
a high energy collision (Bjorken 1997; Amelino-Camelia et~al.\ 1997).

In such conditions, when the quasiparticle mean free path exceeds
or is comparable to the dimensions of the heated bubble,
temperature is not a useful quantity for the description of its
cooling. Most of the analysis of the previous sections is
applicable only if we assume that the reaction energy remains
reasonably well localized while the hot bubble cools through
$T_{\rm c}$. In this case there is no Baked-Alaska effect: No hot
shell will be formed which would separate the interior from the
exterior region. In this situation the exterior region could be
imagined to fix the phase in the cooling bubble, while the phase
front is moving inward, suppressing the formation of new
order-parameter states, which are different from that in the bulk
superfluid outside, and in the same manner suppressing the
formation of vortices. However, it appears that there also exists
another alternative: The influence of the exterior region may not
be decisive if the phase transition front moves sufficiently
rapidly. Which of these alternatives is realized in a particular situation
is still very much a subject of discussion.


For the interpretation of the measurements in neutron irradiation,
a sophisticated understanding of the shape and size of the
constant temperature contours within the heated bubble is not
vitally necessary. In the final results the bubble size does not
enter, since the data can be normalized in terms of the measured
threshold velocity $v_{\rm cn}$.  Its measurement provides an
estimate of the circumference of the bubble, since the largest
possible vortex ring has to be comparable in size to the neutron
bubble.  The diameter of this ring is 1\,--\,2 orders of magnitude
larger than the expected average inter-vortex distance $\xi_{\rm
v}$ in the initial random network which is created by the KZ
mechanism.

\subsubsection{Thermal gradient and velocity of phase front}

For a rough understanding of the superfluid transition in a
temperature gradient let us consider the time dependent
Ginzburg-Landau (TDGL) equation for a one-component order
parameter $\Psi=\Delta/\Delta_0$:
\begin{equation}
        \tau_0{\partial \Psi\over \partial t}= \left(1-{T({\bf r},t)\over
T_{\rm c}}\right) \Psi - \Psi|\Psi|^2 + \xi_0^2 \nabla^2 \Psi ~~,
\label{TDGL}
\end{equation}
where the notations for $\tau_0 \propto 1/\Delta_0$ and $ \xi_0$
are the same as in Sec.~\ref{KW_mechanism}. This equation is the
so-called over-damped limit of the more general TDGL equation
which has a time derivative of second order. The over-damped limit
has been used in numerical simulations
(Laguna and Zurek 1997; Antunes et~al.\ 1999; Aranson et~al.\ 1999) and analytical estimations
(Dziarmaga 1998, 1999) of the density of topological defects
in a homogeneous quench. An extension of the above equation to
superconductivity was used by Ibaceta and Calzetta (1999) to study the
formation of defects after a homogeneous quench in a 2-dimensional
type II superconductor.

If the quench occurs homogeneously in the whole space ${\bf r}$, the
temperature depends only on one parameter, the quench time $\tau_{\rm Q}$:
\begin{equation}
        T(t)=  \left(1-{t\over \tau_{\rm Q}}\right)T_{\rm c}  ~~.
\label{Thomogeneous}
\end{equation}
In the presence of a temperature gradient, say, along $x$, a new
parameter appears, which together with $\tau_{\rm Q}$ characterizes
the evolution of the temperature:
\begin{equation}
        T(x-v_{\rm T}t)=\left(1-{t-x/ v_{\rm T}\over  \tau_{\rm Q}}\right) T_{\rm c}
~~.
\label{TInPropFront}
\end{equation}
Here $ v_{\rm T}$ is the velocity of the temperature front which is
related to the temperature gradient (Kibble and Volovik 1997)
\begin{equation}
        \nabla_x T={T_{\rm c} \over  v_{\rm T}  \tau_{\rm Q}}~~.
\label{TempGrad}
\end{equation}

With this new parameter $ v_{\rm T}$ we may classify transitions
to belong to one of two limiting regimes: slow or fast propagation
of the transition front. At slow velocities, $v_{\rm T}\rightarrow
0$, the order parameter moves in step with the temperature front
and
\begin{equation}
      |\Psi(x,t)|^2 = \left(1-{T(x-v_{\rm T}t)\over T_{\rm c}}\right)
\Theta (1-T(x-v_{\rm T}t)/ T_{\rm c})  ~~.
\label{PsiSlow}
\end{equation}
Here $\Theta$ is the step function. In this case phase coherence
is preserved behind the transition front and no defect formation
is possible.

The opposite limit of large velocities, $v_{\rm T}\rightarrow
\infty$, is more interesting. Here the phase transition front lags
behind the temperature front (Kopnin and Thuneberg 1999). In the space between these
two boundaries the temperature is already below $T_{\rm c}$, but
the phase transition did not yet happen, and the order parameter
has not yet formed, $\Psi=0$. This situation is unstable towards
the formation of blobs of the new phase with $\Psi\neq 0$. This
occurs independently in different regions of space, leading to
vortex formation via the KZ mechanism. At a given point ${\bf r}$
the development of the instability can be found from the
linearized TDGL equation, since during the initial growth of the
order parameter $\Psi$ from zero the cubic term can be neglected:
\begin{equation}
        \tau_0{\partial \Psi\over \partial t}=   { t \over
  \tau_{\rm Q}} \Psi  ~~.
\label{TDGLlinearized}
\end{equation}
This gives an exponentially growing order parameter, which starts from
some seed $\Psi_{\rm fluc}$, caused by fluctuations:
\begin{equation}
        \Psi({\bf r},t)=\Psi_{\rm fluc}({\bf r})\exp {t^2\over  2\tau_{\rm
Q}\tau_0 } ~~.
\label{ExponentialGrowth}
\end{equation}
Because of exponential growth, even if the seed is small, the
modulus of the order parameter reaches its equilibrium value
$|\Psi_{\rm eq}| =\sqrt{1 -T/T_{\rm c}}$ after the Zurek time $t_{\rm
Zurek}$:
\begin{equation}
       t_{\rm Zurek}=\sqrt{ \tau_{\rm
Q}\tau_0 }~~.
\label{ZurekTime}
\end{equation}
This occurs independently in different regions of space and thus
their order-parameter phases are not correlated. The spatial
correlation is lost over distances exceeding $\xi_{\rm v}$ when
the gradient term in Eq. (\ref{TDGL}) becomes comparable to the
other terms at $t=t_{\rm Zurek}$. Equating the gradient term
$\xi_0^2 \nabla^2 \Psi \sim (\xi_0^2 /\xi_{\rm v}^2) \Psi$ to,
say, the term $\tau_0 \partial \Psi/\partial t|_{t_{\rm Zurek}} =
\sqrt{\tau_0 / \tau_{\rm Q}}\Psi$, one obtains the characteristic
Zurek length which determines the initial length scale of defects,
\begin{equation}
  \xi_{\rm v} =
  \xi_0 \; (\tau_{\rm Q}/\tau_0)^{1/4}\;,
\end{equation}
which is the same as in Eq.~(\ref{eq:xi-initial}).

We can estimate the lower limit for the characteristic value of
the fluctuations $\Psi_{\rm fluc}=\Delta_{\rm fluc}/\Delta_0$,
which serve as a seed for vortex formation. If there are no other
sources of fluctuations, caused by external noise for example, the
initial seed is provided by thermal fluctuations of the order
parameter in the volume $\xi_{\rm v}^3$.  The energy of such
fluctuations is $\xi_{\rm v}^3 \Delta_{\rm fluc}^2 N_{\rm F}/E_{\rm F}$, where
$E_{\rm F} \sim k_{\rm B} T_{\rm F}$ is the Fermi energy and $N_{\rm F}$ the
fermionic density of states in normal Fermi liquid. Equating this
energy to temperature $T\approx T_{\rm c}$ one obtains the magnitude of
the thermal fluctuations
\begin{equation}
  {|\Psi_{\rm fluc}|\over |\Psi_{\rm eq}|} \sim  \left({\tau_0\over\tau_{\rm
Q}}\right)^{1/8} {k_{\rm B} T_{\rm c}\over E_{\rm F}}~.
\label{Fluctuations}
\end{equation}
The same small parameter $k_{\rm B}T_{\rm c}/E_{\rm F} \sim a/\xi_0 \sim
10^{-3}-10^{-2}$ enters here, which is responsible for the
extremely narrow temperature region of the critical fluctuations
in $^3$He near $T_{\rm c}$. But in our case it only slightly increases
the Zurek time by the factor $\sqrt{\ln [|\Psi_{\rm
eq}|/|\Psi_{\rm fluc}|]}$ and thus does not influence vortex
formation in a homogeneous quench or in an inhomogeneous quench at
large velocities of the temperature front.

Clearly there must exist a characteristic velocity $ v_{\rm T_{\rm
c}}$ of the propagating temperature front, which separates the
fast and slow regimes, or correspondingly transitions with and
without defect formation.  The criterion for defect formation is
that the Zurek time $t_{\rm Zurek}=\sqrt{ \tau_{\rm Q}\tau_0 }$
should be shorter than the time $t_{\rm sw}$ in which the phase
transition front sweeps through the space between the two
boundaries. The latter time is $t_{\rm sw}=x_0/v_{\rm T}$, where
$x_0$ is the lag -- the distance between the temperature front $T
= T_{\rm c}$ and the region where superfluid coherence starts
(order parameter front). If $t_{\rm Zurek}<t_{\rm sw}$,
instabilities have time to develop. If $t_{\rm Zurek}>t_{\rm sw}$,
both fronts move coherently and the phase is continuous. Let us
consider the latter case, with ``laminar'' motion, and find how
the lag $x_0$ depends on $v_{\rm T}$. From the equation $t_{\rm
Zurek}= x_0(v_{\rm T_{\rm c}})/v_{\rm T_{\rm c}}$ we find the
criterion for the threshold where laminar motion becomes unstable
and defect formation starts.

In steady laminar motion the order parameter depends on $x-v_{\rm
  T}t$. Introducing a dimensionless variable $z $ and a dimensionless
parameter $h$,
\begin{equation}
  z=(x- v_{\rm
T}t)(v_{\rm
T}\tau_{\rm Q}\xi_0^2)^{-1/3}~,~h=\left({v_{\rm
T}\tau_0  \over \xi_0 }\right)^{4/3}  \left({\tau_{\rm Q}\over
\tau_0}\right)^{1/3}~,
\label{za}
\end{equation}
the linearized TDGL equation becomes
\begin{equation}
        {d^2 \Psi\over dz^2} +h {d  \Psi\over dz }-z \Psi=0 ~~,
\label{TDGLlaminar}
\end{equation}
or
\begin{equation}
\Psi(z) = \mbox{const} \cdot e^{-hz/2}\chi(z) ~~,~~{d^2 \chi\over
dz^2}   -\left(z +{h^2\over 4}\right)\chi=0\;.
\label{TransformedEquation}
\end{equation}
This means that $\Psi$ is an Airy function, $\chi(z-z_0)$,
centered at $z = z_0 = -h^2/4$ and attenuated by the exponential
factor $e^{-hz/2}$.

When $h\gg 1$, it follows from Eq.~(\ref{TransformedEquation})
that $\Psi(z)$ quickly vanishes as $z$ increases above $-h^2/4$.
Thus there is a supercooled region $ -h^2/4 <z<0$, where $T<T_{\rm
c}$, but the order parameter is not yet formed: the solution is
essentially $\Psi=0$. The lag between the order parameter and
temperature fronts is $|z_0|= h^2/4$ or in conventional units
\begin{equation}
  x_0={1\over 4} { v_{\rm
T}^3 \tau_{\rm Q} \tau_0^2   \over \xi_0^2} ~~.
\label{x_0}
\end{equation}
Setting $t_{\rm Zurek}= x_0(v_{\rm T_{\rm c}})/v_{\rm
  T_{\rm c}}$ one can estimate the value for the velocity of
the temperature front where laminar propagation becomes unstable:
\begin{equation}
   v_{\rm T_{\rm c}}\sim
{\xi_0\over \tau_0} \left({\tau_0\over\tau_{\rm
Q}}\right)^{1/4}\;.
 \label{FrontCriticalVelocity}
\end{equation}
This result corresponds to $h \sim 1$ and is in agreement with that
obtained from scaling arguments by Kibble and Volovik (1997). Here an exact
numerical value cannot be offered for the threshold where laminar
motion ends, but this can in principle be done using the complete
TDGL equation (Kopnin and Thuneberg 1999). For the neutron bubble we might take
$v_{\rm T} \sim R_{\rm b}/\tau_{\rm Q}$, which gives $v_{\rm T}
\sim 10$ m/s. This value provides also an estimate for the limiting
velocity $v_{\rm T_{\rm c}}$. These considerations suggest that
the thermal gradient should be sufficiently steep in the neutron
bubble such that defect formation is to be expected.

\subsection{Quench of infinite vortex tangle}

\subsubsection{Vorticity on microscopic and macroscopic scales}

\begin{figure}[tb]
\centerline{\includegraphics[width=0.6\columnwidth]{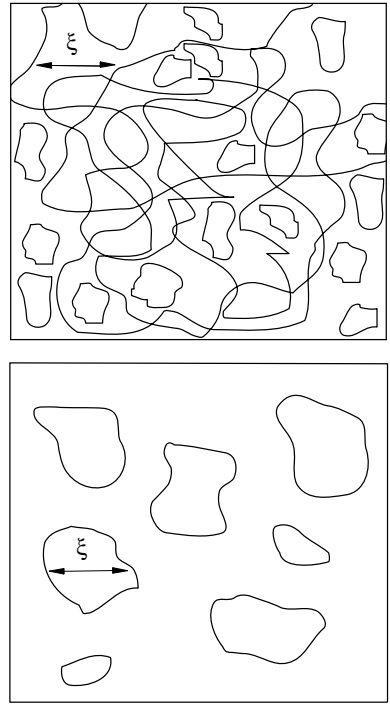}
}
  \bigskip
\caption[]{Normal-to-superfluid transition, when described as a
change-over from an infinite to a finite vortex network. {\em
(Top)} The disordered phase $(T>T_{\rm c})$ is an infinite network
of the defects in the ordered phase -- the vacuum of tangled
closed loops. The coherence length $\xi$ is the mean distance
between the elements in this network. {\em (Bottom)} In the
ordered phase $(T<T_{\rm c})$ the defects also form closed loops,
but the coherence length is now the upper limit in the
distribution of loop sizes.} \label{Infinite-finite-cluster}
\end{figure}

Onsager (1949) was the first to interpret the
$\lambda$-transition in liquid $^4$He from the superfluid to the
normal state in terms of quantized vortices: When the
concentration of the thermally activated quantized vortices
reaches the point where they form a connected tangle throughout
the liquid (and their line tension vanishes), the liquid goes
normal (Fig.~\ref{Infinite-finite-cluster}). This proliferation of
vortices to an infinite network destroys superfluidity, since the
phase slippage processes caused by the back reaction from the
tangle to the superfluid current lead to the decay of this
current.

In Ginzburg-Landau theory vortices are identified as lines of
zeroes in the scalar complex order parameter $\Psi =|\Psi |
e^{i\Phi}$, around which the phase winding number $\nu$ is
nonzero. Above the transition, in the symmetric phase, thermal
fluctuations of the order parameter give rise to an infinite
network of zeroes -- to vortices which exist on the microscopic
scale but are absent on the macroscopic scale (Kleinert 1989).
This is another way of describing the fact that there is no long
range order in the symmetric phase. The properties of such
microscopic vortices -- topologically nontrivial zeroes -- have
been followed across the thermodynamic phase transition in
numerical investigations
(Antunes and Bettencourt 1998; Antunes et~al.\ 1998; Rajantie et~al.\ 1998; Kajantie et~al.\ 1998; Rajantie 1998). The
renormalization group description of the phase transition, based
on the Ginzburg-Landau free energy functional, also contains the
microscopic quantized vortices, but in an implicit form of zeroes
(see e.g. Guillou and  Zinn-Justin 1980 and
Albert 1982).

Some attempts have been made to reformulate the 3-dimensional
phase transition in terms of this vortex picture in the
renormalization group approach
(Williams 1993a,b, 1999; Chattopadhyay et~al.\ 1993), in a manner similar
to the 2-dimensional Berezinskii-Kosterlitz-Thouless transition
(Nelson and  Kosterlitz 1977), i.e. by avoiding consideration of the
Ginzburg-Landau free energy functional altogether. It is possible,
though it has not been proven, that the Ginzburg-Landau model and
the vortex model belong to the same universality class and thus
give the same critical exponents for the heat capacity and
superfluid density. Both ground and space-based measurements of
the critical exponents (Lipa et~al.\ 2003, 1996; Goldner and Ahlers 1992)
have generally been consistent with renormalization-group
estimations
(Str\"{o}sser and  Dohm 2003; Campostrini  et~al.\ 2001; Kleinert and den  Bossche 2001).

In an equilibrium phase transition the infinite network disappears
below $T_{\rm c}$, but in a non-equilibrium phase transition the tangle
of microscopic vortices persists even into the ordered phase, due
to critical slowing down. These vortices finally transform to
conventional macroscopic vortices, when the latter become well
defined. In this language the Kibble-Zurek mechanism corresponds
to a quench of the infinite vortex network across the
non-equilibrium transition from the normal to the superfluid phase
(Yates and Zurek 1998). What is important for us here, is the scaling law
for the distribution of vortices. According to numerical
simulations (Antunes and Bettencourt 1998; Antunes et~al.\ 1998) and the phenomenological
vortex model (Williams 1993a,b, 1999; Chattopadhyay et~al.\ 1993) the
scaling exponent $\delta$, which characterizes the distribution of
vortex loops in Eq.~\ref{DL}), is close to the value
$\delta=2/(d+2)=0.4$ obtained using the concept of Flory
calculations for self-avoiding polymers. It is different from the
value given by the Vachaspati-Vilenkin model for a Brownian-walk
network. On the other hand, this model was applied to the Ginzburg
region of critical fluctuations in $^4$He, which might not be
valid for $^3$He.

\subsubsection{Scaling in equilibrium phase transitions}

Important differences exist between the phase transitions in the
$^4$He and $^3$He liquids. These also are involved in a number of
other phenomena. Let us therefore recall some of these
differences. The temperature region of the critical fluctuations,
where the simple Ginzburg-Landau theory does not work and one must
introduce the thermal renormalization of the Ginzburg-Landau
functional, can be derived from the following simplified
considerations. Let us estimate the length $L_T$ of the thermal
vortex loop, ie. a loop whose energy is comparable to temperature:
\begin{equation}
       \rho_{\rm s}(T){\kappa^2\over 4\pi} L_T \ln \left({L_T\over
\xi(T)}\right)=k_{\rm B} T \approx k_{\rm B}T_{\rm c}~.
\label{ThermalVortex}
\end{equation}
In the broken symmetry phase far below $T_{\rm c}$, this length is
less than the coherence length $\xi(T)$, which means that there
are no real vortices with the energy of order $k_{\rm B}T$, and
real vortices with higher energy are exponentially suppressed.
When the temperature increases, one approaches the point at which
the length $L_T$ becomes comparable to the coherence length $\xi$.
This is the Ginzburg temperature $T_{\rm Gi}$, determined by the
condition $L_T(T=T_{\rm Gi})\sim\xi(T=T_{\rm Gi})$:
\begin{equation}
        \rho_{\rm s}(T_{\rm Gi}) \xi(T_{\rm Gi})  \sim {k_{\rm B} T_{\rm c}\over\kappa^2}~.
\label{GinzburgT1}
\end{equation}
or
\begin{equation}
       \left(1-{T_{\rm Gi}\over T_{\rm c}}\right) \sim  \left({{k_{\rm B} T_{\rm c}}\over
\rho \xi_0\kappa^2}\right)^2 \sim \left({ T_{\rm c}\over
T_{\rm F}}\right)^4~.
\label{GinzburgT2}
\end{equation}
Here  as before  $T_{\rm F} \sim \hbar^2/ma^2\sim 1\,$K is the
degeneracy temperature of the quantum fluid, with $a$ being the
inter-atomic spacing. The region of critical fluctuations -- the
Ginzburg region $T_{\rm Gi}<T<T_{\rm c}$  -- is broad for $^4$He,
where $T_{\rm c}\sim T_{\rm F}$, and extremely small for $^3$He, where $T_{\rm c}
<10^{-2}~T_{\rm F}$.

In the region of Ginzburg fluctuations the scaling exponents for
the thermodynamic quantities, such as $\xi(T)$ and $\rho_{\rm s}(T)$,
are different from those in the Ginzburg-Landau region, $ T_{\rm c} -
T_{\rm Gi}<T_{\rm c}-T \ll T_{\rm c}$:
\begin{eqnarray}
{\xi(T) \over \xi_0} \sim  \left(1-{T\over
T_{\rm c}}\right)^{-\frac12}\hskip -3mm,~~T_{\rm c} -  T_{\rm Gi}<T_{\rm c}-T \ll T_{\rm c}~,
\label{xiBelowGi}
\\
{\xi(T) \over
\xi_0} \sim  \left(1-{T_{\rm Gi}\over T_{\rm c}}\right)^{\nu -\frac12}\left(1-
{T\over
T_{\rm c}}\right)^{-\nu}\hskip -3mm,~~
T_{\rm Gi}<T <T_{\rm c} ~,\label{xiAboveGi}
\end{eqnarray}
\begin{eqnarray}
 {\rho_{\rm s}(T) \over \rho} \sim  \left(1-{T\over
T_{\rm c}}\right),~~ T_{\rm c} -  T_{\rm Gi}<T_{\rm c}-T \ll T_{\rm c} ~,\label{rhosBelowGi}
\\ {\rho_{\rm s}(T) \over \rho} \sim
\left(1-{T_{\rm Gi}\over T_{\rm c}}\right)^{1 - \zeta}\left(1- {T\over
T_{\rm c}}\right)^{ \zeta} ,~~ T_{\rm Gi}<T <T_{\rm c}~. \label{rhosAboveGi}
\end{eqnarray}

There are two relations, the scaling hypotheses, which connect the
critical exponents for $\xi(T)$ and $\rho_{\rm s}(T)$ in the Ginzburg
region, with the exponent of the heat capacity.  In the Ginzburg
region $T_{\rm Gi}<T<T_{\rm c}$, the coherence length is determined by
thermal fluctuations, or which is the same thing, by thermal
vortices. This gives for the relation between the coherence length
$\xi(T)$ and superfluid density $\rho_{\rm s}(T)$ in the Ginzburg
region:
\begin{equation}
        \rho_{\rm s}(T) \xi(T)  \sim {{k_{\rm B} T_{\rm c}}\over\kappa^2}\sim
\rho a \left({ T_{\rm c}\over
T_{\rm F}}\right)~~,~~T_{\rm
Gi}<T<T_{\rm c}~.
\label{Relation1}
\end{equation}
This equation gives the Josephson scaling hypothesis: $\nu=\zeta$.

Another relation between these exponents comes from a
consideration of the free energy, which has the same scaling law
as the kinetic energy of superflow:
\begin{equation}
     F(T)\sim \left(1-{T
\over T_{\rm c}}\right)^{2-\alpha}\! \sim \rho_{\rm s} v_{\rm s}^2 \sim  {\rho_{\rm s}(T)
\over \xi^2(T)} \sim  {\rho \over \xi_0^2} \left(1-{T \over
T_{\rm c}}\right)^{\zeta+2\nu}~. \label{Relation2}
\end{equation}
Here $\alpha$ is the the critical exponent for the heat capacity
in the critical region, $C(T) = - T\partial_T^2 F \sim
(1-T/T_{\rm c})^{-\alpha}$.  Eqs.~(\ref{Relation1}) and
(\ref{Relation2}) give $\nu=\zeta=(2-\alpha)/3$.

\subsubsection{Non-equilibrium phase transitions}

The formation of vortices during a rapid transition into the
broken-symmetry  phase is the subject of dynamic scaling and a
poorly known area in the field of critical phenomena. Dynamic
scaling is characterized by an additional set of critical
exponents, which depend not only on the symmetry and topology of
the order parameter, but also on the interaction of the order
parameter with the different dynamic modes of the normal liquid.
The question first posed by Zurek was the following: What is the
initial density $\xi_{\rm v}$ of macroscopic vortices at the
moment when they become well defined? According to the general
scaling hypothesis one has
\begin{equation}
        \xi_{\rm v}  \sim \xi_0 (\omega
\tau_0)^{-\lambda} ~{\rm and}~ \omega={1\over \tau_{\rm Q}}~,
\label{ScalingForZurek}
\end{equation}
where $\omega$ is the characteristic frequency of the dynamic
process.

In the time-dependent Ginzburg-Landau model (Eq.~(\ref{TDGL})) and
also in its extension based on the renormalization group approach,
the exponent $\lambda$ is determined by the static exponents and
by the exponent for the relaxation time
$\tau=\tau_0(1-T/T_{\rm c})^{-\mu}$:
\begin{equation}
         \lambda ={\nu\over 1+\mu}~.
\label{lambda}
\end{equation}
This follows from the following consideration. When we approach
the critical temperature from the normal phase, at some moment
$t_{\rm Zurek}$ the relaxation time $\tau(t)$ becomes comparable
to the time $t$ which is left until the transition takes place. At
this moment
\begin{equation}
        t_{\rm Zurek} =\tau_0 \left({\tau_{\rm Q}\over \tau_0
}\right)^{\mu/(1+\mu)}~,
\label{ZurekTimeScaling}
\end{equation}
the vortex network is frozen out. After the transition it becomes
``unfrozen'' when $\tau(t)$ again becomes smaller than $t$, the
time after passing the transition. The initial distance between
the vortices is determined as the coherence length
$\xi=\xi_0(t/\tau_{\rm Q})^{-\nu}$ at $t=t_{\rm Zurek}$, which
gives Eq.~(\ref{ScalingForZurek}) with $\lambda$ as in Eq.~(\ref{lambda}).
For the conventional  time-dependent Ginzburg-Landau model in
Eq.~(\ref{TDGL}) one has $\mu=1$, $\nu=1/2$, and thus
$\lambda=1/4$. In the Ginzburg regime, if one assumes that $\mu$
remains the same, while $\nu\approx 2/3$, one obtains
$\lambda\approx 1/3$.

In numerical simulation of a quench in the time-dependent
Ginzburg-Landau model, the main problem becomes how to resolve
between the microscopic vortices (ie. the zeroes in the order
parameter) and real macroscopic vortices. This requires some
coarse-graining procedure, which is not well established. Also
speculations exist (Rivers 2000) that if the quench is limited
to within the region of the Ginzburg fluctuations, then the
network of microscopic vortices might effectively screen out the
real macroscopic vortices, so that the density of the real
vortices after the quench is essentially less than its Zurek
estimate. This could explain the negative result of the recent
pressure-quench experiments in liquid $^4$He (Dodd et~al.\ 1998),
where the final state after the decompression was well within the
region of critical fluctuations.

To summarize, we note that vortices play an important role, not only in the
ordered state, but also in the physics of the broken symmetry phase
transitions. The proliferation of vortex loops with infinite size can be
interpreted to destroy the superfluid long range order above the phase
transition. In a non-equilibrium phase transition from the symmetric normal
phase to the superfluid state, the infinite vortex cluster from the normal
state survives after the rapid quench and becomes the source for the
remanent vorticity in the superfluid state. This is an alternative picture
in which one may understand the formation of vortices in a rapid
non-equilibrium transition.

\subsection{Implications of the quench-cooled experiments} \label{SurfaceMechanism}
\subsubsection{Topological-defect formation} \label{TopDefFormSum}

It is not obvious that a phenomenological model like the KZ mechanism, which is
based on scaling arguments, should work at all: It describes a time-dependent phase
transition in terms of quantities characterizing the equilibrium properties of the
system. Numerical calculations on simple quantum systems, where one studies the
fluctuations in the amplitude of the system wave function while it is quench cooled
below a second order phase transition, have provided much evidence for the KZ model
and appear to agree with its qualitative features. Most attempts of experimental
verification suffer from shortcomings. Measurements on superfluid $^3$He are the
first to test the KZ model more quantitatively. In this case the experimental
deviations from the ideal KZ setup include the presence of a strict boundary
condition and a strong thermal gradient. At present time we can conclude that the
experiments and the model are in reasonable harmony, assuming that open questions
from Sec.~\ref{Simulation} can be answered satisfactorily, as seems likely.

However, even good agreement leaves us with an interesting question: What is the
microscopic basis for the applicability of the KZ mechanism to such experiments? A
rapid quench through the superfluid transition is more amenable to microscopic
analysis in the case of liquid $^3$He than in most other systems, since the freeze
out of order-parameter domains can be demonstrated with physically acceptable
calculations. The consequences from this are exciting and the prospects for a better
understanding of non-equilibrium phase transitions look promising. Although detailed
agreement has not yet been reached between experimental and theoretical work,
nevertheless the effort by Aranson et~al.\ (1999, 2001) illustrates that many aspects of
the neutron absorption event in $^3$He-B can be treated in realistic ways.

Are there implications from such work to cosmological large-scale structure
formation? The combined existing evidence from condensed matter experiments supports
the KZ mechanism at high transition velocities. This says that the KZ mechanism is
an important process in rapid non-equilibrium phase transitions and cannot be
neglected. Whether it was effective in the Early Universe and led to the formation
of large-scale structure, rather than some other phenomenon like inflationary
expansion, is a separate question. It can only be answered by measurements and
investigations on the cosmological scale which directly search for the evidence on
topological defect formation or its exclusion. The fact that no direct traces from
topological defects have been identified, excepting the large-scale structure
itself, is not yet sufficient proof that they can be discarded altogether. As
measurements of the moving first-order AB interface in $^3$He superfluids show, the
interactions of the defects with the phase front and across it are complicated and
no simple general rule exist on the final outcome.

\subsubsection{Phase transitions} \label{PhaseBoundaryFormSum}

As discussed in Secs.~\ref{ABtransition} and \ref{ThresholdVelocity+AB_Interfaces},
the KZ mechanism provides an attractive explanation for the radiation induced first
order transition from supercooled $^3$He-A to $^3$He-B. The nucleation of competing
phases, with different symmetries and local minima of the energy functional, has
been discussed both in superfluid $^3$He (Volovik 2003) and in the cosmological
context (Linde 1990).

Above the critical phase-transition temperature $T_{\rm c}$, liquid $^3$He is in its
symmetric phase: it  has all the symmetries allowed in non-relativistic condensed
matter. The continuous symmetries, whose breaking are relevant for the topological
classification of the defects in the nonsymmetric phases of $^3$He, form the
symmetry group
\begin{equation}
{\bf G}=SO(3)_{\bf L}\times SO(3)_{\bf S}\times U(1)_{\bf N}~. \label{SymmetryG}
\end{equation}
Here $SO(3)_{\bf L}$ is the group of solid rotations  of the coordinate space. The
spin rotations of the group $SO(3)_{\bf S}$  may be considered as a separate
symmetry operation if one neglects the dipolar spin-orbit interaction. The magnetic
dipole interaction between the nuclear spins is tiny in comparison with the energies
characterizing the superfluid transition. The group $U(1)_{\bf N}$ is the {\it
global} symmetry group of gauge transformations, which stems from the conservation
of the particle number $N$ for the $^3$He atoms in their ground states. $U(1)$ is an
exact symmetry if one neglects extremely  rare processes of excitations and
ionization of the $^3$He atoms, as well as the transformation of $^3$He nuclei, in
neutron radiation.

Below $T_{\rm c}$  the ``unified GUT symmetry'' $SO(3)_{\bf L} \times SO(3)_{\bf S}
\times U(1)_{\bf N}$ can be broken to the $U(1) \times U(1)$ symmetry of the A phase
or to the $ SO(3)$ symmetry of the B-phase, with a small energy difference between
these two states, but separated by an extremely high energy barrier ($\sim 10^6
k_{\rm B}T$) from each other. In this situation a thermally activated
A$\rightarrow$B transition becomes impossible. For the radiation-induced
A$\rightarrow$B transition two solutions have been suggested: 1) the baked Alaska
configuration where the transition takes place inside a cool bubble isolated by a
warm shell from the surrounding A-phase bath (Leggett 1984), and 2) the KZ
mechanism, first proposed by Volovik (1996) and then put on quantitative ground
by Bunkov and Timofeevskaya (1998a,b).

This situation can be compared to that in the early Universe. It is believed that
the $SU(3)\times SU(2) \times U(1)$ symmetries of the strong, weak, and
electromagnetic interactions (respectively) were united at high energies (or at high
temperatures). The underlying Grand Unification (GUT) symmetry ($SU(5)$, $SO(10)$,
or a larger group) was broken at an early stage during the cooling of the Universe.
Even the simplest GUT symmetry $SU(5)$ can be broken in different ways: into the
phase $SU(3)\times SU(2) \times U(1)$, which is our world, and into $SU(4)\times
U(1)$, which apparently corresponds to a higher energy state. In supersymmetric
models both phases represent local minima of almost equal depth, but are separated
from each other by a high energy barrier.

In the cosmological scenario, after the symmetry break of the $SU(5)$ GUT state,
both new phases are created simultaneously with domain walls between them. The
calculated probability for the creation of our world --- the $SU(3)\times SU(2)
\times U(1)$ state --- appears to be smaller than that of the false vacuum state of
$SU(4)\times U(1)$ symmetry at higher energy.  Thus initially the state
corresponding to our world appears to have occupied only a fraction of the total
volume. Later bubbles of this energetically preferred state grew at the expense of
the false vacuum state and finally completely expelled it. However, before that
interfaces between the two states were created in those places where blobs of the
two phases met. Such an interface is an additional topologically stable defect,
which is formed in the transition process and interacts with other topological
defects.

\section{Vortex dynamics and quantum field theory analogues}
\label{SecOtherAnalogs}

From the phenomenological point of view, vortex dynamics in $^3$He-B is similar to
$^4$He-II, but the properties which govern the dynamics arise from the quite
different microscopics of the p-wave Cooper-paired fermion superfluid. These
properties turn out to have interesting analogies with various problems in quantum
field theory (Volovik 2003). It is these connections which are in the focus of
the discussion in this section.

The fundamental starting point is the notion that the bosonic and fermionic
excitations in superfluid $^3$He are in many respects similar to the excitations of
the energetic physical vacuum of elementary particle physics -- the modern ether.
This similarity allows us to model, with concepts borrowed from $^3$He physics, the
interactions of elementary particles with the evolving strings and domain walls,
which are formed in a rapid phase transition, for instance. Such processes become
important after defect formation in the initial quench and give rise to the
cosmological consequences which we are measuring today.

The quantum physical vacuum -- the former empty space -- is in reality a richly
structured and asymmetric medium.  Because the new quantum ether is such a
complicated material with many degrees of freedom, one can learn to analyze it by
studying other materials, {\it ie.} condensed matter (Wilczek 1998).  Fermi
superfluids, especially $^3$He-A, are the best examples, which provide opportunities
for such modelling. The most pronounced property of $^3$He-A is that, in addition to
the numerous bosonic fields (collective modes of the order parameter which play the
part of gauge fields in electromagnetic, weak, and strong interactions) it contains
gapless fermionic quasiparticles, which are similar to the elementary excitations of
the quantum physical vacuum (leptons and quarks).

It is important that the quantum physical vacuum belongs to the same class
of fermionic condensed matter as $^3$He-A: both contain topologically
stable nodes in the energy spectrum of the fermionic excitations. As a
result both of these fermionic systems display, for example, the
gravitational and gauge fields as collective bosonic modes. Other fermionic
systems belong either to a class, which is characterized by Fermi surfaces
(such as normal metals and the normal $^3$He liquid), or to a class with a
gap in the fermionic spectrum (such as conventional superconductors and
$^3$He-B). High-temperature superconductors seem to belong to a marginal
class, with topologically unstable lines of gap nodes.
Thus $^3$He-A (together with $^3$He-A$_1$) is the condensed
matter system in which the properties of the physical vacuum can, in
principle, be probed with laboratory experiments.

\subsection{Three topological forces acting on a vortex and their
  analogues} \label{CosmForces}

Here we consider the experimentally observed forces which act on a moving vortex in
superfluid $^3$He. As listed in Sec.~\ref{EvolSimSec}, there are 3 different
topological contributions to the total force. The more familiar Magnus force arises
when the vortex moves with respect to the superfluid vacuum. In the case of a
relativistic cosmic string this force is absent, since the corresponding superfluid
density of the quantum physical vacuum is zero. However, the analog of this force
appears if the cosmic string moves in a uniform background charge density
(Davis and Shellard 1989; Lee 1994). The other two forces of topological origin -- the
Iordanskii force and spectral flow force -- also have analogues in the case of a
cosmic string
(Volovik and  Vachaspati 1996; Volovik 1998; Sonin 1997; Wexler 1997; Shelankov 1998a,b).

We start with the Iordanskii force (Iordanskii 1964, 1965; Sonin 1975), which
arises when the vortex moves with respect to the heat bath represented by the normal
component of the liquid, or the quasiparticle excitations. The latter corresponds to
the matter of particle physics.  The interaction of quasiparticles with the velocity
field of the vortex resembles the interaction of matter with the gravitational field
induced by such a cosmic string, which has an angular momentum, -- the so-called
spinning cosmic string (Mazur 1986). The spinning string induces a
peculiar space-time metric, which leads to a difference in the time which a particle
needs to orbit around the string at same speed, but in opposite directions
(Harari and Polychronakos 1988). This gives rise to the quantum gravitational Aharonov-Bohm effect
(Mazur 1986, 1987, 1996). We discuss how the same
effect leads to an asymmetry in the scattering of particles from the spinning string
and to the Iordanskii lifting force which acts on the spinning string or on the
moving vortex.

The spectral flow force, which also arises when the vortex moves with respect to the
heat bath, is a direct consequence from the chiral anomaly effect. The latter
violates the conservation of fermionic charge.  The anomalous generation by the
moving vortex of fermionic charge or momentum (called ``momentogenesis'') leads to a
net force acting on the vortex. The existence of this force was experimentally
confirmed in the $^3$He measurements of Bevan et~al.\ (1997b). This phenomenon is
based on the same physics as the anomalous generation of matter in particle physics
and bears directly on the cosmological problem of baryonic asymmetry of our
Universe: the question why the Universe contains so much more matter than antimatter
(``baryogenesis'').

The experimental observation of the opposite effect to momentogenesis has been
reported by Krusius et~al.\ (1998): the conversion of quasiparticle momentum into
a non-trivial order parameter configuration or ``texture''.  The corresponding
process in a cosmological setting would be the creation of a primordial magnetic
field due to changes in the matter content.  Processes, in which magnetic fields are
generated, are very relevant to cosmology since magnetic fields are ubiquitous now
in the Universe. Our Milky Way and other galaxies, as well as clusters of galaxies,
are observed to have a magnetic field whose generation is still not understood. One
possible mechanism is that a seed field was amplified by the complex motions
associated with galaxies and clusters of galaxies. The seed field itself is usually
assumed to be of cosmological origin.

It has been noted that the two problems of cosmological genesis -- baryo- and
magnetogenesis -- may be related to each other (Roberge 1989; Vachaspati and Field 1994, 1995; Vachaspati 1994).
More recently an even stronger reason for a possible connection was proposed
(Joyce and  Shaposhnikov 1997; Giovannini and  Shaposhnikov 1997). In this same way their analogs are
related in $^3$He, where the order parameter texture is the analog of magnetic
field, while the normal component of the superfluid represents the matter: the
moving vortex texture leads to an anomalous production of quasiparticles, while an
excess of the quasiparticle momentum -- the net quasiparticle current of the normal
component -- leads to the formation of textures. This mapping of cosmology to
condensed matter is not simply a picture: the corresponding effects in the two
systems are described by the same equations in the low-energy regime, by quantum
field theory and the axial anomaly.

\subsection{Iordanskii force} \label{IordanskiiForceSec}

\subsubsection{Superfluid vortex vs spinning cosmic string} \label{RelatHe}

To clarify the analogy between the Iordanskii force and the Aharonov-Bohm effect,
let us consider the simplest case of phonons propagating in the velocity field of
the quantized vortex in the Bose superfluid $^4$He-II. According to the Landau
theory of superfluidity, the energy of a quasiparticle moving in the superfluid
velocity field ${\bf v}_{\rm s}({\bf r})$ is Doppler shifted: $E({\bf p})= \epsilon({\bf
p})+ {\bf p}\cdot{\bf
  v}_{\rm s}({\bf r})$. In the case of the phonons with the spectrum
$\epsilon({\bf p})=cp$, where $c$ is the sound velocity, the
energy-momentum relation is thus
\begin{equation}
\left(E- {\bf p}\cdot{\bf v}_{\rm s}({\bf r})\right)^2=c^2p^2
 ~.
\label{PhononSpectrum}
\end{equation}
Eq.~(\ref{PhononSpectrum}) can be written in the general Lorentzian form with
$p_{\mu}=(-E,{\bf p})$:
\begin{equation}
g^{\mu\nu}p_\mu p_\nu=0 ~~
\label{LoretzianSpectrumPhonons}
\end{equation}
where the metric is
\begin{equation}
g^{00}=1,~~~~g^{0i}=-v_{s}^i,~~
g^{ik}=- c^2 \delta^{ik} +v_{\rm s}^iv_{\rm s}^k ~~.
\label{gikPhonons}
\end{equation}
We use the convention to denote indices in the 0\,--\,3 range by Greek
letters and indices in the 1\,--\,3 range by Latin letters.
Thus the dynamics of phonons in the presence of the velocity field is
the same as the dynamics of photons in the gravity field
(Unruh 1976): both are described by the light-cone equation $ds=0$. The
interval $ds$ for phonons is given by the inverse metric
$g_{\mu\nu}$ which determines the geometry of the effective space:
\begin{equation}
 ds^2=g_{\mu\nu}dx^\mu dx^\nu  ~~,
\label{IntervalGeneral}
\end{equation}
where $x = (t,{\bf r})$ are physical (Galilean) coordinates in the laboratory
frame.

A similar relativistic equation holds for the fermionic quasiparticles in superfluid
$^3$He-A in the linear approximation close to the gap nodes. In general, i.e. far
from the gap nodes, the spectrum of quasiparticle in $^3$He-A is not relativistic:
\begin{equation}
  \epsilon^2({\bf p}) = v_{\rm F}^2 (p-p_{\rm F})^2 +
  \frac{\Delta_{\rm A}^2}{p_{\rm F}^2} (\hat{{\bf l}} \times {\bf p})^2.
  \label{epsA}
\end{equation}
Here $v_{\rm F}(p-p_{\rm F})$ is the quasiparticle energy in the normal Fermi liquid state above
the transition, with $p_{\rm F}$ the Fermi momentum and $v_{\rm F}=p_{\rm F}/m^*$; $m^*$ is the
effective mass, which is of order of the mass $m_3$ of the $^3$He atom; $\Delta_{\rm A}$
is the so-called gap amplitude and the unit vector $\hat{{\bf l}}$ points in the
direction of the gap nodes.

The energy in Eq.~(\ref{epsA}) is zero at two points ${\bf p} = e {\bf A}$ with
${\bf A} = p_{\rm F}\hat{{\bf l}}$ and $e=\pm 1$. Close to the two zeroes of the energy
spectrum one can expand the equation $\left(E- {\bf p}\cdot{\bf v}_{\rm s}({\bf
r})\right)^2=\epsilon^2({\bf p})$ in ${\bf p} - e {\bf A}$ and write it in a form
similar to the propagation equation for a massless relativistic particle in curved
spacetime in the presence of an electromagnetic vector potential:
\begin{equation}
g^{\mu\nu}(p_\mu -e A_\mu)(p_\nu-eA_\nu)=0\,.
\label{LoretzianSpectrumAphase}
\end{equation}
Here $A_0 = p_{\rm F} (\hat{{\bf l}}\cdot{\bf v}_{\rm s})$ and
the metric is anisotropic with the anisotropy axis along
the ${\hat{\bf l}}$-vector:
\begin{equation}
  \begin{array}{c}
g^{00}=1,\quad g^{0i}=-v_{s}^i,\\[1.2ex]
g^{ik}=- c_\perp^2 (\delta^{ik}- \hat l^i\hat l^k)  - c_\parallel^2
\hat l^i\hat l^k+v_{\rm s}^iv_{\rm s}^k,\\[1.2ex]
c_\parallel = v_{\rm F}, \quad c_\perp = \Delta_{\rm A}/v_{\rm F}.
\end{array}
\label{gikAphase}
\end{equation}
The quantities $c_\parallel$ and $c_\perp$ correspond to the speeds of light
propagating along or transverse to $\hat{{\bf l}}$.

For simplicity, let us turn back to the case of phonons and vortices in $^4$He-II
which is described by Eqs.~(\ref{LoretzianSpectrumPhonons}), (\ref{gikPhonons}). If
the velocity field is generated by one vortex with $\nu$ quanta of circulation,
${\bf v}_{\rm s}=\nu\kappa\hat{\mbox{\boldmath$\phi$}}/2\pi r$, then the interval
(\ref{IntervalGeneral}) in the effective space, where the phonon is propagating
along geodesic curves, becomes:
\begin{eqnarray}
  ds^2=\left(1-{v_{\rm s}^2\over c^2}  \right)\left(dt +{\nu\kappa\,
d\phi\over 2\pi( c^2-v_{\rm s}^2)} \right)^2 \hspace{5em}\nonumber\\
\hfill -{dr^2\over c^2}-{dz^2\over c^2}
-{ r^2d\phi^2\over c^2-v_{\rm s}^2}
 ~.
\label{PhononInterval}
\end{eqnarray}

The origin of the Iordanskii force lies in the scattering of
quasiparticles for small angles, so large distances from the vortex
core are important.
Far from the vortex $v_{\rm s}^2/c^2$ is small and can be neglected,
and one has
\begin{equation}
ds^2=\left( dt + {d\phi\over \omega}\right)^2 -{1\over c^2}(dz^2+ dr^2
+r^2d\phi^2),~~ \omega={2\pi  c^2 \over  \nu \kappa}\,,
 \label{IntervalVortexAsymp}
\end{equation}
The connection between time and the azimuthal angle $\phi$ in the interval suggests
that there is a characteristic angular velocity $\omega$.  A similar metric with
rotation was obtained for the so-called spinning cosmic string in $3+1$ space-time,
which has the rotational angular momentum $J$ concentrated in the string core, and
for the spinning particle in the 2+1 gravity
(Mazur 1986, 1996; Staruszkievicz 1963; Deser et~al.\ 1984):
\begin{equation}
ds^2=\left( dt + {d\phi\over \omega}\right)^2 -{1\over c^2}(dz^2+ dr^2
+r^2d\phi^2),~~\omega={1\over 4JG}
\label{IntervalSpinningStringAsymp}
\end{equation}
where $G$ is the gravitational constant.  This gives the following
correspondence between the circulation $\nu\kappa$ around the vortex and
the angular momentum $J$ of the spinning string
\begin{equation}
\kappa \nu = 8\pi JG~.
\label{JvsN}
\end{equation}
Although we here consider the analogy between the spinning string and vortices in
$^4$He-II, there exists a general statement that vortices in any superfluid have the
properties of spinning cosmic strings (Davis and Shellard 1989). In particular, the
spinning string generates a density of angular momentum in the vacuum outside the
string (Jensen and Kuvcera 1993). The density of angular momentum in the superfluid vacuum
outside the vortex is also nonzero and equals at $T=0$
\begin{equation}
  {\bf r} \times \rho  {\bf v}_{\rm s} = \hbar \nu n_{\rm B} \hat{{\bf z}},
\end{equation}
where $n_{\rm B}$ is the density of elementary bosons in the superfluid vacuum: the
density $\rho/m_4$ of $^4$He atoms in superfluid $^4$He-II or the density
$\rho/2m_3$ of Cooper pairs in superfluid $^3$He.

\subsubsection{Gravitational Aharonov-Bohm effect}

For the spinning string the gravitational Aharonov-Bohm effect is a
peculiar topological phenomenon (Mazur 1986) which can be
modeled in condensed matter. On the classical level the propagation of
particles is described by the relativistic equation $ds^2=0$.  Outside the
string the space metric, which enters the interval $ds$, is flat,
Eq.~(\ref{IntervalSpinningStringAsymp}). But there is a difference in the
travel time for particles along closed paths around the spinning string in
opposite directions. As can be seen from
Eq.~(\ref{IntervalSpinningStringAsymp}), this time difference is
(Harari and Polychronakos 1988)
\begin{equation}
2\tau={4\pi \over \omega}~~.
\label{TimeDelay}
\end{equation}
At large distances form the core the same equation is
approximately valid due to the equivalence of the metrics in
Eqs.~(\ref{IntervalVortexAsymp}) and
(\ref{IntervalSpinningStringAsymp}).  The asymmetry between the
particles orbiting in different directions around the vortex implies that,
in addition to the symmetric part of the cross section,
\begin{equation}
  \sigma_\parallel = \int_0^{2\pi} d\theta\, (1-\cos\theta)\, |a(\theta)|^2,
\end{equation}
where $a(\theta)$ is a scattering amplitude, there should be an asymmetric part
of the scattering cross section,
\begin{equation}
\sigma_\perp =\int_0^{2\pi} d\theta ~\sin\theta ~|a(\theta)|^2.
\label{sigmaPerpGeneral}
\end{equation}
The latter is the origin of the Iordanskii
force acting on the vortex in the presence of a net momentum from the
quasiparticles. Another consequence of
Eqs.~(\ref{IntervalVortexAsymp}), (\ref{IntervalSpinningStringAsymp})
is displayed on the quantum level: the
connection between the time variable $t$ and the angular variable $\phi$ in
Eqs.~(\ref{IntervalVortexAsymp}), (\ref{IntervalSpinningStringAsymp}) implies
that the scattering cross sections of phonons (photons) from the vortex (string)
should be periodic functions of energy, with the period equal to $\hbar
\omega$.

Calculations which allow us to find both symmetric and
asymmetric contributions to the scattering of quasiparticles in
the velocity field of the vortex have been performed by Sonin (1997) for phonons
and rotons in $^4$He-II  and by Cleary (1968) for
the Bogoliubov-Nambu quasiparticles in conventional superconductors.
In the case of phonons the propagation is described by the Lorentzian
equation for the scalar field $\Phi$:
$g^{\mu\nu}\partial_\mu\partial_\nu ~\Phi=0$, with $g^{\mu\nu}$ from
Eq.~(\ref{gikPhonons}). We are interested in large distances from the
vortex core. Thus the quadratic terms ${\bf v}_{\rm s}^2/c^2$ can be neglected
and the equation can be rewritten as (Sonin 1997)
\begin{equation}
E^2\Phi -c^2\left(-i\nabla + {E\over c}{\bf v}_{\rm s}({\bf r})\right)^2\Phi=0 ~~.
\label{ModifiedScalarField}
\end{equation}
This equation maps the problem under discussion to the Aharonov-Bohm
(AB) problem for the magnetic flux tube (Aharonov and Bohm 1959) with the effective vector
potential ${\bf A}={\bf v}_{\rm s}$, where the electric charge $e$ is
substituted by the mass $E/c^2$ of the particle
(Mazur 1987; Jensen and Kuvcera 1993; Gal'tsov and Letelier 1993). Actually ${\bf v}_{\rm s}$ plays the
part of the vector potential of the so called gravimagnetic field (Volovik 2003). Because of the mapping between the electric charge
and the mass of the particle, one obtains the AB expression
(Aharonov and Bohm 1959) for the symmetric differential cross section
\begin{equation}
{d \sigma_\parallel\over d\theta} = {\hbar c\over 2\pi E \sin^2(\theta/2)}~
\sin^2   {\pi E\over
\hbar \omega}~.
 \label{DiffCrossSectionString}
\end{equation}
This expression was obtained for the scattering of particles with energy $E$
in the background of a spinning string with zero
mass (Mazur 1987, 1996) and represented the gravitational
AB effect. Note the singularity at $\theta \rightarrow 0$.

Returning back to vortices, one finds that the analogue with the spinning string
is not exact. In more accurate
calculations one should take into account that as distinct from the
charged particles in the AB effect, the current in the case of phonons
is not gauge invariant. As a result the scattering of the phonon with
momentum $p$ and energy $E$ from the vortex is somewhat
different (Sonin 1997):
\begin{equation}
{d \sigma_\parallel\over d\theta} = {\hbar c\over 2\pi E }\cot^2{\theta\over 2}
~  \sin^2   {\pi E\over
\hbar \omega}~.
 \label{DiffCrossSectionVortex}
\end{equation}
The algebraic difference between the AB results in
Eqs.~(\ref{DiffCrossSectionString}) and (\ref{DiffCrossSectionVortex}) is $(\hbar c/ 2\pi E ) ~ \sin^2 (\pi E/ \omega)$, which is independent of the scattering angle $\theta$ and
thus is not important for the singularity at small scattering angles,
which is present in Eq.~(\ref{DiffCrossSectionVortex}) as well.  For small
$E$ the result in Eq.~(\ref{DiffCrossSectionVortex}) was obtained by
Fetter (1964). The generalization of the Fetter result for
quasiparticles with arbitrary spectrum $\epsilon({\bf p})$ (rotons in
$^4$He-II and the Bogoliubov-Nambu fermions in superconductors) was
recently suggested by Demircan et~al.\ (1995): In our notations it is
$(\nu\kappa^2 p/8\pi v_G^2) \cot^2(\theta/ 2)$, where $v_G=d\epsilon/dp$
is the group velocity of a quasiparticle.

\subsubsection{ Asymmetric cross section of scattering from a vortex}

The Lorentz force, which acts on the flux tube in the presence of an electric
current, has its counterpart -- the Iordanskii force, which acts on the
vortex in the presence of a mass current of the normal component.
The Lorentz-type Iordanskii force comes from the asymmetric contribution to
the cross section (Sonin 1997; Shelankov 1998a,b), which has the same
origin as the singularity at small angles in the symmetric cross section
and leads to a non-zero transverse cross section.

For the phonons in $^4$He-II with the spectrum $E({\bf p})=cp$ the transverse
cross section is (Sonin 1997)
\begin{equation}
\sigma_\perp ={\hbar \over p} ~ \sin  {2\pi
 E\over \hbar \omega}\;.
\label{sigmaPerpVortex}
\end{equation}
At low $E \ll \hbar \omega$ the result becomes classical: $\sigma_\perp = 2\pi
c/\omega$ does not contain the Planck constant $\hbar$. This means that in the
low energy limit the asymmetric cross section can be obtained from the
classical theory of scattering. In this case it can be  generalized for an arbitrary
spectrum $E({\bf p})$ of scattering particles (Sonin 1997).  Let us consider
a particle with the spectrum  $E({\bf p})$ moving in the background of the
velocity field ${\bf v}_{\rm s}({\bf r})$ around the vortex. The velocity field
modifies the particle energy due to the Doppler shift, $E({\bf p},{\bf
r})=E({\bf p}) + {\bf p}\cdot{\bf v}_{\rm s}({\bf r})$.  Far from the vortex, where
the circulating velocity is small, the trajectory of the quasiparticle is
almost a straight line parallel to, say, the axis $y$, with the distance from the
vortex line being the impact parameter $x$. It moves along this line with
almost constant momentum $p_y\approx p$ and almost constant group velocity
$dy/dt=v_G=d\epsilon/dp$. The change in transverse momentum during this
motion is determined by the Hamiltonian equation $dp_x/dt=-\partial
E/\partial x=-p_y \partial v_{sy}/\partial x$, or
$dp_x/dy=-(p/v_G)\partial v_{sy}/\partial x$. The transverse cross
section is obtained by integration of $\Delta p_x/p$ over the impact
parameter $x$:
\begin{equation}
\sigma_\perp = \int_{-\infty}^{+\infty}{dx\over
v_G}\int_{-\infty}^{+\infty}dy {\partial v_{sy}\over \partial x}
={
\nu\kappa\over v_G}~.
\label{sigmaPerpLinear}
\end{equation}
This result is purely classical: Planckús constant $\hbar$
drops out (it enters only via the quantized circulation $\nu\kappa$ which characterizes the
vortex).

\subsubsection{Iordanskii force: quantized vortex and spinning string}

The asymmetric part of scattering, which describes the momentum
transfer in the transverse direction, after integration over the
distribution of excitations gives rise to the transverse force acting
on the vortex if the vortex moves with respect to the normal
component. This is the Iordanskii force:
\begin{eqnarray}
\nonumber
{\bf f}_{\rm Iordanskii}=\int {d^3p\over (2\pi)^3}\sigma_\perp(p)
v_G n({\bf p})
 {\bf p}\times {\hat{\bf z}} =\\
=- \nu\kappa  {\hat{\bf z}}\times  \int
{d^3p\over (2\pi)^3}  n({\bf p})
 {\bf p}
= \nu\kappa  {\bf P}_{\rm n}\times {\hat{\bf z}}\,.
\label{IordanskiiForce}
\end{eqnarray}
It is proportional to the density of mass current ${\bf P}_{\rm n}$ carried by
excitations (matter). Of all the parameters describing the vortex, it depends only on the
circulation $\nu\kappa$. This confirms the topological origin of this force. In the
case of the equilibrium distribution of quasiparticles one has ${\bf
  P}_{\rm n}=\rho_{\rm n}{\bf v}_{\rm n}$, where $\rho_{\rm n}$ and ${\bf v}_{\rm n}$ are the
density and velocity of the normal component of the liquid. Thus one
obtains Eq.~(\ref{IordanskiiForceGeneral}). To avoid the
conventional Magnus force in this derivation, we assumed that the
asymptotic velocity of the superfluid component of the liquid is zero
in the frame of the vortex.

Eq.~(\ref{IordanskiiForce}) was obtained using the asymptotic
behavior of the flow field ${\bf v}_{\rm s}$, which induces the same
effective metric (\ref{IntervalVortexAsymp}) as the metric around the
spinning string (\ref{IntervalSpinningStringAsymp}). We can thus
apply this result directly to the spinning string.  The asymmetric
scattering cross-section of relativistic particles from the
spinning string is given by Eq.~(\ref{sigmaPerpVortex}). This means
that in the presence of momentum from matter the spinning cosmic
string experiences a lifting force, which corresponds to
the Iordanskii force in superfluids.  This force can be obtained from a
relativistic generalization of Eq.~(\ref{IordanskiiForce}).  The
momentum density ${\bf P}_{\rm n}$ of quasiparticles should be substituted
by the component $T_0^i$ of the energy-momentum tensor. As a result,
for 2+1 space-time and for low energy $E$, which corresponds to
low temperatures $T$ for matter, the Iordanskii force on a spinning
string moving with respect to the matter is
\begin{equation}
f_{\rm Iordanskii}^{\alpha}=8\pi JG  \varepsilon^{\alpha\beta\gamma}
u_\beta u_\mu
T^\mu_\gamma~.
\label{RelativisticIordanskiiForce}
\end{equation}
Here $u_\alpha$ is the 3-velocity of the string and $T^\mu_\gamma$ is
the asymptotic value of the energy-momentum tensor of the matter at
the site of the string. Using the Einstein equations one can rewrite
this as
\begin{equation}
f_{\rm Iordanskii}^{\alpha}=  J   \varepsilon^{\alpha\beta\gamma} u_\beta u_\mu
R^\mu_\gamma~,
\label{CurvatureIordanskiiForce}
\end{equation}
where $R^\mu_\gamma$ is the Riemannian curvature at the location of
the string.  This corresponds to the force which acts on a particle with
spin $J$ from the gravitational field owing to the interaction of the spin
with the Riemann tensor (Thorne and Hartle 1985; Mino et~al.\ 1997).

Note that previously we have shown how ideas first developed in
the cosmological context (such as nucleation of bubbles of different
broken symmetry phases during a rapid phase transition) could be applied to
superfluid helium. Here we have the opposite case when ideas
originating in helium physics have found application in other systems
described by quantum field theory.


\begin{figure}[tb]
  \centerline{\includegraphics[width=0.9\columnwidth]{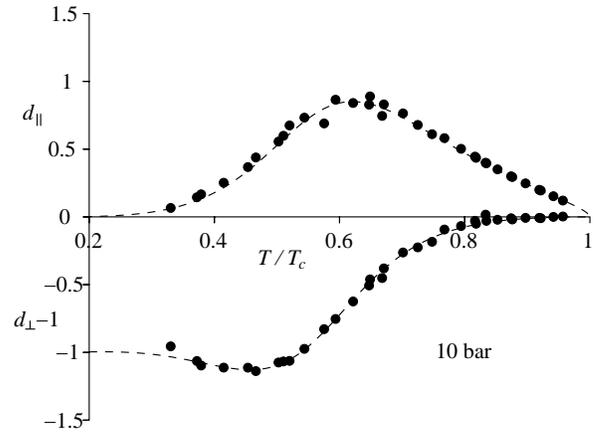}}
  \bigskip
  \caption{Mutual friction parameters $d_{\|,\perp}$ as a function of
  temperature in $^3$He-B at
    10 bar. The negative sign of $d_{\perp}-1$ at $T/T_{\rm c}\approx 0.4$--$0.5$
    constitutes experimental verification for the existence of the Iordanskii
    force. The curves are fits to measured data. (From
    Bevan et~al.\ 1997a).}
  \label{dperp-he3b-lowT}
\end{figure}


The Iordanskii force has been experimentally identified in rotating
$^3$He-B (Bevan et~al.\ 1997a). Eq.~(\ref{dperp}) describes the temperature
dependence of the mutual friction coefficient $d_\perp$. Here the term $m_3
{\cal C}(T)$ arises from the spectral flow force which is discussed in
Sec.~\ref{SpectralFlowSingular} and vanishes at low temperatures. Thus at
low temperatures $d_\perp \approx - \rho _{\rm n}/\rho$
(Kopnin et~al.\ 1995), where $\rho$ is the total density of the liquid.
The negative value of $d_\perp$ arises entirely from the existence of the
Iordanskii force. This is in accordance with the experimental data, which
show that $d_\perp$ does have negative values at low $T$
(Fig.~\ref{dperp-he3b-lowT}). At higher $T$ the spectral flow force
dominates, which leads to the sign reversal of $d_\perp$. This fact can be
interpreted to represent experimental verification of the analog of the
gravitational Aharonov-Bohm effect for a spinning cosmic string.

\subsection{Spectral flow force and chiral anomaly}

\subsubsection{Chiral anomaly} \label{ChiralAnomaly}

In the standard model of electroweak interactions there are certain quantities, like
the baryon number $Q_{\rm B}$, which are classically conserved but can be violated by
quantum mechanical effects known generically as ``chiral anomalies''. (Each of the
quarks is assigned $Q_{\rm B}=1/3$ while the leptons (neutrinos and the electron) have
$Q_{\rm B}=0$.) The process leading to particle creation is called ``spectral flow'', and
can be pictured as a process in which fermions flow under an external perturbation
from negative energy levels towards positive energy levels.  Some fermions therefore
cross zero energy and move from the Dirac sea into the observable positive energy
world.

\begin{figure}[!!!!tb]
  \centerline{\includegraphics[width=0.9\columnwidth]{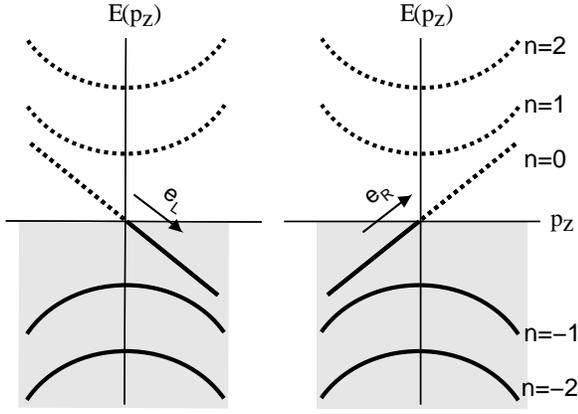}}
  \bigskip
  \caption{The energy spectrum of a chiral particle in constant
    magnetic fields along $\hat z$ (Landau levels). The plots on left
    and right show spectra for left and right particles,
    respectively.}
  \label{ChiralAnomalyFig}
\end{figure}

The origin for the axial anomaly can be seen from the behavior of a chiral particle
in constant magnetic field, ${\bf A}=(1/2){\bf
  B}\times {\bf r}$. A chiral particle we call a particle without mass,
but with spin $\vec\sigma$ [=1/2]. It can be classified as a right or left particle,
depending on whether its spin is parallel or antiparallel to its momentum. The
Hamiltonians for the right particle with the electric charge $e_{\rm R}$ and for the left
particle with the electric charge $e_{\rm L}$ are
\begin{equation}
{\cal H}= c\vec\sigma\cdot({\bf p}- e_{\rm R}{\bf A}),\quad {\cal H}=
-c\vec\sigma\cdot({\bf
p}- e_{\rm L}{\bf A}) ~.
\label{WeylForLeftRight}
\end{equation}
As usual, the motion of the particles in the plane perpendicular to ${\bf B}\| \hat
{\bf z}$ is quantized in Landau levels. Thus the free motion is effectively reduced
to one-dimensional motion along ${\bf B}$ with momentum $p_z$.
Fig.~\ref{ChiralAnomalyFig} shows the energy spectrum where the thick lines
represent the occupied negative-energy states.  The peculiar feature of the spectrum
is that, because of the chirality of the particles, the lowest ($n=0$) Landau level
is asymmetric. It crosses zero only in one direction: $E=cp_z$ for the right
particle and $E=-cp_z$ for the left. If we now apply an electric field ${\bf E}$
along $z$, particles are pushed from negative to positive energy levels according to
the equation of motion $\dot p_z =e_{R} E_z$ ($\dot p_z =e_{L} E_z$) and the whole
Dirac sea moves up (down) creating particles and electric charge from the vacuum.
This motion of particles along the ``anomalous'' branch of the spectrum is called
{\em spectral flow}. The rate of particle production is proportional to the density
of states at the Landau level, which is
\begin{equation}
N_{\rm R}(0)= {\vert e_{\rm R}{\bf B}\vert \over 4\pi^2}  ~,~N_{\rm L}(0)= {\vert e_{\rm L}{\bf B}\vert
\over 4\pi^2}\;.
\label{DOSMagneticField}
\end{equation}
The production rate of particle number $n=n_{\rm R}+n_{\rm L}$ and of charge $Q=n_{\rm R}e_{\rm R}+n_{\rm L}e_{\rm L}$
from vacuum is
\begin{equation}
\dot{n} ={1\over {4\pi^2}} (e_{\rm R}^2-e_{\rm L}^2){\bf E} \cdot {\bf
B},\quad
\dot{Q}={1\over {4\pi^2}} (e_{\rm R}^3-e_{\rm L}^3){\bf E} \cdot
{\bf B}  ~.
\label{ChargeParticlProduction}
\end{equation}


This is an anomaly equation for the production of particles from vacuum of the type
found by Adler (1969) and by Bell and Jackiw (1969) in the context of neutral
pion decay. We see that for particle or charge creation, without creation of
corresponding antiparticles, it is necessary to have an asymmetric branch in the
dispersion relations $E(p)$, which crosses the axis from negative to positive
energy. Additionally, the symmetry between the left and right particles has to be
violated: $e_{\rm R}\neq e_{\rm L}$ for charge creation and $e_{\rm R}^2\neq e_{\rm L}^2$ for particle
creation.

In the electroweak model there are two gauge fields whose ``electric'' and
``magnetic'' fields may become a source for baryoproduction: The hypercharge field
$U(1)$ and the weak field $SU(2)$. Anomalous zero mode branches exist in the core of
a Z-string, where quarks, electrons and neutrinos are all chiral particles with
known hypercharge and weak charges. If we consider a process in which one electron,
two $u$-quarks and one $d$-quark are created, then lepton and baryon numbers are
changed by one unit while electric charge is conserved~(Bevan et~al.\ 1997b). If we sum
appropriate charges for all particles according to
Eq.~(\ref{ChargeParticlProduction}) the rate of this process is
\begin{equation}
\dot n_{\rm bar} = \dot n_{\rm lept}
= {{N_{\rm F}} \over {8 \pi^2}} \left (
-  {\bf B}^a_W\cdot {\bf E}_{aW} +
   {\bf B}_Y\cdot {\bf E}_{Y} \right)
\label{2}
\end{equation}
where $N_{\rm F}=3$ is the number of families (generations) of fermions,
${\bf B}^a_W$ and ${\bf E}_{aW}$ are the colored $SU(2)$ magnetic and
electric fields, while ${\bf B}_Y$ and ${\bf E}_{Y}$ are the magnetic
and electric fields of the $U(1)$ hypercharge. While a color and
hypercharge magnetic flux is always present in the Z-string core, a
color and hypercharge electric field can also be present along the
string if the string is moving across a background electromagnetic
field (Witten 1985) or in certain other processes such as the
decoupling of two linked loops (Vachaspati and Field 1994, 1995; Garriga and Vachaspati 1995).  Thus parallel
electric and magnetic fields in the string change the baryonic charge
and can lead to cosmological baryogenesis (Barriola 1995) and to the
presence of antimatter in cosmic rays (Starkman 1996).

In superconductors and in superfluid $^3$He an anomalous zero mode branch exists for
fermions in the core of quantized vortices. For electrons in superconductors it was
first found by Caroli et~al.\ (1964), and for vortices in superfluid $^3$He by
Kopnin (1993). One of the physically important fermionic charges in
$^3$He-A, $^3$He-B and superconductors which, like baryonic charge in the standard
model, is not conserved due to the anomaly, is linear momentum. The spectral flow of
momentum along the zero mode branch leads to an additional ``lift'' force which acts
on a moving vortex.


\begin{figure}
  \centerline{\includegraphics[width=0.9\columnwidth]{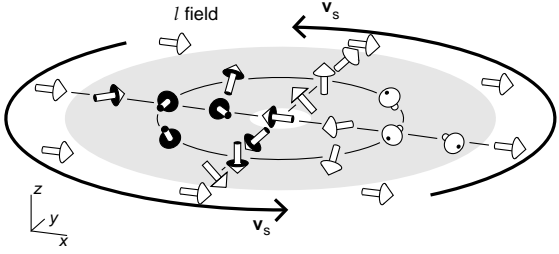}}
  \bigskip
  \caption{The order
    parameter ${\hat{\bf l}}$-texture in the soft core of the
    continuous Anderson-Toulouse-Chechetkin vortex.}
  \label{ContVortexFig}
\end{figure}


The analogy is clearest for the continuous vortex in $^3$He-A, which has two quanta
of superfluid circulation, $\nu=2$ (Blaauwgeers  et~al.\ 2000).  This vortex is
similar to the continuous Z-vortex in electroweak theory: it is characterized by a
continuous distribution of the order parameter vector ${\hat{\bf l}}$, which denotes
the direction of the angular momentum of the Cooper pairs
(Fig.~\ref{ContVortexFig}). When multiplied by the Fermi wave number
$k_{\rm F}=p_{\rm F}/\hbar$, this vector acts on the quasiparticles like an effective
``electromagnetic'' vector potential ${\bf A}=k_{\rm F}{\hat {\bf l}}$. The quasiparticles
in $^3$He-A, which are close to the gap nodes, are chiral: they are either left or
right handed (Volovik and  Vachaspati 1996). As follows from the BCS theory of $^3$He-A
the sign of the ``electric'' charge $e$, introduced in Sec.~\ref{RelatHe},
simultaneously determines the chirality of the fermions. This is clearly seen from a
simple isotropic example (with $c_\parallel=c_\perp=c$):
\begin{equation}
{\cal H}=-e c\vec\sigma\cdot({\bf p}- e{\bf A}) ~.
\label{WeylIsotropic}
\end{equation}
A particle with positive (negative) $e$ at the north (south) pole is left-handed
(right-handed). Here $\vec\sigma$ is a Bogoliubov spin.

For such gapless chiral fermions the
Adler-Bell-Jackiw anomaly applies and the momentum (``chiral charge'') of
quasiparticles is not conserved in the presence of ``electric'' and
``magnetic'' fields, which are defined by
\begin{equation}
{\bf E}=k_{\rm F} \partial_t {\hat {\bf l}},\quad
{\bf B}=k_{\rm F}{\bf \nabla}\times {\hat {\bf l}}.
\label{HeEB}
\end{equation}
In $^3$He-A each right-handed quasiparticle carries the momentum ${\bf p}_{\rm R} =
p_{\rm F}{\hat {\bf l}}$ (we reverse the sign of momentum when it is used as fermionic
charge), and a left-handed quasiparticle has ${\bf p}_{\rm L} = -{\bf
  p}_{\rm R}$. According to Eq.~(\ref{ChargeParticlProduction}) the production rates
for right and left handed quasiparticles are (since $e_{\rm R}^2 = e_{\rm L}^2 = 1$ in
this case)
\begin{equation}
 \dot n_{\rm R} = -\dot n_{\rm L} = \frac{1}{4\pi} {\bf E} \cdot {\bf B}.
\end{equation}
As a result there is a net creation of quasiparticle momentum
${\bf P}$ in a time-dependent texture:
\begin{equation}
\partial_t {\bf P}=\int d^3r~( {\bf p}_{\rm R} \dot n_{\rm R} + {\bf p}_{\rm L} \dot n_{\rm L}) =
{1\over {2\pi^2}}\int d^3r~ p_{\rm F}\hat {\bf l} ~( {\bf E}
\cdot   {\bf B} \, \, ) ~~.
\label{3}
\end{equation}
What we know for sure is that the total linear momentum is conserved. Then
Eq.~(\ref{3}) means that, in the presence of the time-dependent texture, the
momentum is transferred from the superfluid motion of {\it vacuum} to {\it matter}
({\it ie} to the heat bath of quasiparticles which form the normal component).

\subsubsection{Anomalous force acting on a continuous vortex and baryogenesis
  from textures}

Let us take as an example the simplest model, namely the vortex with a {\em
continuous soft} core structure in superfluid $^3$He-A. The core has the following
distribution of the unit vector $\hat {\bf l}({\bf r})$, which points in the
direction of the point-like gap nodes in the smooth core
\begin{equation}
\hat {\bf l}({\bf r})=\hat {\bf z}\cos\eta(r) + \hat {\bf
r}\sin\eta(r) ~~,
\label{l}
\end{equation}
where $z,r,\phi$ are the cylindrical coordinates. Here $ \eta(r)$ is a function
which ensures that the $\hat {\bf l}$-vector orientation changes within the smooth
core from $\hat {\bf l}(0)=-\hat {\bf z}$ to $\hat {\bf l}(\infty)=\hat {\bf z}$.
The circulation of the superfluid velocity along a path far outside the soft core
corresponds to $\nu=2$: $\oint d{\bf r}\cdot {\bf v}_{\rm s}=2\kappa$. Such a continuous
${\hat {\bf l}}$ texture thus represents a doubly quantized vortex without
singularities. In practice the vortex measured in the NMR experiments is subject to
an external magnetic field (${\bf B}\| \hat {\bf z}$) and the $\hat {\bf l}$-vector
at infinity is fixed to the $(x,y)$ plane (see Fig.~\ref{ContVortexFig} with $\hat
{\bf l}(\infty)=\hat {\bf x}$), but this does not change the topological structure
of the vortex.

When the continuous vortex moves in $^3$He-A with a velocity ${\bf
  v}_{\rm L}$ it generates a time dependent $\hat{\bf l}$ texture $\hat{{\bf l}} =
\hat{{\bf l}}({\bf r} - {\bf v}_{\rm L} t)$.  Hence both an ``electric'' and a
``magnetic'' field from Eq.~(\ref{HeEB}) exist and this leads to ``momentogenesis''.
Integration of the anomalous momentum transfer in Eq.~(\ref{3}) over the
cross-section of the soft core of the moving vortex gives an additional force which
acts on the vortex due to spectral flow:
\begin{equation}
{\bf f}_{\rm sf }= \partial_t {\bf P}=\pi \hbar \nu {\cal C}_0 \hat{\bf z}\times
({\bf v}_{\rm L} - {\bf v}_{\rm n}) ~. \label{4}
\end{equation}
Here $\hat {\bf z}$ is the direction of the vortex, ${\cal C}_0= k_{\rm F}^3/3\pi^2$, and
${\bf v}_{\rm n}$ is the heat bath velocity. Thus we have obtained
Eq.~(\ref{SpFlForceGeneral}) with a temperature independent parameter ${\cal
C}(T)={\cal C}_0$.


\begin{figure}
  \centerline{\includegraphics[width=0.9\linewidth]{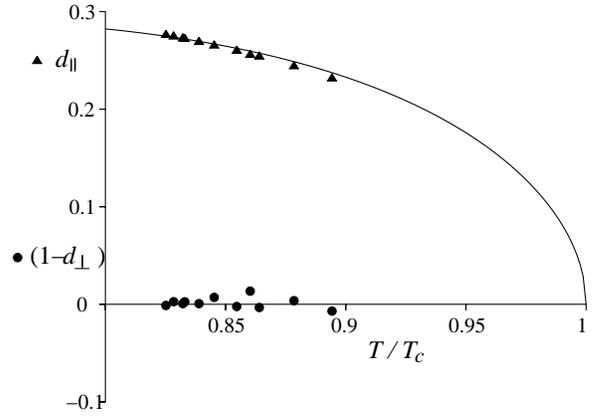}}
  \bigskip
  \caption{Mutual friction parameters $d_\parallel$ and $(1-d_\perp)$
    in $^3$He-A at 29.3
    bar. The curve is a theoretical fit to the data points. (From
    Bevan et~al.\ 1997a).}
  \label{dperpAphaseFig}
\end{figure}


Measurements of the mutual friction coefficients in $^3$He-A with the continuous
vortex  (Bevan et~al.\ 1997a) provide experimental verification for the spectral flow
force. According to Eq.~(\ref{dperp}) it should be that $d_\perp = (m_3{\cal
C}_0-\rho_{\rm n})/\rho_{\rm s}$. The value of $m_3{\cal C}_0$ is the total mass density $\rho$
in the normal phase. Its difference from $\rho$ in the superfluid phase is thus
determined by the effect of superfluidity on the particle density which is extremely
small: $\rho - m_3 {\cal C}_0\sim \rho (\Delta_0/v_{\rm F}p_{\rm F})^2=\rho
(c_\perp/c_\parallel)^2 \ll \rho$. Thus one must have $d_\perp\approx 1$ for all
practical temperatures. $^3$He-A experiments at 29.3\,bar and $T>0.82\,T_{\rm c}$
are consistent with this conclusion  within experimental uncertainty: it was found
that $|1-d_\perp|<0.005$, as demonstrated in Fig.~\ref{dperpAphaseFig}.

\subsubsection{Anomalous force acting on a singular vortex and baryogenesis
  with strings} \label{SpectralFlowSingular}

The discussion of spectral flow in the previous sections and particularly
Eqs.~(\ref{ChargeParticlProduction}) cannot be directly applied to the singular
vortex structures which are found in $^3$He-B and superconductors. The reason is
that the deviation in the magnitude of the order parameter from its equilibrium
value in the cores of such vortices create a potential well for the core
quasiparticles. In this well quasiparticles have discrete energy levels with some
characteristic separation $\hbar \omega_0$ instead of continuous spectra (as a
function of $p_z$), as was considered above. Thus the theory of spectral flow
becomes more complicated, but can still be constructed (Kopnin et~al.\ 1995; Kopnin 2002).

The basic idea is that discrete levels have some broadening $\hbar/\tau$, resulting
from the scattering of core excitations by the free excitations in the heat bath
outside the core (or by impurities in superconductors). At low temperatures, when
the width of levels is much less than their separation, i.e. $\omega_0 \tau \gg 1$,
spectral flow is essentially suppressed, ${\cal C}(T) = 0$. In the opposite case,
$\omega_0 \tau \ll 1$, the levels overlap and we have a situation similar to
spectral flow in a continuous spectrum: ${\cal C}(T) \approx {\cal C}_0$. We may
construct an interpolation formula between these two cases:
\begin{equation}
 {\cal C}(T) \sim {{\cal C}_0\over 1+\omega_0^2\tau^2}~.
\label{C_=Renormalized}
\end{equation}
In fact, both the $d_\parallel$ and $d_\perp$ mutual friction coefficients are
affected by this renormalization of the spectral flow force
(Kopnin et~al.\ 1995; Stone 1996):
\begin{equation}
d_\parallel-i(1-d_\perp)=
\frac{\rho}{\rho_{\rm s}}\frac{\omega_0\tau}{1+i\omega_0
\tau}\tanh\frac{\Delta(T)}{2k_{\rm B}T}~~.
\label{6}
\end{equation}

Let us derive this equation using the Landau-type phenomenological description for
fermions in the vortex core, as developed by Stone (1996). For simplicity we
consider the axisymmetric vortex core; the general case of the asymmetric core is
discussed by Kopnin and  Volovik (1998). The low-energy spectrum of
Caroli-de-Gennes-Matricon quasiparticles around a vortex contains an anomalous
branch of fermionic zero modes. In the case of the axisymmetric vortex, excitations
on this branch are characterized by the angular momentum $L_z$
\begin{equation}
E(L_z,p_z)=  -\omega_0(p_z) L_z ~.
\label{SingCore1}
\end{equation}
For superconductors with a coherence length $\xi$ much larger than the inverse Fermi
momentum, $p_{\rm F}\xi \gg 1$, the electron wavelength is short compared with the core
size, and the quasiclassical approximation is relevant. The quasiclassical angular
momentum $L_z$ is a continuous variable; thus the anomalous branch crosses zero as a
function of $L_z$ at $L_z=0$ and spectral flow can occur along this branch between
the vacuum states with $E<0$ and the excited states with $E>0$.  Such spectral flow
occurs during the motion of the vortex with respect to the normal component, where
it is caused by the interaction with impurities in superconductors or with the
thermal scattering states in superfluids. In the quasiclassical approximation, the
Doppler shifted spectrum of the fermions in the moving vortex has the form
\begin{equation}
E(L_z,{\bf p})=  -\omega_0(p_z) L_z + ({\bf v}_{\rm s} - {\bf v}_{\rm L})\cdot{\bf p}~.
\label{SingCore2}
\end{equation}
Here the momentum ${\bf p}$ is assumed to be at the Fermi surface: ${\bf
  p}=(p_{\rm F}\sin\theta \sin\varphi, p_{\rm F}\sin\theta \cos\varphi, p_{\rm F}\cos\theta) $.
The azimuthal angle $\varphi$ is canonically conjugated to the angular momentum
$L_z$. This allows us to write the Boltzmann equation for the distribution function
$n(L_z,\varphi)$ at fixed $p_z=p_{\rm F}\cos\theta$:
\begin{eqnarray}
&\partial_t n-\omega_0 \partial_\varphi n -\partial_\varphi (({\bf v}_{\rm s} - {\bf
v}_{\rm L})\cdot{\bf p})
~\partial_{L_z} n&\nonumber\\
&\displaystyle= -{n(L_z,\varphi)-  n_{\rm eq}(L_z,\varphi)\over \tau}\;,&
\label{SingCore3}
\end{eqnarray}
where the collision time $\tau$ characterizes the interaction of the bound state
fermions with impurities or with the thermal fermions in the normal component
outside the vortex core. The equilibrium distribution function is:
\begin{eqnarray}
  &n_{\rm eq}(L_z,\varphi)=& \nonumber\\
  &f(E- ({\bf v}_{\rm n} - {\bf v}_{\rm L})\cdot{\bf
p})= f(-\omega_0 L_z + ({\bf v}_{\rm s} - {\bf v}_{\rm n})\cdot{\bf p})\,,& \label{SingCore4}
\end{eqnarray}
where $f(E)=(1+\exp(E/T))^{-1}$ is the Fermi-function.

Introducing the shifted variable
\begin{equation}
l=L_z - ({\bf v}_{\rm s} - {\bf v}_{\rm n})\cdot{\bf p}/\omega_0~,
\label{SingCore5}
\end{equation}
one obtains the equation for $n(l,\varphi)$
\begin{eqnarray}
\partial_t n-\omega_0 \partial_\varphi n -\partial_\varphi (({\bf v}_{\rm n} - {\bf
v}_{\rm L})\cdot{\bf k}) ~
\partial_l n\nonumber\\
= -{n(l,\varphi)-  f(-\omega_0 l))\over \tau}~,
\label{SingCore6}
\end{eqnarray}
which does not contain ${\bf v}_{\rm s}$.  To find the force acting on the vortex
from the heat bath environment, we are interested in the evolution of the
total momentum of quasiparticles in the vortex core:
\begin{equation}
{\bf P}=\sum {\bf p}=\int_{-p_{\rm F}}^{p_{\rm F}} {dp_z\over 2\pi} {\bf P} (p_z)~,~ {\bf P}
(p_z)=  {1\over 2}\int dl {d\varphi\over 2\pi}  n(l,\varphi) {\bf p}\,.
\label{SingCore7}
\end{equation}
It appears that the equation for ${\bf P} (p_z)$ can be written in closed form
\begin{eqnarray}
\nonumber
\partial_t {\bf P} (p_z)-\omega_0 \hat z\times {\bf P} (p_z) +{{\bf P}
(p_z) \over
\tau}=\\-{1\over
4}p_{\rm F}^2\sin^2\theta
\hat z\times ({\bf v}_{\rm n} - {\bf v}_{\rm L}) (f(\Delta(T))-f(-\Delta(T)))  ~.
\label{SingCore8}
\end{eqnarray}
Here we take into account that $\int dl \partial_l n$ is limited by the bound states
below the gap $\Delta(T)$ of bulk liquid: above the gap $\Delta(T)$ the spectrum of
fermions is continuous, {\em ie.} the inter-level distance $\omega_0=0$.

In steady state vortex motion one has $\partial_t  {\bf P}(p_z)=0$. Then, since
$f(\Delta(T))-f(-\Delta(T))=\tanh (\Delta(T)/2T)$, one obtains the following
contribution to the momentum from the heat bath to the core fermions due to the
spectral flow of bound states below $\Delta(T)$
\begin{eqnarray}
\nonumber
{\bf F}_{\rm bsf}=\int_{-p_{\rm F}}^{p_{\rm F}} {dp_z\over 2\pi}  {{\bf P}
(p_z) \over
\tau}
={\kappa\over
4} \tanh {\Delta(T)\over 2T} \times \\
\int_{-p_{\rm F}}^{p_{\rm F}} {dp_z\over 2\pi}  { p_{\rm F}^2 -
p_z^2\over 1+
\omega_0^2\tau^2} [({\bf v}_{\rm L} - {\bf v}_{\rm n})
\omega_0\tau +\hat z\times({\bf v}_{\rm L} - {\bf v}_{\rm n})]  ~.
\label{SingCore9}
\end{eqnarray}

The spectral flow of unbound states above $\Delta(T)$ is not suppressed, the
corresponding $\omega_0\tau=0$, since the distance between the levels in the
continuous spectrum is $\omega_0=0$. This gives
\begin{eqnarray}
&&{\bf F}_{\rm usf} = \nonumber\\
&&{\kappa\over
4}p_{\rm F}^2\int_{-p_{\rm F}}^{p_{\rm F}} {dp_z\over 2\pi} \sin^2\theta \left(1- \tanh
{\Delta(T)\over 2T}\right)  \hat z\times({\bf v}_{\rm L} - {\bf v}_{\rm n})  ~.
\label{SingCore10}
\end{eqnarray}
Thus the total nondissipative (transverse) and frictional (longitudinal)
parts of
the spectral-flow force are
\begin{eqnarray}
\nonumber
&&{\bf F}_{\rm sf}^\perp={\kappa\over
4} \int_{-p_{\rm F}}^{p_{\rm F}} {dp_z\over 2\pi} (p_{\rm F}^2-p_z^2) \times\\
&&~~~~\left[1 - \tanh {\Delta(T)\over 2T}{\omega_0^2\tau^2 \over 1+
\omega_0^2\tau^2} \right]\hat z\times({\bf v}_{\rm L} - {\bf
v}_{\rm n}),
\label{TransverseForce}
\end{eqnarray}
\begin{eqnarray}
&&{\bf F}_{\rm sf}^\parallel=
({\bf v}_{\rm L} - {\bf v}_{\rm n}){\kappa\over
4}\tanh {\Delta(T)\over 2T}\times\nonumber\\
&&~~~~\int_{-p_{\rm F}}^{p_{\rm F}} {dp_z\over 2\pi} (p_{\rm F}^2 -
p_z^2){\omega_0\tau\over 1+
\omega_0^2\tau^2}
\label{SingCore11}
\end{eqnarray}
In the limit $\omega_0\tau \ll 1$, the friction force disappears, while the
transverse spectral flow force is maximal and coincides with that obtained for the
continuous vortex in Eq.~(\ref{4}) with the same value of ${\cal
  C}_0=(1/4) \int_{-p_{\rm F}}^{p_{\rm F}} dp_z (p_{\rm F}^2-p_z^2)=p_{\rm F}^3/3\pi^2$. In the general
case the result in Eqs.~(\ref{TransverseForce}) and (\ref{SingCore11}) is
complicated since both $\omega_0$ and $\tau$ depend on $p_z$. However, if one
neglects this dependence and adds the Magnus and Iordanskii forces, one obtains
Eq.~(\ref{6}) for all $T$. If the temperature is not too high, so that $\tanh
\Delta(T)/ 2T \approx 1$, one obtains Eq.~(\ref{C_=Renormalized}) for the
renormalized spectral flow parameter.

\begin{figure}
  \centerline{\includegraphics[width=0.9\columnwidth]{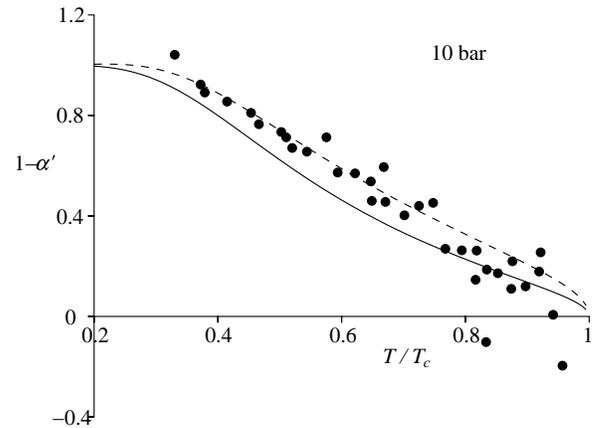}}
  \bigskip
  \caption{Experimental values of $1-\alpha'$ in $^3$He-B at 10 bar
    compared with the theoretical result in Eq.~(\protect\ref{dcomb}). The
    full curve is for the theoretical value of $\Delta(T)$ and the broken
    curve is the fit with reduced $\Delta$. (From
    Bevan et~al.\ 1997a).}
  \label{d-B-compar}
\end{figure}

Comparison of Eq.~(\ref{6}) with experiment should take into account that neither
$\omega_0$ nor $\tau$ are known with good precision. However, the following
combination of mutual friction coefficients does not depend on these parameters
explicitly:
\begin{equation}
  1 - \alpha' = \frac{1-d_\perp}{d_\parallel^2 + (1-d_\perp)^2} =
  \frac{\rho_{\rm s}}{\rho} \left[ \tanh \frac{\Delta(T)}{2k_{\rm B} T}
  \right]^{-1}.
  \label{dcomb}
\end{equation}
This equation (Kopnin et~al.\ 1995) is compared to experimental data on mutual friction in $^3$He-B (Bevan et~al.\ 1997a) in Fig.~\ref{d-B-compar}. The agreement is excellent in view of the
approximations in the theory.


This shows that the chiral anomaly is relevant for the interactions of condensed
matter vortices (analogues of strings) with fermionic quasiparticles (analogues of
quarks and leptons). For continuous vortices in $^3$He-A the spectral flow of
fermions between the superfluid ground state (vacuum) and the heat bath of positive
energy particles forming the normal component (matter) dominates at any relevant
temperature.  For singular vortices in $^3$He-B it is important at $T \sim T_{\rm
c}$ and vanishes at $T\ll T_{\rm c}$.

It is interesting to note that in a homogeneous ground state (vacuum) momentum
(fermionic charge) is conserved separately in the ground state (vacuum) and in the
heat bath of excitations (matter).  Topological defects are thus the mediators for
the transfer of momentum between these two subsystems. The motion of a vortex across
the flow changes both the topological charge of the vacuum (say, the winding number
of superflow in a torus geometry) and the fermionic charge (angular momentum in the
same geometry). All these processes are similar to those involved in the
cosmological production of baryons and can thus be investigated in detail.

However, it is important to note that although spectral flow leads to anomalous
creation of fermionic charge from vacuum, the total fermionic charge of vacuum plus
matter remains conserved. There is also no intrinsic bias in this process for the
direction of charge transfer: from vacuum to matter or in the opposite direction.
The necessary conditions for this mechanism to start operating  are that the system
is in a non-equilibrium state and under the influence of symmetry breaking which
biases the direction of charge transfer. How such conditions were realized in the
early Universe is not clear at present -- if we interpret the positive baryonic
charge of matter in our Universe to arise from spectral flow. But it is commonly
believed that broken P and CP invariances play crucial roles. In contrast in the
$^3$He case the situation is completely under control: The relevant symmetry
breaking is achieved by applying rotation or a magnetic field, while the
non-equilibrium state is generated by applying external superflow.

The experimental verification of momentogenesis in superfluid $^3$He can thus be
viewed to support these ideas of cosmological baryogenesis via spectral flow and
points to a future where several cosmological problems are modelled and studied in
light of the superfluid $^3$He example (Volovik 2003). Now let us consider the
opposite effect which leads to the analogue of magnetogenesis.

\subsection{Analog of magnetogenesis: vortex textures generated in
  normal-superfluid counterflow}

From Eq.~(\ref{ChargeParticlProduction}) it follows that the magnetic field
configuration can absorb fermionic charge. If this magnetic field has helicity, it
acquires an excess of right-moving particles over left-moving particles:
\begin{equation}
(n_{\rm R}-n_{\rm L})_{\bf A}={1\over 2\pi^2} {\bf A}\cdot ({\bf\nabla}\times {\bf A})~~.
\label{anomaly}
\end{equation}
The right-hand side is the so called Chern-Simons (or topological)
charge of the magnetic field.

The transformation of particles into a magnetic field configuration opens the
possibility to examine the cosmological origin of galactic magnetic fields from a
system of fermions. This is the essential step in the scenario described by
Joyce and  Shaposhnikov (1997). In this model an initial excess of right-handed
electrons, $e_{\rm R}$, was assumed to have been generated in the early Universe. This
excess would then have survived until the electroweak phase transition (at about
$10^{-10}$ s after the big bang) at which point anomalous lepton (and baryon) number
violating processes became efficient enough to erase the excess.  However, it
appears that well before the electroweak transition an instability developed, where
the excess of the right electrons transformed into a hypermagnetic field. Then, when
the electroweak transition took place, it transformed part of the hypermagnetic
field into electromagnetic field, so that the universe was bestowed with a
primordial (electro-) magnetic field.

We can now discuss the corresponding process in $^3$He-A -- the transformation of
fermionic charge to magnetic field. In our case it is the quasiparticle momentum
which plays the part of the relevant fermionic charge. The net quasiparticle
momentum is generated by the relative flow of the normal and superfluid components.
This fermionic charge is transformed via the chiral anomaly to order parameter
texture which, as we have seen, plays the part of the magnetic field. The
transformation occurs in the form of an instability which transfers the excess in
momentum to the formation of ${\hat {\bf l}}$-textures. The process corresponds to
the counterflow instability observed in $^3$He-A, which has been discussed both
experimentally and theoretically (Ruutu et~al.\ 1997b). Thus the $^3$He-A analogy
closely follows the cosmological scenario described by Joyce and  Shaposhnikov (1997).

Let us discuss this instability. In the presence of counterflow, ${\bf
  v} ={\bf v}_{\rm n}-{\bf v}_{\rm s}$, of the normal component of $^3$He-A liquid
with respect to the superfluid, the ${\bf {\hat l}}$-vector is oriented along the
flow, ${\bf {\hat l}}_0\parallel {\bf v}$. We are interested in the stability
condition for such homogeneous counterflow with respect to the generation of
inhomogeneous perturbations $\delta {\bf {\hat l}}$,
\begin{equation}
{\bf {\hat l}} = {\bf {\hat l}}_0 + \delta {\bf {\hat l}}({\bf r},t)\ ,
\label{oscillating-l}
\end{equation}
keeping in mind that the space and time dependences of $\delta {\hat{\bf l}}$
correspond to ``hyperelectric field'' ${\bf E}=k_{\rm F} \partial_t \delta {\hat {\bf l}}$
and ``hypermagnetic field'' ${\bf B}=k_{\rm F}{\bf \nabla}\times \delta {\hat {\bf l}}$.

It is important for our consideration that the $^3$He-A liquid is anisotropic in the
same manner as a nematic liquid crystal. For the relativistic fermions this means
that their motion is determined by the geometry of some effective spacetime which in
$^3$He-A is described by the metric tensor in Eq.~(\ref{gikAphase}).  As we have
already discussed above, in the presence of counterflow the energy of quasiparticles
is Doppler shifted by the amount ${\bf p}\cdot{\bf v}$. Since the quasiparticles are
concentrated near the gap nodes, this energy shift is constant and opposite for the
two gap nodes: ${\bf p}\cdot{\bf v}\approx \pm p_{\rm F}({\hat{\bf l}}_0\cdot{\bf v})$.
The counterflow therefore produces what would be an effective chemical potential in
particle physics, which has opposite sign for the right- and left-handed particles:
\begin{equation}
 \mu_{\rm R}= -\mu_{\rm L} = p_{\rm F}({\hat{\bf l}}_0\cdot{\bf v}).
 \label{effmu}
\end{equation}
The kinetic energy of the counterflow is
\begin{equation}
 E_{\rm kin}={1\over 2}{\bf v}  \rho_{n \parallel}{\bf v}  ~.
\label{he3ernr}
\end{equation}
Here the density of the normal component is a tensor in the anisotropic $^3$He-A
liquid, and only the longitudinal component $\rho_{n
\parallel}$ is involved.

Let us consider the low-temperature limit $T\ll T_{\rm c}$, where $T_{\rm c} \sim
\Delta_0$ is the superfluid transition temperature. Then using the expression for
the longitudinal density of the normal component of $^3$He-A (Vollhardt and W{\"o}lfle 1990)
\begin{equation}
  \rho_{n\parallel} = \pi^2 \rho \frac{m_3^*}{m_3}
  \left( \frac{k_{\rm B} T}{\Delta_0} \right)^2
\end{equation}
and also the $^3$He-A equivalent of the
chemical potential (\ref{effmu}) one obtains
\begin{equation}
 E_{\rm kin}
\approx {1\over 6} m_3^* k_{\rm F}^3
{T^2\over \Delta_0^2}   ({\hat{\bf l}}_0\cdot {\bf v} ) ^2
\equiv {1\over 6} \sqrt{-g}~T^2 \mu_{\rm R}^2 ~. \label{lowternr}
\end{equation}
In the last equality an over-all constant appears to be the square root of the
determinant of an effective metric in $^3$He-A: $\sqrt{-g}=1/c_\parallel
c_\perp^2=m_3^* k_{\rm F}/\Delta_0^2$.  In relativistic theories the rhs of
Eq.~(\ref{lowternr}) is exactly the energy density of the massless right-handed
electrons in the presence of the chemical potential $\mu_{\rm R}$. Thus the kinetic
energy, stored in the counterflow, is exactly analogous to the energy stored in the
right-handed electrons.  The same analogy occurs between the net quasiparticle
linear momentum, ${\bf
  P}=\rho_{\rm n} {\bf v}$, along ${\bf{\hat l}}_0$ and the chiral charge of the
right electrons,
\begin{equation}
 n_{\rm R}\equiv {1\over p_{\rm F}}{\bf
P}\cdot {\bf{\hat l}}_0 ~.
\label{ChargeVsMomentum}
\end{equation}
The inhomogeneity which absorbs the fermionic charge, is represented by a magnetic
field configuration in real physical vacuum and by a $\delta {\hat{\bf l}}$-texture
in $^3$He-A. However, Eq.  (\ref{anomaly}) applies in both cases, if in $^3$He-A we
use the standard identification ${\bf A} = k_{\rm F} \delta {\bf {\hat l}}$.

Just as in the particle physics case, we now consider the instability
towards the production of the ``magnetic'' texture due to the excess of
chiral particles. This instability can be seen by considering the gradient
energy of the inhomogeneous texture on the background of the superflow. In
the geometry of the superflow, the textural contribution to the free energy
of the $\delta{\hat{\bf l}}$-vector is completely equivalent to the
conventional energy of the hypermagnetic field (Volovik 1992)
\begin{eqnarray}
F_{\rm grad}= {\rm ln}~ \left ( {\Delta_0^2\over T^2 }~ \right )
{{p_{\rm F}^2v_{\rm F}}\over {24\pi^2\hbar}}~
( \hat {\bf l}_0\times ({\bf\nabla} \times  \delta \hat {\bf l}))^2
\nonumber\\
~~\equiv
{{\sqrt{-g}}\over {4\pi}e_{\rm eff}^2} g^{ij}g^{kl}F_{ik}F_{jl}
= F_{\rm magn}~~.
\label{magenergy}
\end{eqnarray}
Here $F_{ik}= \nabla_i A_k -\nabla_k A_i$\label{fik} and we again have included
the effective anisotropic metric in Eq.~(\ref{gikAphase}) appropriate
for $^3$He-A.


\begin{figure}
  \centerline{\includegraphics[width=0.8\linewidth]{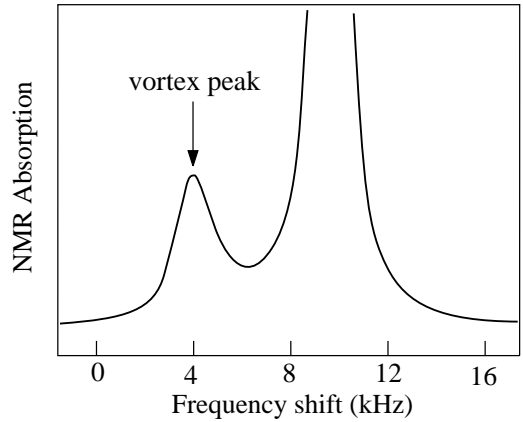}}
  \bigskip
  \caption{NMR signal from an array of continuous ATC vortices, with the frequency
    shift from the Larmor value plotted on the horizontal axis. The large
    truncated peak on the right is generated by the bulk order-parameter
    texture, while the small satellite peak on the left is produced by the
    continuous soft-core vortex textures.  The frequency shift of the satellite
    peak from the Larmor frequency indicates the type of vortex, while the
    intensity of the peak is proportional to the number of vortices of this
    particular type.}
  \label{ContVortexNMR}
\end{figure}


It is interesting that the logarithmic factor in the gradient energy plays
the part of the running coupling $e_{\rm eff}^{-2} =(1/3\pi \hbar c)\ln
(\Delta_0/T)$ in particle physics, where $e_{\rm eff}$ is the effective
hyperelectric charge; while the gap amplitude $\Delta_0$, the ultraviolet
cutoff, plays the part of the Planck energy scale.  Now if one has the
counterflow in $^3$He-A, or its equivalent -- an excess of chiral charge
produced by the chemical potential $\mu_{\rm R}$ -- the anomaly gives rise to an
additional effective term in the magnetic energy, corresponding to the
interaction of the charge absorbed by the magnetic field with the chemical
potential. This effective energy term is:
\begin{eqnarray}
F_{CS}=(n_{R}-n_{\rm L})\mu_{\rm R}=
{1\over 2\pi^2} \mu_{\rm R}  {\bf A}\cdot ({\bf\nabla}\times {\bf A})
\nonumber\\
={3\hbar\over 2m}\rho ({\hat{\bf l}}_0\cdot{\bf v}) (\delta{\hat{\bf l}}\cdot
{\bf\nabla}\times \delta{\hat{\bf l}})~. \label{csenergy}
\end{eqnarray}
The rhs corresponds to the well known anomalous interaction of the counterflow with
the ${\hat{\bf l}}$-texture in $^3$He-A, where $\rho$ is the mass density of $^3$He
(Volovik 1992) (the additional factor of 3/2 enters due to nonlinear effects).

For us the most important property of this term is that it is linear in the
derivatives of $\delta {\bf {\hat l}}$. Its sign thus can be negative,
while its magnitude can exceed the positive quadratic term in
Eq.~(\ref{magenergy}). This leads to the helical instability where the
inhomogeneous $\delta {\bf {\hat l}}$-field is formed. During this
instability the kinetic energy of the quasiparticles in the counterflow
(analog of the energy stored in the fermionic degrees of freedom) is
converted into the energy of inhomogeneity $ {\bf\nabla}\times
\delta{\hat{\bf l}}$, which is the analog of the magnetic energy of the
hypercharge field.

When the helical instability develops in $^3$He-A, the final result is the formation
of a ${\hat{\bf l}}$-texture, which represents a large jump from the vortex-free
counterflow texture towards the global free energy minimum, the equilibrium vortex
texture. The new texture contains a central part with a periodic ${\hat{\bf
l}}$-texture, where the elementary lattice cell represents a so-called
Anderson-Toulouse-Chechetkin (ATC) continuous vortex (Fig.~\ref{ContVortexFig}). ATC
vortices give rise to a characteristic satellite peak in the NMR absorption spectrum
where their number is directly proportional to the height of the satellite peak,
Fig.~\ref{ContVortexNMR} (Blaauwgeers  et~al.\ 2000).


\begin{figure}
  \centerline{\includegraphics[width=0.9\linewidth]{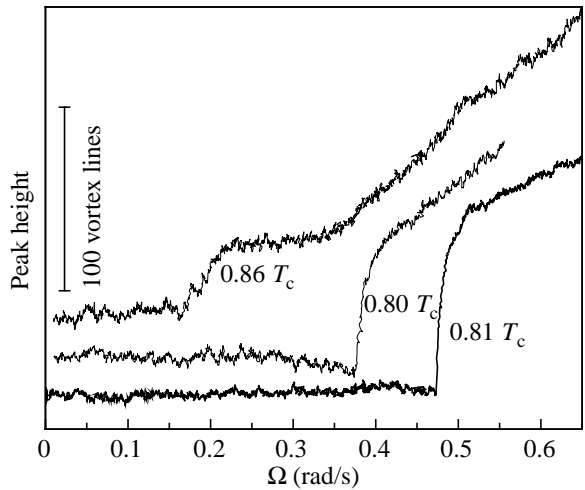}}
  \bigskip
  \caption{The height of the satellite peak, which is generated by
    continuous vortex lines, is here monitored during a slow increase of $\Omega$. Initially there are no vortex lines in the sample and the peak height is zero.
    When the velocity of the counterflow ${\bf v}$ in the
    ${\hat{\bf l}}_0$ direction (corresponding to the chemical
    potential $\mu_{\rm R}$ of the chiral electrons) exceeds some critical
    value, a textural instability occurs and the texture transforms into a configuration with a lower critical velocity. Simultaneously a larger number of vortex lines is formed. This means that the center of the sample cylinder becomes suddenly filled
    with the ${\hat{\bf l}}$-texture (or hypermagnetic field) which forms a
    central cluster of continuous vortex lines. (From Ruutu et~al.\ (1996c)).}
  \label{experiment2}
\end{figure}


Experimentally this is observed when the rotation velocity is slowly increased,
starting from a state with vortex-free counterflow ({\it ie.} with fermionic charge,
but no hypermagnetic field).  The helical instability is then observed as a
discontinuity, when the vortex satellite is suddenly formed
(Fig.~\ref{experiment2}). Its peak height jumps from zero to a magnitude which
approaches that in the equilibrium vortex state, which means that the counterflow is
essentially reduced and the number of vortex lines is close to that in equilibrium.
Most of the counterflow (fermionic charge) thus becomes converted into the vortex
texture (magnetic field).

Together with the results of Bevan et~al.\ (1997b), the textural superflow instability
shows that the chiral anomaly is an important mechanism in the interaction of vortex
textures (the analogue of the hypercharge magnetic fields and cosmic strings) with
fermionic excitations (analogue of quarks and leptons). These two experiments
verified both processes which are induced by the anomaly: the nucleation of
fermionic charge from vacuum, observed by Bevan et~al.\ (1997b), and the inverse
process of the nucleation of an effective magnetic field from the fermion current as
observed by Ruutu et~al.\ (1997b).

\subsection{Vortex mass: chiral fermions in strong magnetic field}

Until now we have assumed that the mass $M_{\rm V}$ of a vortex can be neglected in experimental vortex dynamics. How well justified is this approach? The term $M_{\rm V} \partial_t {\bf v}_{\rm L}$ in the force balance equation for the vortex contains the time derivative and thus at very low frequencies of vortex motion it can be neglected compared to other forces, which are directly proportional to ${\bf v}_{\rm L}$.  To be more quantitative, we estimate the vortex mass in the BCS superfluids and superconductors. There are several contributions to the vortex mass. It will also become clear that these are related to some most peculiar phenomena in quantum field theory. We start from the contribution, which is relevant for the Bose superfluid $^4$He.

\subsubsection{``Relativistic'' mass of the vortex}

In the hydrodynamic theory the mass of the vortex is nonzero owing to the
compressibility of the liquid. Since sound propagation in fluids is similar to light
propagation in vacuum, the hydrodynamic energy of the vortex, soliton or other
extended object moving in the liquid is connected with its hydrodynamic mass per
unit length by the ``relativistic'' equation $E=ms^2$, where the speed of sound $s$
substitutes the speed of light (Davis 1992; Duan 1995; Kao and Lee 1995; Iengo and Jug 1995; Wexler and Thouless 1996). Thus the
hydrodynamic mass of a vortex loop of length $L$ at $T=0$ is according to
Eq.~(\ref{eq11}):
\begin{equation}
M_{\rm compr}={E_{\rm kin}\over s^2}={{\rho
\kappa^2 L} \over {4\pi s^2}}
\; \ln {L\over {\xi}}~.
\label{HydroMass}
\end{equation}
For Fermi superfluids $s$ is on the order of the Fermi velocity $v_{\rm F}\sim p_{\rm F}/m$ ($m$
is the mass of the electron in metals or of the $^3$He atom), and the estimate for
the hydrodynamic mass of a small vortex loop of length $L$ is $M_{\rm compr}\sim \rho
a^2 L \ln L/\xi$, where $a$ is the interatomic distance.  For superfluid $^4$He, where
the core size $\xi \sim a$, we can roughly speaking associate the hydrodynamic vortex mass $\sim \rho a^2 L $ with the ``mass of the liquid in the vortex core volume''.  However for
$^3$He superfluids and for superconductors, where $\xi \gg a$, the hydrodynamic vortex mass is much less than $\rho \xi^2 L $ and other contributions become more important.

\subsubsection{Contribution from bound states to the  mass of a singular vortex}

It appears that the most important contribution to vortex mass originates from the
quasiparticles occupying the bound states in the vortex core and thus forming the
normal component concentrated in the core. For the vortices in conventional
superconductors and in $^3$He-B this contribution to the vortex mass depends on
$\omega_0\tau$ and in the clean-limit case it is proportional to the mass of the
liquid in the vortex core, as was first found by Kopnin (1978) for
superconductors and by Kopnin and Salomaa (1991) for superfluid $^3$He-B: $M_{\rm
Kopnin} \sim \rho \xi^2 L$. This core mass is essentially larger than the
logarithmically divergent contribution, which comes from the compressibility. In
spite of the logarithmic divergence, the latter contains the speed of sound in the
denominator and thus is smaller by the factor $(a/\xi)^2 \ll 1$, where $a$ is the
interatomic distance. The compressibility mass of the vortex dominates in Bose
superfluids, where the core size is small, $\xi \sim a$.

According to Kopnin's theory the core mass comes from the fermions trapped in the
vortex core
(Kopnin 1978; Kopnin and Salomaa 1991; van Otterlo et~al.\ 1995; Volovik 1997; Kopnin and Vinokur 1998).
This   Kopnin mass of the vortex can be derived using a phenomenological approach.
Let us consider the limit of  low $T$  in  the superclean regime $\omega_0\tau\gg 1$
(Volovik 1997). If the vortex moves with  velocity ${\bf v}_{\rm L}$ with
respect to the superfluid component, the fermionic energy spectrum in the vortex
frame is Doppler shifted and has in Eq.~(\ref{SingCore2}) the form: $E=E_0(q)-
{\bf p}\cdot {\bf v}_{\rm L}$, where $q$ represents the fermionic degrees of freedom in
the stationary vortex.  The summation over the fermionic degrees of freedom  leads to
the extra linear momentum of the vortex $\propto{\bf v}_{\rm L}$:
\begin{eqnarray}
&&{\bf P}=\sum_q {\bf p}\, \theta (-E)=\sum_q {\bf p} ({\bf k}\cdot
{\bf v}_{\rm L}) \delta (E_0) =M_{\rm Kopnin}{\bf v}_{\rm L} ,\nonumber\\
&&M_{\rm Kopnin} ={1\over 2} \sum_q {\bf p}_\perp^2  \delta (E_0)~.
\label{DOS}
\end{eqnarray}
For the axisymmetric vortex, where $E_0=-L_z \omega_0(p_z)$ and  $\sum_q =
\int dL_z dp_z dz/2\pi$, one has
\begin{equation}
M_{\rm Kopnin} =  L\int_{-p_{\rm F}}^{p_{\rm F}} {dp_z\over 4\pi}{p_{\rm F}^2 -p_z^2 \over
\omega_0(p_z)}\;. \label{GeneralKopninMass}
\end{equation}
Eq.~(\ref{GeneralKopninMass}) can also be obtained from the time dependent kinetic
equation (\ref{SingCore3}).  It is the coefficient in the contribution to the
longitudinal force, which is linear in external frequency $\omega$: ${\bf F}_{\rm
sf}^\parallel=-i\omega M_{\rm Kopnin} {\bf v}_{\rm L}$.

Note that this vortex mass is determined in essentially the same way as the normal
component density in the bulk system. The Kopnin vortex mass is nonzero if the
density of fermionic states is finite in the vortex core, which is determined by the
inter-level spacing $\omega_0$ in the core: $N(0)\propto 1/\omega_0$. That gives for
the Kopnin vortex mass per unit length an estimation: $M_{\rm
  Kopnin} \sim p_{\rm F}^3/\omega_0 \sim \rho \xi^2$. A method to measure this contribution for the $^3$He-B vortex with spontaneously broken axisymmetry of the vortex core, by driving the core into rapid rotation with an NMR excitation field, has been suggested by Kopnin and  Volovik (1998). The much stronger connection to the normal component fraction in the core of a vortex with continuous
structure, like the dominant vortex line in $^3$He-A, will be examined in the next
subsection.

\subsubsection{Kopnin vortex mass in the continuous-core model: connection to chiral
fermions in magnetic field}

\begin{figure}[!!!!tb]
  \centerline{\includegraphics[width=0.9\linewidth]{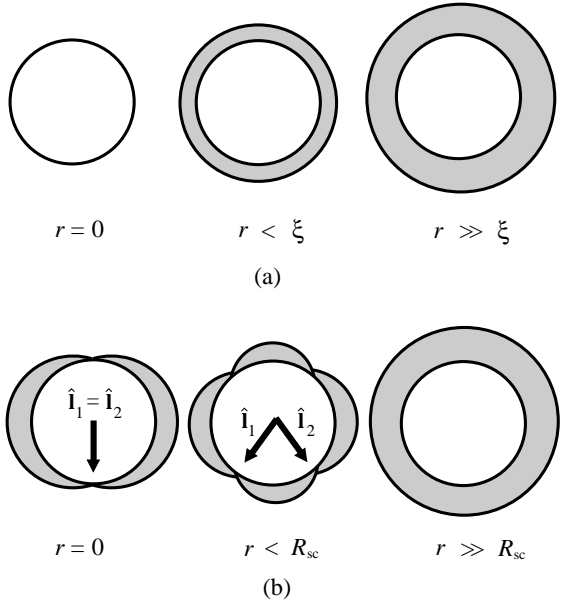}}
  \bigskip
  \caption{Smoothing operation of the singular vortex core. (a) In the singular
    vortex the gap is continuously reduced and becomes zero exactly on the vortex
    axis (at $r=0$). (b) For some vortices it is energetically favorable to
    avoid the vanishing order parameter at $r=0$. Instead,
    within the smooth core, $r<R_{\rm sc}$, point-like gap nodes appear in the
    spectrum of fermions (Volovik and  Mineev 1982). The unit vectors
    $\hat{\bf l}_1$ and $\hat{\bf l}_2$ show the directions to the nodes at
    different $r$. Close to the gap nodes the spectrum of the fermions is
    similar to that in $^3$He-A (example at $r=0$).}
  \label{nodes}
\end{figure}

The continuous-core vortex in $^3$He-A is the best example -- a model case -- which
helps to understand the vortex core mass. The continuous-core model can also be
applied to other Fermi superfluids and superconductors: the singular core can
thereby be smoothed, so that the $1/r$-singularity of the superfluid velocity is
removed, by introducing point nodes in the superfluid energy gap in the core region
(Fig.~\ref{nodes}). As a result the superfluid/superconducting state in the vortex
core acquires the properties of $^3$He-A with
its continuous vorticity and point gap nodes
(Volovik and  Mineev 1982; Salomaa and Volovik 1987).  After that one can easily separate
different contributions to the vortex mass.  Actually this is not only a model:
Spontaneous smoothing of the velocity singularity occurs in the cores of both types
of $^3$He-B vortices (Salomaa and Volovik 1987), while in heavy fermion and high-$T_{\rm c}$
superconductors such smoothing can occur due to the admixture of different pairing
states in the vortex core.

For the smoothed singly quantized vortices of $^3$He-B and superconductors
one has two $\hat {\bf l}$-vectors: $\hat {\bf l}_1$ and $\hat {\bf l}_2$,
each for one of the two spin projections. The simplest distribution of both
fields is given by Eq.~(\ref{l}) with such $\eta(r)$ that $\hat {\bf
  l}_1(0)=\hat {\bf l}_2(0)=-\hat {\bf z}$ and $\hat {\bf
  l}_1(\infty)=-\hat {\bf l}_2(\infty)=\hat {\bf r}$ (Salomaa and Volovik 1987).
The region of radius $R_{\rm sc}$, where the texture of $\hat{\bf
  l}$-vectors is concentrated, represents the smoothed {\it nonsingular core} of
the vortex.

In the continuous vortex, the normal component in the core can be considered to be a
local quantity, determined at each point in the vortex core. This consideration is
valid for a smooth core with radius $R_{\rm sc}\gg \xi$, where the local classical
description of the fermionic spectrum can be applied. The main contribution comes
from the point-like gap nodes, where the classical spectrum has the form
$E_0=\sqrt{v_{\rm F}^2(p-p_{\rm F})^2 + \Delta_0^2(\hat{\bf p}\times \hat {\bf l})^2}$ and
$\Delta_0$ is the gap amplitude. In the presence of a gradient in the $\hat {\bf
l}$-field, which acts on the quasiparticles as an effective magnetic field, this
gapless spectrum leads to the nonzero local DOS, discussed in
Sec.~\ref{ChiralAnomaly} for the relativistic chiral fermions. To apply the DOS in
Eq.~(\ref{DOSMagneticField}) to the case of anisotropic $^3$He-A, one must make the
covariant generalization of the DOS, by introducing the general metric tensor and
then substituting it by the effective $^3$He-A metrics which describes the
anisotropy of $^3$He-A. The general form of the DOS of the chiral fermions in the
curved space is
\begin{equation}
N (0)= {\vert e\vert \over 2\pi^2}\sqrt{-g}\sqrt{{1\over 2}g^{ij}
g^{kl}F_{ik}F_{jl}},
\label{GeneralDOSMagneticField}
\end{equation}
where $F_{ik}$ is defined on p.~\pageref{fik}. One now can apply this to $^3$He-A
where $\vert e\vert=1$, and the metric tensor is given by Eq.~(\ref{gikAphase}).
Neglecting the dependence of the small velocity field $v_{\rm s}$ in the smooth core, one
obtains the following local DOS for the fermions in the $\hat{\bf l}$-texture at
$T=0$:
\begin{equation}
N (0,{\bf r})= {p_{\rm F}^2\over 2\pi^2 \Delta_{\rm A}}  \vert (\hat {\bf l}\times(
\nabla\times \hat {\bf l})\vert  \;.
\label{DOS-A-phase}
\end{equation}
This DOS can be inserted in the expression for the local density of the normal
component at $T=0$ (see Eq.~(5.24) in the review by Volovik (1990)):
\begin{equation}
(\rho_{\rm n})_{ij}({\bf r})=  \hat  l_i\hat
l_j p_{\rm F}^2 N(0,{\bf r})~.
\label{NormalComponent}
\end{equation}
For the axisymmetric continuous vortex (Eq.~(\ref{l})) one has
\begin{equation}
N(0,{\bf r})={p_{\rm F}^2 \over 2\pi^2 \Delta_{\rm A}}
\sin\eta |\partial_r \eta|.
\label{NormalComponentInVortex}
\end{equation}

The integral of this normal density tensor over the cross section of
the soft core gives the Kopnin mass of the vortex in
the local density representation
\begin{eqnarray}
M_{\rm Kopnin}&=& \int d^2r ({\hat {\bf v}}_{\rm L} \cdot \hat  {\bf l})^2  p_{\rm F}^2
N(0,{\bf r}) \nonumber\\
&=& {p_{\rm F}^2 \over 4\pi^2 \Delta_{\rm A}} \int d^2r \sin^3\eta(r)|\partial_r \eta| \;,
\label{KopninMass}
\end{eqnarray}
where ${\hat {\bf v}}_{\rm L}$ is the unit vector in the direction of the vortex velocity
${\bf v}_{\rm L}$ . The same equation for the mass can be obtained  from
Eq.~(\ref{GeneralKopninMass}), using the exact expression for the inter-level
distance $\omega_0(p_z)$ (Volovik 1997).

Since $v_{\rm F}/\Delta_0 \sim \xi$ one obtains that the Kopnin mass of the continuous
vortex  $\sim  \rho \xi R_{\rm sc} $, i.e. it is linear in the dimension $R_{\rm
sc}$ of the core (Kopnin 1995). Thus it follows  that the area law for the
vortex mass is valid only for vortices with a core size of order $\xi$ (i.e. $R_{\rm
sc}\sim \xi$).

Note that the vortex mass discussed above comes from the normal component trapped in
the vortex and which thus moves with the vortex velocity, ${\bf v}_{\rm n}={\bf v}_{\rm L}$. In
this consideration it is assumed that $\omega_0\tau\gg 1$. In this limit the normal
component in the core and the normal component in the heat bath do not interact with
each other and thus can move with different velocities. The local hydrodynamic
energy of the normal component trapped by the vortex is
\begin{equation}
F={1\over 2} (\rho_{\rm n})_{ij}({\bf r})({\bf v}_{\rm L}-{\bf v}_{\rm s})_i({\bf v}_{\rm L}-{\bf v}_{\rm s})_j
\;.
\label{NormalComponentEnergy}
\end{equation}
This can be rewritten in a form, which is valid also for chiral fermions:
\begin{equation}
F={\mu_{\rm R}^2 + \mu_{\rm L}^2\over 8\pi^2} \sqrt{-g}\sqrt{{1\over 2}g^{ij}
g^{kl}F_{ik}F_{jl}} \;,
\label{FermionsHighMagneticField}
\end{equation}
where, as before in Eq.~(\ref{effmu}), the chemical potential of the left and right
fermions in $^3$He-A are expressed in terms of the counterflow: $\mu_{\rm R}=-\mu_{\rm L}= p_{\rm F}
({\hat{\bf l}\cdot}{\bf v})$. Eq.~(\ref{FermionsHighMagneticField}) represents the
magnetic energy of the chiral particles with finite chemical potential in a strong
magnetic field $B\gg \mu^2$.

\subsubsection{Associated hydrodynamic mass}

Recently the problem of another vortex mass of hydrodynamic origin was raised by
Sonin et~al.\ (1998). It is the so-called backflow mass discussed by Baym and Chandler (1983),
which also can be proportional to the core area.  Here we compare these two
contributions in the superclean regime and at low $T\ll T_{\rm c}$ using the model of a
continuous core.  The associated (or induced) mass  appears when, say, an external
body moves in the superfluid. This mass depends on the geometry of the body. For the
moving cylinder of radius $R$ it is the mass of the liquid displaced by the
cylinder,
\begin{equation}
M_{\rm associated}=\pi R^2
\rho ~~,
\label{AssociatedMassCylinder}
\end{equation}
which is to be added to the actual mass of the cylinder to obtain the total inertial
mass of the body. In superfluids this part of the superfluid component moves with
the external body and thus can be associated with the normal component.  A similar
mass is responsible for the normal component in porous materials, in aerogel for
instance, where some part of the superfluid  is hydrodynamically trapped by the
pores. It is removed from the overall superfluid motion and thus becomes part of the
normal component.

In the case when  a vortex is trapped on a wire of radius $R_{\rm wc}\gg \xi$, such
that the wire replaces the vortex core, Eq.~(\ref{AssociatedMassCylinder}) gives the
vortex mass due to the backflow around the moving core.  This is the simplest
realization of the   backflow mass of the vortex discussed by Baym and Chandler (1983).
For this vortex with a wire-core the Baym-Chandler mass is the dominant mass of the
vortex. The Kopnin mass, which can result from normal excitations trapped near the
surface of the wire, is essentially less.

Let us now consider the Baym-Chandler mass for the free vortex at $T=0$, using again
the continuous-core model. In the wire-core vortex this mass arises from the
backflow caused by the inhomogeneity in $\rho_{\rm s}$: $\rho_{\rm s}(r>R_{\rm wc})=\rho$ and
$\rho_{\rm s}(r<R_{\rm wc})=0$. Similar but less severe inhomogeneity of $\rho_{\rm s}=\rho
-\rho_{\rm n}$ occurs in the continuous-core vortex due to the nonzero local normal
density in Eq.~(\ref{NormalComponent}). Owing to the profile of the local superfluid
density, the external flow  is disturbed near the core according to the continuity
equation
\begin{equation}
{\bf \nabla}\cdot (\rho_{\rm s} {\bf v}_{\rm s})=0 ~~.
\label{ContinuityEquation1}
\end{equation}
If the smooth core is large, $R_{\rm sc}\gg \xi$, the deviation  of the superfluid
component inside the smooth core is small from its asymptotic value outside the
core: $\delta \rho_{\rm s}=\rho -\rho_{\rm s} \sim (\xi/R_{\rm sc}) \rho \ll \rho$ and can be
considered as a perturbation. Thus if the asymptotic value of the velocity  of the
superfluid component with respect of the core is ${\bf v}_{s0}=-{\bf v}_{\rm L}$, the
disturbance $\delta {\bf v}_{s}={\bf \nabla}\Phi$ of the superflow in the smooth
core is given by:
\begin{equation}
\rho {\bf \nabla}^2 \Phi=  v_{s0}^i\nabla^j (\rho_{\rm n})_{ij} ~~.
\label{ContinuityEquation2}
\end{equation}
The kinetic  energy of the backflow gives the  Baym-Chandler mass
of the vortex
\begin{equation}
M_{\rm BC}={\rho \over  v_{s0}^2} \int d^2r \left({\bf \nabla}  \Phi \right)^2 ~~.
\label{BCmass1}
\end{equation}
In the simple approximation that the normal component in Eq.~(\ref{NormalComponent})
is considered isotropic,   one obtains
\begin{equation}
M_{\rm BC}={1\over 2 \rho} \int d^2r \rho_{\rm n}^2(r) \sim \rho \xi^2 ~~.
\label{BCmass2}
\end{equation}

The Baym-Chandler mass does not depend on the core radius $R_{\rm sc}$, since the
large area $R_{\rm sc}^2$ of integration in Eq.~(\ref{BCmass2}) is compensated by
the small value of the normal component in the rarefied core, $\rho_{\rm n} \sim \rho
(\xi/R_{\rm sc})$. That is why this mass is   parametrically smaller than the Kopnin
mass in Eq.~(\ref{KopninMass}), if the smooth core is large: $R_{\rm sc}\gg \xi$.

In conclusion, both contributions to the mass of the vortex result from the mass of
the normal component trapped by the vortex.  The difference between Kopnin mass and
Baym-Chandler backflow mass  is only in the origin of the normal component trapped
by the vortex. The relative importance of the two masses depends on the vortex core
structure: (1) For the free continuous vortex with a large core size $R_{\rm sc}\gg
\xi$, the Kopnin mass dominates: $M_{\rm Kopnin} \sim \rho R_{\rm sc}\xi \gg M_{\rm
BC} \sim \rho \xi^2$. (2) For the circulation trapped around a wire of radius
$R_{\rm wc}\gg \xi$, the Baym-Chandler mass is proportional to the core area,
$M_{\rm BC} \sim \rho R_{\rm wc}^2$, and is parametrically larger than the Kopnin
mass. (3) For the free vortex core with a core radius $R\sim \xi$ the situation is
not clear since the continuous core approximation does not work any more. But
extrapolation of the result in Eq.~(\ref{BCmass2}) to $R\sim \xi$ suggests that the
Baym-Chandler mass can be comparable with the Kopnin mass.

\subsubsection{Topology of the energy spectrum: gap nodes and their ramifications}

Zeroes in the fermionic spectrum, such as Fermi surfaces (surfaces of zeroes) and
Fermi points (point zeroes) play an extremely important role in condensed matter and
in analogue models of the low-energy physics of the quantum vacuum
(Volovik 2003). In condensed matter, the gapless fermions interacting with the
Bose  fields of the order parameter lead to the anomalous behavior of superfluids
and superconductors at low temperatures, $T\ll T_{\rm c}$, such as spectral flow in vortex
dynamics, non-analytic behavior of the current and gradient energy, nonlinear and
nonlocal Meissner effect, etc. The counterpart of this behavior in high energy
physics manifests itself in the axial anomaly, baryogenesis, zero charge effect,
running coupling constants, photon mass, etc. It is the zeroes in the fermionic
spectrum, through which the conversion of the vacuum degrees of freedom into that of
matter takes place.

Similar zeroes, but in real space, exist in the cores of topological defects,
especially in quantized vortices (or cosmic strings in high energy physics).
Actually the real-space zeroes and the momentum-space zeroes are described by the
same topology extended to the 8=4+4 dimensional space. For example from the
topological point of view, the Fermi-surface represents the vortex singularity of
the Green's function in the $\omega,{\bf p}$  space, where $\omega$ is the Matsubara
frequency.  The Green's function $G(\omega,p)= 1/(i\omega+v_{\rm F}(p-p_{\rm F}))$ displays a
vortex in the $\omega, p$ plane with the winding number $\nu=1$. This makes the
Fermi surface topologically stable and robust under perturbations of the Fermi
system. Even though the pole in the  Green's function can disappear under some
perturbations, the Fermi surface will survive in the marginal and Luttinger
superfluids. The latter thus belong to the same class of Fermi systems as the Landau
Fermi-liquid.

In the same manner, superfluids and superconductors with a non-vanishing gap behave
in the vicinity of the vortex core like superfluid $^3$He-A in bulk -- point-like
gap nodes appear to be the common feature. Owing to the common topological origin of
the point nodes, the fermions near the gap nodes in gapless Fermi superfluids and
superconductors and the low energy fermions localized in the cores of vortices in
conventional gapped superconductors produce similar anomalous effects.

\section{Final remarks}

We have examined here a few cases in which superfluid $^3$He has been used as an analogue system, to model new concepts in quantum field theory. The focus in Chap.~\ref{ExpChap} is on the question whether the observed vortex formation in a rapid 2nd order phase transition is described correctly by the Kibble-Zurek scaling theory in $^3$He-B. In $^3$He-A vortex formation (Ruutu et~al.\ 1997b) and vortex dynamics (Eltsov et~al.\ 2002) are very different. And so are the measuring techniques which are required to study these features. In fact, the rotating measurements on the KZ mechanism in Chap.~\ref{ExpChap} cannot be repeated in $^3$He-A, because vortex-free flow can be reached only up to $\lesssim 1\,$mm/s while, as seen in Fig.~\ref{CritVel&NeutrThreshold}, the required threshold velocity $v_{\rm cn}$ is at least twice higher. No such fundamental limitations restrict calorimetric measurements on neutron absorption events in $^3$He-A, but present measuring techniques are not sufficiently developed to tackle this question. The consequences from spectral flow and chiral anomaly in the vortex dynamics of especially $^3$He-A have been described in Chap.~\ref{SecOtherAnalogs}. A number of other analogue models involving both $^3$He-A and B are described in the monograph by Volovik (2003).

What is the role of analogue studies in examining the validity of a new concept in physics? If such a test gives a positive result, then good grounds may exist to claim that the concept has fundamental value as a physical model. However, even if a particular model has been shown to work in superfluid $^3$He this does not imply universal applicability. Analogue studies are not a replacement for  cosmological observations or a means to spare investments in high-energy accelerators! For instance, from a condensed matter experiment no direct conclusions can be drawn on the presence or absence of the KZ mechanism and its consequences in the Early Universe. Although superfluid  $^3$He experiments in Chap.~\ref{ExpChap} support the validity of the KZ mechanism, at present this scenario is not a popular explanation for the anisotropy of the cosmic background radiation, which was precipitated by the inhomogeneity of the Early Universe when it had cooled to the point where it became transparent to radiation. Theories based on the presence of an epoch with exponentially accelerated inflationary expansion appear to provide better fits. Nevertheless, defects are formed in abundance in non-equilibrium phase transitions in condensed matter physics and one is left wondering why cosmological transitions should be any different. In this sense it remains an open mystery why defects would not have formed,  or if indeed they were formed, what was their role in shaping the Early Universe.

Irrespective of its cosmological origin, in condensed matter physics the KZ mechanism has today become an important new concept in rapid non-equilibrium phase transitions. It is one of the few intrinsic processes of the bulk material which accounts for defect formation in a wide spectrum of different systems in the absence of boundaries, impurities, or other non-ideal constraints. The strongest experimental support to date for the KZ mechanism comes from the neutron absorption measurements in $^3$He-B where different types of defects are observed to form (usual mass-flow vortices, spin-mass vortices, AB interfaces). In these measurements it is also directly seen that later phase transition fronts, which sweep through the system, interact with the existing defects and that the final outcome depends on the relative velocity of the phase front with respect to the defects. The amount of work performed on the KZ mechanism in the form of simulations, measurements, and analysis by far exceeds other examples of analogue studies. It serves as a good example of what can be achieved with analogue studies in the unification of physical principles.

 \section*{Acknowledgements}

We would like to dedicate this work to Tom Kibble and Wojciech Zurek. We are
indebted to Yu.~Bunkov, A.P. Finne, A.~Gill, H.~Godfrin, H.E.~Hall, J.R.~Hook,
T.~Jacobson, N.~Kopnin, A.~Leggett, Yu.~Makhlin, P.~Mazur, B.~Pla\c{c}ais,
J.~Ruohio, V.~Ruutu, E.~Thuneberg, T.~Vachaspati, G.~Williams, Wen Xu,and X.~Zhang.
Much of the cowork with these and many other colleagues was made possible by the
ULTI visitor programs under the EU Human Capital and Mobility program (contract no.
CHGECT94-0069) and its continuation in the EU program Improving the Human Research
Potential (no. HPRI-CT-1999-00050). This collaborative work has been inspired by the
European Science Foundation programs Cosmology in the Laboratory (COSLAB) and Vortex
Matter in Superconductors at Extreme Scales and Conditions (VORTEX).

\end{document}